Methods and Measures for Analyzing Complex Street Networks and Urban Form

By

Geoffrey D. Boeing

A dissertation submitted in partial satisfaction of the

requirements for the degree of

Doctor of Philosophy

in

City and Regional Planning

in the

Graduate Division

of the

University of California, Berkeley

Committee in Charge:

Professor Paul Waddell, Chair
Professor Robert Cervero
Professor Elizabeth Macdonald
Professor David O'Sullivan

Spring 2017

Methods and Measures for Analyzing Complex Street Networks and Urban Form

Abstract

Methods and Measures for Analyzing Complex Street Networks and Urban Form

by

Geoffrey D. Boeing

Doctor of Philosophy in City and Regional Planning

University of California, Berkeley

Professor Paul Waddell, Chair

Recent years have witnessed an explosion in the science of networks. Much of this research has been stimulated by advances in statistical physics and the study of complex systems – that is, systems that comprise many interrelated components whose interactions produce unpredictable large-scale emergent behavior. Cities are complex systems formed both through decentralized, bottom-up, self-organizing processes as well as through top-down planning interventions. Humans shape their urban ecosystems (the built environment, institutions, cultures, etc.) and are in turn shaped by them. Cities comprise numerous interdependent components that interact through networks – social, virtual, and physical – such as street networks.

This dissertation examines urban street networks, their structural complexity (emphasizing density, connectedness, and resilience), and how planning eras and design paradigms shape them. Interventions into a complex system often have unpredictable outcomes, even if the intervention is minor, as effects compound or dampen nonlinearly over time. Such systems' capacity for novelty, through emergent features that arise from their components' interactions, also makes them unpredictable. These interactions and the structure of connections within a system are the subject of network science. In cities, the structural characteristics of circulation networks influence how a city's physical links organize its human dynamics. Urban morphologists have long studied the built form's complexity and, following from scholars such as Jane Jacobs and Christopher Alexander, various urban design paradigms today speak both directly and indirectly to the value of complexity in the built environment. However, these claims are often made loosely, without formally connecting with theory, implications, or evaluation frameworks.

This dissertation develops an interdisciplinary typology of measures for assessing the complexity of urban form and design, particularly emphasizing street network analytic measures. Street network analysis has held a prominent place in network science ever since Leonhard Euler presented his famous Seven Bridges of Königsberg problem in 1736. The past 15 years have been no exception as the growth of interdisciplinary network science has included numerous applications to cities and their street networks. These studies have yielded new understandings of urban form and design, transportation flows and access, and the topology and resilience of urban street networks. However, current limitations of data availability, consistency, and technology have resulted in four substantial shortcomings: small sample sizes, excessive network simplification, difficult reproducibility, and the lack of consistent, easy-to-use research tools. While these shortcomings are by no means fatal, their presence can limit the scalability, generalizability, and interpretability of empirical street network research.

To address these challenges, this dissertation presents OSMnx, a new tool to download and analyze street networks and other geospatial data from OpenStreetMap for any study site in the world. OSMnx contributes five capabilities for researchers and practitioners: first, the downloading of political boundaries, building footprints, and elevation data; second, the scalable retrieval and construction of street networks from OpenStreetMap; third, the algorithmic correction of network topology; fourth, the ability to save street networks as shapefiles, GraphML, or SVG files; and fifth, the ability to analyze street networks, including projecting and visualizing networks, routing, and calculating metric and topological measures. These measures include those common in urban design and transportation studies, as well as measures of the structure and topology of the network. This study illustrates the use of OSMnx and OpenStreetMap to consistently conduct street network analysis with extremely large sample sizes, with clearly defined network definitions and extents for reproducibility, and using non-planar, directed graphs.

This study collects and analyzes 27,000 U.S. street networks from OpenStreetMap at metropolitan, municipal, and neighborhood scales – namely, every U.S. city and town, census urbanized area, and Zillow-defined neighborhood. It presents wide-ranging empirical findings on U.S. urban form and street network characteristics, emphasizing measures relevant to graph theory, urban design, and morphology such as structural complexity, connectedness, density, centrality, and resilience. We find that the typical American urban area has approximately 26 intersections/km$^2$, 2.8 streets connected to the average node, 160m average street segment lengths, and a network that is 7.4% more circuitous than straight-line streets would be. The typical city has approximately 25 intersections/km$^2$, 2.9 streets connected to the average node, 145m average street segment lengths, and a network that is 5.5% more circuitous than straight-line streets would be. The typical Zillow neighborhood has approximately 46 intersections/km$^2$, 2.9 streets connected to the average node, 135m average street segment lengths, and a network that is 4.4% more circuitous than straight-line streets would be. At all three scales, 3-way intersections are by far the most prevalent intersection type across the U.S.

We find a strong linear relationship, invariant across scales, between total street length and the number of nodes in a network. This contradicts some previous findings in the literature that relied on smaller sample sizes and different geographic contexts. We also find that most networks demonstrate a lognormal distribution of street segment lengths. However, an obvious exception to lognormal distribution lies in those networks that exhibit substantial uniformity network-wide. At the neighborhood scale, examples include downtown neighborhoods with consistent orthogonal grids, such as that of Portland, Oregon. At the municipal scale, examples include towns in the Great Plains that have orthogonal grids with consistent block sizes, platted at one time, and never subjected to sprawl. These spatial signatures of the Homestead Act, successive land use regulations, urban design paradigms, and planning instruments remain etched into these cities' urban forms and street networks today. Nebraska's cities have the lowest circuity, the highest average number of streets per node, the second shortest average street segment length, and the second highest intersection density. These findings illustrate how street networks across the Great Plains developed all at once and grew little afterwards – unlike, for instance, cities in California that were settled in the same era but were later subjected to substantial sprawl.

The characteristics of a city street network fundamentally depend on what "city" means: municipal boundaries, urbanized areas, or certain core neighborhoods? The first is a political/legal definition, but it captures the scope of city planning jurisdiction and decision-making for top-down interventions into a street network. The second captures the wider self-organized human system and its emergent built form, but tends to aggregate multiple heterogeneous built forms together into a single unit of analysis. The third captures the nature of the local built environment and lived experience, but at the expense of a broader view of the urban system and metropolitan-scale trip-taking. In short, multiple scales in concert provide planners a clearer view of the urban form and the topological and metric complexity of the street network than any single scale can.

The emerging methods of computational data science, visualization, network science, and big data analysis have broadened the scope of urban design's traditional toolbox. Such methods may yield new insights and rigor in urban form/design research, but they may also promulgate the weaknesses of reductionism and scientism by ignoring the theory, complexity, and qualitative nuance of human experience crucial to urbanism. The tools we use shape the kinds of questions we can even ask about cities. Today, the dissemination of quantitative network science into the social sciences offers an exciting opportunity to study the dynamics and structure of cities and urban form, but paths forward must consider cities as uniquely human complex systems, inextricably bound up with politics, privilege, power relations, and planning decisions.

This dissertation comprises six substantive chapters bookended by introductory and concluding chapters. As a whole, the dissertation is divided into two primary parts. The first comprises chapters 2 and 3 and develops the theoretical framework. Chapter 2 introduces the background of the nonlinear paradigm by discussing systems, dynamics, self-similarity, and the nature of prediction in the presence of nonlinearity. These foundations set up the complexity theories of cities and the study of networks presented in chapter 3. This first part of the dissertation emphasizes the dynamics of complex urban systems before we turn our attention to their structure in the second part. Urban circulation networks serve as a physical substrate that underlies and organizes the city's complex human interactions. Chapter 4 collates various indicators of complexity from multiple research literatures into a typology of measures of the complexity of urban form, emphasizing the scale of urban design practice. In particular, it presents several measures of network complexity and structure that we then operationalize in chapters 5, 6, and 7. Methodologically, chapter 5 introduces OSMnx, a new tool to acquire, construct, correct, visualize, and analyze complex urban street networks. Chapter 6 applies OSMnx empirically in a small case study of street networks in Portland, Oregon. Chapter 7 then expands the empirical application of OSMnx to a large study of 27,000 urban street networks at various scales across the U.S. These street networks and measures data sets have been shared in a public repository for other researchers to re-purpose.

For Tracy and Alden:
forsan et haec olim meminisse iuvabit

Uneducated people who have no experience of true reality will never adequately govern a city, and neither will people who have been allowed to spend their whole lives in education. The former fail because they do not have a single goal in life at which all their actions, public and private, inevitably aim; the latter because they would refuse to act, thinking they had emigrated, while still alive, to the Isles of the Blessed.

—Plato, *Republic*

# Table of Contents

# List of Figures

# List of Tables

# Acknowledgements

The only thing harder than having to read a dissertation is having to write one. Fortunately, I did not have to write it alone. The entire intellectual process of a doctoral program involves an endless array of stimulating conversations, challenges, critiques, and the development of a broad support network. I can unequivocally say that I would not be here today without the help and support of countless other people.

First and foremost, I need to thank my dissertation committee for their unwavering assistance, guidance, and constructive criticism over the years. Robert Cervero shaped my research methods and approach to studying urban form through our many conversations. Elizabeth Macdonald always used her expertise in urban design and the human experience of the built environment to push my research in more thoughtful and nuanced directions. David O'Sullivan took my inchoate babblings about complexity and cities and directed them into something far more useful and interesting – your arrival at Berkeley could not have come at a better time. Most of all, I owe a deep debt of gratitude to my dissertation chair, Paul Waddell. Doctoral students have no closer or more impactful academic relationship than the one they share with their primary advisor. I have been lucky to have a dissertation chair who is always thoughtful, supportive, and encouraging. Five years ago, I sent you a rambling email inquiring how I might fit in at the Department of City and Regional Planning. Considering how busy you and your inbox tend to be, it remains a miracle that you responded so quickly and took the time to speak with me on the phone and in person. The rest is history. Thank you.

One significant perk of attending UC Berkeley has been my privileged access to a smorgasbord of incredible faculty and courses. In particular, my experience here was shaped by courses taught by Dan Chatman, Ananya Roy, Steven Weber, Laurel Larsen, and the professors on my committee. I am also deeply indebted to the conversations, critiques, and advice I have received from Karen Frick, Nicholas de Monchaux, Carolina Reid, Karen Chapple, Daniel Rodríguez, Charisma Acey, and Teresa Caldeira. Finally, teaching a subject is the best way to learn it (or so the old cliché goes), and one of my most cherished experiences at Berkeley has been teaching Urban Informatics with Paul Waddell over the past few years. My dissertation could not have happened without the skills I developed giving these lectures and holding office hours. I am grateful to each of my students for forcing me to figure out the why's and the how's until I could really explain them.

While we all like to kvetch about the lonely, soul-crushing experience of grad school, I truly had no such experience. I remain eternally grateful for the friends I have made over the years at Berkeley: Matt Wade, Jesus Barajas, Andrea Broaddus, Lizzy Mattiuzzi, Jake Wegmann, Sophie Gonick, Nicola Szibbo, Troy Reinhalter, Fletcher Foti, Aksel Olsen, Keith Lee, Alice Sverdlik, Heather Arata, Pedro Peterson, Ariel Bierbaum, Nat Decker, Dave Amos, Rocio Sanchez-Moyano, Sam Maurer, Chris Mizes, Aaron Young, Lana Salman, my cohort-mates – Mukul Kumar, Julia Tierney, Lisa Rayle, and Miriam Solis – and the many, many others I need to mention here but cannot because I have to get this dissertation filed and it would take far too long to list you all. Your intellects, humility, and willingness to engage in endless conversation have served as both a scholarly model and a welcome refuge. You made Wurster Hall my home.

Six years ago, I was sitting on my couch in San Diego with Joe Smyser and Ernesto Ramirez. We were playing Super Nintendo (probably NBA Jam), surrounded by dozens of empty beer bottles, and talking about the lives of doctoral students. Without your insights and encouragement, I surely would have never returned to school. Thank you for pushing me toward life's next stage. But then again, if not for you two, I would probably still be hanging out on the beach all day in San Diego instead of sitting here writing a dissertation all day in Berkeley. But with that in mind, I do need to thank San Diego and Berkeley, as well as Rancho Cucamonga, Phoenix, Portland, Brooklyn, and London: the cities I have lived in set me directly on this path to becoming an urban scholar.

Lastly, I must thank my family. My parents, Dennis and Rosemary, and my brothers, John and Nathan – you spent years shaping me into who I am now, and I would not be here today if not for the trajectory you placed me on in childhood. Tracy – no one could ask for a more patient and loving partner. Thank you for helping me be me, and for being so tolerant when I lock myself in my room to work on the dissertation all night. A saner person would have long since fled from my absurdities, so thank you for liking the weird. And of course, thank you for having an awesome job and earning all our money over the past five years. Love you. Alden – thanks for being a good eater and sleeper. You arrived just in time to sneak into these acknowledgments, but the drudgery of endlessly writing and rewriting a dissertation is so much more tolerable when I have you to play with the rest of the time. Stay-at-home parenting has been more fun than I ever would have imagined. Thanks buddy.

This dissertation research was supported by funding from the University of California Regents' Fellowship, the Doctoral Completion Fellowship, and additional grants from the UC Berkeley Graduate Division. I could not have finished this project without the support of these programs. The Bay Area, it turns out, is a pretty expensive place to live.

# Chapter 1: Introduction

## 1.1. Abstract

This chapter introduces the context of and motivation for the study presented in this dissertation. It then summarizes the organization and contribution of each of the subsequent chapters. Complex systems have been widely studied by social and natural scientists in terms of their dynamics and their structure. Scholars of cities and urban planning have incorporated complexity theories from qualitative and quantitative perspectives. From a structural standpoint, the urban form may be characterized by the morphological complexity of its circulation networks – particularly their density, resilience, centrality, and connectedness. This dissertation unpacks theories of nonlinearity and complex systems, then develops a framework for assessing the complexity of urban form and street networks. It introduces a new tool, OSMnx, to collect street network and other urban form data for anywhere in the world, then analyze and visualize it. Finally, it presents a large empirical study of 27,000 street networks, examining their metric and topological complexity relevant to urban design, transportation research, and the human experience of the built environment.

## 1.2. Introduction

In his twelfth epistle, published in 20 BC, the ancient Roman poet Horace asked: *quid uelit et possit rerum concordia discors*? His question – which wonders at the purpose and power of the world's "discordant harmony" – hints at the doctrine of the Greek philosophers Pythagoras, Heraclitus, and Empedocles (Gordon 2007). The philosophy underlying *concordia discors* held that the cosmos was shaped by an endless struggle between nature's elements, and that out of this ongoing dissonance arose the comprehensible order of the perceivable world (Scholtz 2008). Horace's turn of phrase lives on, borrowed throughout history to refer to large-scale harmony, structure, and function emerging unpredictably from smaller-scale disordered interactions (Stegenga 2012). Today, in the scientific study of large interacting systems and their rich emergent behavior, *concordia discors* goes by another name: complexity.

*Complexity* is perhaps most simply expressed by the familiar phrase, "the whole is greater than the sum of its parts." Complex systems comprise many interrelated parts and – through nonlinear interactions and feedback – can adapt, become resilient to perturbations and system shocks, and evolve large-scale, emergent phenomena that could not have been predicted or understood simply by examining the system's interdependent subcomponents. There is no single, unified complexity theory but rather a wide array of theoretical concepts and tools that can be applied to the study of complex systems across numerous disciplines (Manson and O'Sullivan 2006; Haken 2012).

Scholars argue that complexity's comprehensive framework can help link quantitative *space* studies with qualitative *place* studies and thus has significant implications for city planning (Portugali 1999; 2006). It problematizes prediction and situates uncertainty – of knowledge, interventions, and forecasts – at the center of studying systems. Complexity calls for a reset of positivism and a new world view that embraces uncertainty and unpredictability (de Roo 2010a; 2010b; Silva 2010). The urban complexity literature argues that cities are shaped through bottom-up, self-organizing processes as well as top-down planning interventions that are analogous to perturbations of the complex urban system (e.g., Bretagnolle et al. 2009; Barthélemy et al. 2013; Mansury 2015). However, the relationships between top-down and bottom-up – and complexity and simplicity – are not simply binary but rather a spectrum on which arguments and practices may be situated.

With regards to city planning, complexity has implications for rationality, predictability, uncertainty, optimization, and collaboration. It offers another way to understand and examine the processes and patterns of urban form and social systems. It also presents a role for planning through debates around bounded rationality, externalities, and social justice. Complexity theorists argue that self-organized emergence does not necessarily lead to desirable features – planning can be thought of as a top-down perturbation of the complex system, guiding it toward desirable system attractors (Portugali 1999; Webster 2010; Allen 2012; Marshall 2012a; Uitermark 2015). Complex systems tend to exhibit leverage points that offer targeted intervention opportunities, though their existence is not always apparent (Meadows 2008).

As discussed in the upcoming chapters, many concepts from complexity studies have been applied to cities and planning – some metaphorically, others more deeply – including bifurcation, basins of attraction and alternative stable states, self-organization, emergence, equifinality, feedback, limit cycles and attractors, fractals and spatial pattern formation, resilience, path dependence, self-organized criticality, and complex networks (Sengupta et al. 2016; Batty 2017). The next section introduces this conceptual context.

## 1.3. Context of the Study

Perhaps the most effective instances of complexity percolating into urban planning scholarship have focused on understanding processes and patterns and informing practice. Unfortunately, the flip side of this has too often been an emphasis on simply demonstrating how predictions and planning itself may be ill-suited to accomplishing their goals, without providing practical, politically-feasible guidelines for how society should proceed. Thus, this dissertation proposes that the most important elements of complexity theory for informing planning scholarship are those that blend the *pragmatic* with the *theoretical* – that is, elements that offer 1) a useful toolkit for empirical research to inform and aid practicing planners and 2) a theoretical lens to re-conceptualize oversimplifications of cities.

Complex network analysis is one such example. For instance, Barthélemy's work (2011; Barthélemy et al. 2013) provides straightforward ties to planning practice by examining historical road networks and using novel methods to analyze their topology over time. In turn, complexity's lessons for urban design are important for informing planning scholarship. Complexity is commonly invoked in terms of livability, but the ends and the means do not always conform to a robust understanding of the implications of complex systems. Marshall (2012a) points out that the desirable complexity of traditional cities may not be best served by attempts to mimic it through the large-scale, top-down master planning sometimes embraced by movements like the New Urbanism (see also Ellis 2002; Rodriguez et al. 2006; Banai and Rapino 2009). Complexity theory may assist in the ongoing reconciliation of the aims of such design with their means.

Complexity theory can also inform future planning scholarship through its critique of certainty. The perceived infallibility of planners and their latest ideas resulted in countless urban planning disasters during the twentieth century (Hall 1982; Scott 1998). Complexity calls instead for flexibility, small steps, collaboration, and a planning philosophy aimed at creating organic urban ecosystems for the way humans actually live. This is a critical connection between complexity and planning scholarship as it provides a useful grounding framework for safe, pleasant, equitable, and enjoyable urban environments. Scholarly discussions of urban form and street networks have long used the vocabulary of complexity, with theorists such as Jane Jacobs (1961) – who famously argued for urban "organized complexity" – and Christopher Alexander (1965; cf. Harary and Rockey 1976; Levinson and Huang 2012; Marshall 2012b) – who famously argued that "a city is not a tree" – serving as intellectual forerunners to today's urban complexity theorists.

Urban design theorists have long considered the neighborhood in the context of complexity, bottom-up organicism, and top-down intervention. Since the earliest days of cities, neighborhoods would form organically around important points like temples (for discussions of the "organic" analogy, see Herbert 1963 and Marshall 2008). These landmarks were often situated by central authorities but neighborhoods would self-organize around them later (Mumford 1961; cf. Braun and Hogenberg 2011). Following Weaver (1948), Jacobs (1961) suggested that cities are problems of *organized complexity* and embraced the organic agglomeration, diversity, and proximity of such traditional neighborhoods. Rather than designing monolithic single-use zones, she contended that planners should provide what is lacking in a neighborhood to maximize diversity. Likewise, Alexander's pattern language presented flexible frameworks for diversity, opportunities for social mixing, and other complexity-flavored principles (Alexander et al. 1977; 2002; 2008; Vitins and Axhausen 2014; Park 2015). His patterns' descendants, such as form-based codes, attempt to regulate urban form by balancing bottom-up flexibility (i.e., emergence and self-organization) with top-down predictability (Talen 2011).

Related to these theories of design and complexity, researchers have long probed the complex nature of the transportation-land use connection, one of the most studied relationships in the planning literature (Ewing and Cervero 2010). Transportation options and investments influence land use/urban form (and vice versa) through a complex set of feedback loops, individual agents' ever-shifting preferences, hysteresis, path dependence, and nonlinear dynamics. Major top-down planning interventions into the transportation-land use connection include neighborhood supply controls (Levine 2006), parking policy and feedback loops for automobility (Shoup 2002), and the decision to price congestion and/or build more roads (Downs 2004; Cervero 2013).

Marshall (2012a) claims that such interventions can be conceptualized as an artificial selection used for the public good to overrule the natural selection of the market which may prioritize individual utility at the expense of public utility. Market distortions and externalities indicate a role for planners (Dahlman 1979; Adams and Tiesdell 2010; Holcombe 2013). In effect, these interventions seek to perturb an urban ecosystem that has emerged out of some combination of decentralized bottom-up organization and top-down decision-making. Boarnet (2011) suggests that researchers have not thoroughly studied how these different types of policy might interact nonlinearly to produce amplified or diminished effects – a key trait of complex systems.

At the intersection (no pun intended) of the study of transportation and urban form and design lies our cities' circulation networks. This dissertation presents new methods for assessing the complexity of the urban built form, particularly through its street networks. Urban street networks are complex spatial networks that evolve through planning decisions and self-organization, and in turn shape human connections and interactions within the city. This study is situated between urban form and design, street network analysis, and complexity studies. It presents new techniques and measures for collecting and analyzing street networks through the lens of complexity – particularly focusing on density, connectedness, and resilience. These structural attributes of the urban form influence the human interactions and dynamics that play out through space along the network.

## 1.4. Motivation

Street network analysis has become prevalent in the past few years in the urban planning and transportation literature. Some studies have focused on the urban form, others on transportation and flow, and others on the topology, complexity, and resilience of street networks. However, the current literature suffers from some shortcomings, discussed in detail in chapter 5 but summarized briefly here.

First, the sample sizes in cross-sectional studies tend to be quite small, typically around 5 to 50 networks, yet these studies often make claims of generalizability to cities at large. Second, studies usually simplify the representation of the street network to a planar or undirected primal graph for tractability. This may be both unnecessary and undesirable, as we shall discuss. Third, the dozens of decisions that go into analysis – such as spatial extents, topological simplification and correction, definitions of nodes and edges, etc. – are often ad hoc or only partly reported, make reproducibility challenging. Fourth, the current landscape of tools and methods offers no ideal technique that balances usability, customizability, reproducibility, and scalability in acquiring, constructing, and analyzing network data.

To address these challenges, this study's primary methodological contribution is OSMnx, a new research platform developed by this author to download political boundary geometries and street network data from OpenStreetMap, then construct, project, analyze, map, and visualize the networks. This functionality is discussed in section 1.5.5 and in detail in chapter 5. This new research tool offers scholars, designers, and engineers the ability to analyze street networks, calculate routes, project and visualize the networks, and calculate network metrics and statistics. These metrics and statistics include both those common in urban design and transportation studies, as well as complexity measures of the structure and topology of the network. In particular, this dissertation situates these methods in the context of complexity. It aims to democratize and disseminate the application of advanced complex network theoretic measures to social scientists and urban designers through a simple intuitive tool. It also seeks to make these studies reproducible by formalizing and simplifying the many ad hoc decisions that went into network acquisition and analysis in the past. Finally, it addresses the longstanding sample size limitation by conducting a preliminary empirical study that explores trends in the structure of 27,000 U.S. street networks at multiple scales.

## 1.5. Organization and Contribution by Chapter

This dissertation begins and ends with introductory and concluding chapters that book-end its six central substantive chapters. These six chapters unpack the foundations of the nonlinear paradigm, contextualize urban street network analysis within theories of complexity, create a typology for measuring the complexity of urban form, present a new method for acquiring, constructing, analyzing and visualizing street networks, and conduct a multi-scale analysis of urban street networks across the United States. These chapters constitute the two primary parts of the dissertation. The first is a deep dive into nonlinearity, complex systems, network analysis, and complexity theories of cities. It focuses on dynamics and process, but suggests a bridge to structure. The second builds on this foundation to explore applications of street network analysis methodologically and empirically.

### 1.5.1. Chapter 1 – Introduction

This chapter has thus far introduced the motivation for and context of the study presented in this dissertation. The remainder of this chapter summarizes the organization and contribution of each of the subsequent chapters in this dissertation.

### 1.5.2. Chapter 2 – Foundations of the Nonlinear Paradigm

Chapter 2 provides a background for the rest of the dissertation and has two primary aims. First it lays the foundation underlying the complexity theories of cities presented in chapter 3 by introducing the fundamentals of nonlinear dynamics, chaos, fractals, self-similarity, and the limits of prediction. It does so in an interdisciplinary way through several visualization methods to analyze and understand system behavior. Second it presents Pynamical, a new tool developed by this author to visualize and explore nonlinear dynamical systems' behavior. Nearly all nontrivial real-world systems are nonlinear dynamical systems. The modern study of complexity partly evolved from initial explorations of the surprising behavior of such systems. Although the social sciences are increasingly studying society and cities through this lens, seminal concepts remain murky or loosely adopted in the literature.

This chapter introduces systems, dynamics, self-similarity, and prediction to set up the discussion in chapter 3 of complexity, cities, and the study of networks. It makes two primary contributions: one theoretical, one methodological. First, it reviews the qualitative analysis of nonlinear dynamical systems' behavior for an interdisciplinary body of urban scholars and planners. Most formal treatments of chaos and nonlinear dynamics in the scholarly literature are densely technical and geared toward an audience of mathematicians and physicists. For this chapter, rather, readers require only a familiarity with algebra. Second, this chapter makes a methodological contribution by presenting Pynamical, a new tool developed by this author as part of this study, to visualize and explore nonlinear dynamical systems' behavior. Comparable tools usually must be developed from scratch or rely on expensive commercial software such as MATLAB. Developing tools for exploring, understanding, and visualizing dynamical systems in Python makes them available to a wider audience of systems analysts, researchers, and students. Pynamical provides a fast, simple, reusable, extensible, free, and open-source new means for exploring system behavior – particularly for the qualitative analysis of such systems in research and pedagogy.

### 1.5.3. Chapter 3 – Complexity and Cities

Building on the background of chapter 2, chapter 3 presents the theoretical framework of complex systems and cities, culminating in network theory and analysis – the primary lens this study uses in all subsequent chapters. Discussions of complexity and complex systems have appeared throughout the planning literature for years. These principles have been applied everywhere from the communicative turn and collaborative rationality, to cellular automata and agent-based urban models, to the design of resilient, livable neighborhoods. However, the interdisciplinary appeal and trendiness of complexity in the social sciences has resulted in a morass of ambiguous terminology, internal inconsistencies, and overloaded concepts open to multiple interpretations.

Unlike the other substantive chapters in this dissertation, this chapter makes neither an empirical nor a methodological contribution. However, it offers a theoretical contribution to the urban planning literature by unpacking the key foundational concepts of complex systems and network science in a straightforward manner. It provides explanatory examples of these concepts familiar to scholars and practitioners not already versed in the technical science of complexity. Most relevant to this present study, this chapter presents the theory of networks and the methods of network analysis that form the foundation of the remaining chapters. In doing so, it addresses the transition from focusing on dynamics to focusing on structure, and suggests a bridge between the two.

### 1.5.4. Chapter 4 – Measuring the Complexity of Urban Form and Design

Building on the theories of complexity and networks presented in chapter 3, chapter 4 develops a typology of measures for assessing the complexity of the urban built form. In particular, it extends quantitative methods from network science, ecosystems studies, fractal geometry, information theory, and urban planning to the practice of neighborhood-scale urban design and the analysis of its qualitative human experience. Metrics at multiple scales are scattered throughout these bodies of literature and have useful applications in analyzing the built form that results from local planning and design processes. Rich linkages between complexity theory and urban design have been underexplored by researchers at the neighborhood and street scales – the scales of daily human experience. The urban design literature frequently cites the value of *complexity* in neighborhood design, but these arguments often lack the formalism found in complex systems science. If neighborhood complexity is considered important, how might we interpret it and how might it be assessed?

This chapter unpacks the connections between neighborhood-scale built form and measures of its complexity. It contributes a new typology of tools and metrics from different scientific disciplines to assess measures of complexity that apply to urban form and particularly at urban design's scale of intervention. In particular, the measures of network structure characterize the complexity of the circulation network in terms of density, resilience, and connectedness. These attributes influence the way an urban system's physical links can structure complex interactions and dynamics. The analytical framework developed here is generalizable to empirical research of multiple neighborhood types and design standards. In particular, network-analytic measures in this typology are operationalized in the next chapter, and applied empirically in the subsequent two empirical chapters.

### 1.5.5. Chapter 5 – Acquiring, Analyzing, and Visualizing Street Networks

Scholars have studied street networks in various ways. However, there are some limitations to the current urban planning/street network analysis literature. To address these challenges, this study presents a new tool developed by this author as part of this study, to make the collection of data and creation and analysis of street networks simple, consistent, and automatable. OSMnx is a new Python package that downloads political boundary geometries, street networks, and building footprints from OpenStreetMap.

OSMnx contributes five significant new capabilities for researchers and city planners: first, the automatic downloading of place boundaries and building footprints; second, the tailored and automated downloading and construction of street networks from OpenStreetMap; third, the automated correction and simplification of network topology;

fourth, the ability to save street networks to disk as shapefiles, GraphML, or SVG files; and fifth, the ability to analyze street networks, calculate routes, project and visualize the networks, and calculate network metrics and statistics. These metrics and statistics include both those common in urban design and transportation studies, and metrics that measure the structure and topology of the network.

This chapter makes two primary methodological contributions. First, it presents new methods and tools for acquiring, constructing, correcting, projecting, analyzing, and visualizing street networks. Second, it adapts measures from traditional network analysis to make them better-suited to accurately describing the physical form of street intersections and network connectivity.

### 1.5.6. Chapter 6 – Case Study: Portland, Oregon

This short chapter presents a small case study to simply but plainly demonstrate the use of OSMnx for research. It collects three small half-kilometer sections of the street network in different neighborhoods in Portland, Oregon to perform a cross-sectional analysis. The scale of analysis and sample size are small, but they provide simple, comprehensible examples to illustrate the network concepts presented in chapter 3, the network measures presented in chapter 4, and the methodological tool presented in chapter 5. This chapter thus serves to tie these preceding threads together empirically.

Accordingly, this chapter has two aims. First, as discussed, it demonstrates the functionality of OSMnx with a simple case study. Second, it presents empirical findings of three street networks in Portland, Oregon and uses the quantitative measures to compare and contrast these network sections. It introduces these neighborhoods first from a brief qualitative and historical perspective, then explores their comparative quantitative measures of network complexity and structure. Then it discusses these empirical findings and insights that may be drawn from them. It identifies significant chokepoints in the suburban network and demonstrates how there could be substantial gains in network resilience if one-way streets in the dense, orthogonal downtown were converted to two-way streets. This small case study demonstrates OSMnx simply before embarking on the large multi-scale analysis of 27,000 street networks in the next chapter.

### 1.5.7. Chapter 7 – A Multi-Scale Analysis of Urban Street Networks

Following on the small case study in chapter 6, chapter 7 presents a large multi-scale analysis of 27,000 street networks. The empirical literature on street network analysis is growing ever richer, but suffers from some limitations. First, sample sizes tend to be fairly small due to data availability, gathering, and processing constraints. Second, reproducibility is difficult when the dozens of decisions that go into analysis – such as spatial extents, topological simplification and correction, definitions of nodes and edges, etc. – are ad hoc or only partly reported. Third, and related to the first two, studies frequently oversimplify to planar or undirected primal graphs for tractability, or use dual graphs despite the loss of geographic and metric information. Fourth, the current landscape of tools and methods offers no ideal technique that balances usability, customizability, reproducibility, and scalability in acquiring, constructing, and analyzing network data.

This fourth limitation above was addressed by introducing OSMnx and demonstrating its use in a small case study of Portland, Oregon in chapters 5 and 6. Chapter 7 addresses the first three limitations by conducting an analysis of street networks at multiple scales, with large sample sizes, with clearly defined network definitions and extents for reproducibility, and using non-planar, directed graphs. In particular, it examines urban street networks – represented as primal, non-planar, weighted multidigraphs with possible self-loops – through the framework of complexity developed in chapters 3 and 4, focusing on structure, density, connectedness, centrality, and resilience.

Most studies in the street network literature that conduct topological and/or metric analysis tend to have sample sizes ranging around 5 to 50 networks. This chapter instead conducts a large analysis of 27,000 urban street networks at multiple overlapping scales across the United States. Namely, it examines the street networks of every U.S. incorporated city and town, urbanized area, and Zillow-defined neighborhood. In total, we use OSMnx to download, construct, and analyze 497 urbanized areas' street networks, 19,655 cities' and towns' street networks, and 6,857 neighborhoods' street networks. It uses these street networks to conduct four analyses: at the metropolitan scale, at the municipal scale, at the neighborhood scale, and a case study looking deeper at the neighborhood-scale street networks in the city of San Francisco, California.

This chapter presents preliminary empirical findings that emphasize street network complexity in terms of density, resilience, and connectedness. The orthogonal grid we see in the downtowns of Portland and San Francisco have high density (i.e., intersection and street densities), connectedness (i.e., average number of streets per node), and order (based on circuity and statistical dispersion of node types), but low resilience in the presence of one-way streets, measured by maximum betweenness centralities and average node connectivity increases when switching from one-way to bi-directional edges. Sprawling, disconnected suburban neighborhoods rank low on all measures of complexity, with the exception that their high circuity can lend itself to disorder. This discussion argues that street networks can be complex either inherently because of their form, or indirectly through how that form structures human dynamics. This shaping of human interactions and connections by the urban form links the theory of dynamical complexity in the early chapters with the empirical analysis of form and structural complexity in the latter chapters.

We also find that *scale* is critically important in analyses of street networks. However, invariant to scale, we find a strong linear relationship between total street length and the number of nodes in a network. This provides new evidence that contradicts some previous findings in the literature that relied on purely theoretical models or small sample sizes. We also find that most networks empirically demonstrate a lognormal distribution of street segment lengths. An obvious exception to lognormal distribution lies in those networks that exhibit substantial uniformity across the entire network, such as the consistent orthogonal grid of downtown Portland, Oregon. At the municipal scale, towns in the Great Plains typically have orthogonal grids with consistent block sizes, platted at one time, and never subjected to sprawl. Similarly, comparing median street networks of each state, Nebraska has the lowest circuity, the highest average number of streets per node, the second shortest average street segment length, and the second highest intersection density for similar reasons. Various spatial signatures of bygone planning instruments and design paradigms remain etched into the urban form and street networks of cities across the United States through path dependence – a hallmark of complex systems.

Finally, this study has made these network datasets and their attribute datasets available in a public online repository for other researchers to study and re-purpose (see Appendix).

### 1.5.8. Chapter 8 – Conclusion

The dissertation concludes with a brief synopsis of the study, a summary of its key findings, a discussion of their contribution to the academic literature and to planning practice, and potential trajectories for future research. In particular, future work remains in exploring the links between structural complexity and the complexity of human dynamics. OSMnx provides opportunities to do so by providing a rich new basket of built form variables to model individual and collective human behavior through urban circulation networks.

# Chapter 2: Foundations of the Nonlinear Paradigm

## 2.1. Abstract

Nearly all nontrivial real-world systems are nonlinear dynamical systems. Chaos describes certain nonlinear dynamical systems that have a sensitive dependence on initial conditions. Chaotic systems are always deterministic and may be very simple, yet produce completely unpredictable and divergent behavior. The modern study of complex systems evolved from these initial explorations, and although the social sciences are increasingly studying these types of systems, seminal concepts remain murky or loosely adopted. This chapter has two primary aims. First it introduces the foundations of nonlinear dynamics, chaos, fractals, self-similarity, and the limits of prediction through several visualization methods to analyze and understand system behavior. Second it presents Pynamical, a new tool that created by this author as part of this study, to visualize and explore nonlinear dynamical systems' behavior.

## 2.2. Introduction

Chaos theory is a branch of mathematics that deals with nonlinear dynamical systems. A *system* is a set of interacting components that form a larger whole. *Nonlinear* means that this system's outputs are disproportional to its inputs: due to feedback or multiplicative effects between the components, the whole becomes something greater than the mere sum of its individual parts. Lastly, *dynamical* means the system changes over time based on its current state. Nearly every nontrivial real-world system is a nonlinear dynamical system. Chaotic systems are a type of nonlinear dynamical systems that may contain very few interacting parts and may follow simple rules, but all have a sensitive dependence on their initial conditions (Hastings et al. 1993; Rickles et al. 2007).

One might expect that any simple deterministic system would produce easily predictable behavior. Yet despite their deterministic simplicity, over time these systems can produce wildly unpredictable, divergent, and fractal (i.e., infinitely detailed and self-similar without ever actually repeating) behavior due to that sensitivity. Forecasting such systems' futures thus requires an impossible precision of measurement and computation. Chaos fundamentally indicates that there are limits to knowledge and prediction because some futures may be unknowable with any precision. Further, interventions into a system may have unpredictable outcomes even if the intervention is minor, as tiny effects can compound (or be damped) nonlinearly over time.

Real-world chaotic and fractal systems span the spectrum from leaky faucets (Suetani et al. 2012), to plants (Singh et al. 2012; Walker 2012), to heart rates (Glass 2009; Hoshi et al. 2013; Babbs 2014), to cryptography (Hong and Dong 2010; Makris and Antoniou 2012). Many scholars have studied the implications of nonlinearity, chaos, and fractals for the social sciences, including sociology (Richards 1996; Guastello 2013), urban studies (Batty 1991; 2008b; Batty and Longley 1994; Batty and Xie 1999; Benguigui et al. 2000; Shen 2002; Chen and Zhou 2008), economics (Rosser 1996; Oxley and George 2007; Chen 2008; Guégan 2009; Puu 2013), architecture (Hamouche 2009; Ostwald 2013), and city planning (Cartwright 1991; Innes and Booher 2010; Batty and Marshall 2012; Batty 2013c; Chettiparamb 2014; Narh et al. 2016).

One constant throughout the interdisciplinary history of nonlinear dynamical systems study is that nonlinear systems are extremely difficult to solve analytically because they cannot be broken down into constituent parts, solved individually, then recombined as a solution (Hofstadter 1985). Scientists have instead relied heavily on visual and qualitative approaches – a perspective first developed by Henri Poincaré in the late 1800s – to discover and analyze the dynamics of nonlinearity (Alpigini 2004; Layek 2015). Information visualization helps analysts detect and examine hidden structure in complex data sets (Chen 2006). In particular, few fields have drawn as heavily from visualization as nonlinear dynamics and chaos have for their pivotal discoveries, from Lorenz's first visualization of strange attractors (Lorenz 1963), to May's groundbreaking bifurcation diagrams (May 1976), to phase diagrams for discerning higher-dimensional hidden structures in data (Packard et al. 1980). Such nonlinear analysis is particularly useful, yet underutilized, for exploring time series (Bradley 2003; Bradley and Kantz 2015). These methods in turn have broad applicability to visual information analysis and the interdisciplinary study of nonlinear and complex systems.

This chapter introduces nonlinearity through the methods of data visualization, using a logistic model to dissect the terminology, illustrate pertinent features of chaos and fractals, and discuss wide-ranging implications for knowledge and prediction. It has two primary aims. First, it introduces the foundations of nonlinear dynamics, chaos, fractals, self-similarity, and the limits of prediction. Although the social sciences are increasingly studying these types of systems, some of the seminal concepts remain murky or loosely adopted in the theoretical literature (Chettiparamb 2006). Most *formal* treatments of chaos and nonlinear dynamics in the scholarly literature are densely technical and geared toward an audience of mathematicians and natural scientists. For this chapter, rather, readers require only a familiarity with algebra. We thus do not cover the rigorous mathematical underpinnings of chaos and nonlinear dynamics, but the references throughout cite both the original foundational publications in this field as well as recent scholarly developments. Second, this chapter presents Pynamical, a new tool created by this author to visualize and explore nonlinear dynamical systems' behavior. Comparable tools usually must be developed from scratch or rely on expensive commercial software such as MATLAB (Tomida 2008). Pynamical provides a fast, simple, reusable, extensible, free, and open-source new means for exploring system behavior – particularly for the qualitative analysis of such systems in research and pedagogy.

The following section provides a background to the logistic map and the concepts of system dynamics and attractors. Then we introduce several information visualization techniques to explore qualitative system behavior, bifurcations, the path to chaos, fractals, and strange attractors. We investigate the difference between chaos and randomness. Finally, we visualize the famous butterfly effect and conclude with a discussion of its implications for scientific prediction and complexity. All of these models and visualizations are developed in Python using Pynamical; for readability, we reserve the technical details of its functionality for the discussion section and appendix.

## 2.3. Background and Model

The meteorologist Edward Lorenz is widely considered the father of chaos theory (Stewart 2000). Danforth (2013) relates an anecdote in which Lorenz describes chaos as "when the present determines the future, but the approximate present does not approximately determine the future." Lorenz first discovered chaos by accident while developing a simple mathematical model of atmospheric convection, using three ordinary differential equations (Lorenz 1963). He found that nearly indistinguishable initial conditions could produce completely divergent outcomes, rendering weather prediction impossible beyond a time horizon of about a fortnight (Gleick 1991).

How can this possibly happen with a simple deterministic system? We will explore an example using the *logistic map*, a model based on the common *s*-curve logistic function that shows how a population grows slowly, then rapidly, before tapering off as it reaches its environment's carrying capacity (May 1974; Li et al. 2011). The logistic function uses a differential equation that treats time as continuous. The logistic map instead uses a difference equation to look at discrete time steps (Pastijn 2006; Strogatz 2014). It is called the logistic *map* because it maps the population value at any time step to its value at the next time step: $x_{t+1} = r\,x_t\,(1-x_t)$. This nonlinear equation defines the rules, or *dynamics*, of our system: $x$ represents the population at some time $t$, and $r$ represents the growth rate. Thus, the population level at any given time is a function of the growth rate parameter value and the previous time step's population level. If the growth rate is set too low, the population will die out and go extinct. Higher growth rates might settle toward a stable value or fluctuate across a series of population booms and busts.

| Generation | $r = 0.5$ | $r = 1.0$ | $r = 1.5$ | $r = 2.0$ | $r = 2.5$ | $r = 3.0$ | $r = 3.5$ |
|---|---|---|---|---|---|---|---|
| 1 | 0.500 | 0.500 | 0.500 | 0.500 | 0.500 | 0.500 | 0.500 |
| 2 | 0.125 | 0.250 | 0.375 | 0.500 | 0.625 | 0.750 | 0.875 |
| 3 | 0.055 | 0.188 | 0.352 | 0.500 | 0.586 | 0.562 | 0.383 |
| 4 | 0.026 | 0.152 | 0.342 | 0.500 | 0.607 | 0.738 | 0.827 |
| 5 | 0.013 | 0.129 | 0.338 | 0.500 | 0.597 | 0.580 | 0.501 |
| 6 | 0.006 | 0.112 | 0.335 | 0.500 | 0.602 | 0.731 | 0.875 |
| 7 | 0.003 | 0.100 | 0.334 | 0.500 | 0.599 | 0.590 | 0.383 |
| 8 | 0.002 | 0.090 | 0.334 | 0.500 | 0.600 | 0.726 | 0.827 |
| 9 | 0.001 | 0.082 | 0.334 | 0.500 | 0.600 | 0.597 | 0.501 |
| 10 | 0.000 | 0.075 | 0.333 | 0.500 | 0.600 | 0.722 | 0.875 |
| 11 | 0.000 | 0.069 | 0.333 | 0.500 | 0.600 | 0.603 | 0.383 |
| 12 | 0.000 | 0.065 | 0.333 | 0.500 | 0.600 | 0.718 | 0.827 |
| 13 | 0.000 | 0.060 | 0.333 | 0.500 | 0.600 | 0.607 | 0.501 |
| 14 | 0.000 | 0.057 | 0.333 | 0.500 | 0.600 | 0.716 | 0.875 |
| 15 | 0.000 | 0.054 | 0.333 | 0.500 | 0.600 | 0.610 | 0.383 |
| 16 | 0.000 | 0.051 | 0.333 | 0.500 | 0.600 | 0.713 | 0.827 |
| 17 | 0.000 | 0.048 | 0.333 | 0.500 | 0.600 | 0.613 | 0.501 |
| 18 | 0.000 | 0.046 | 0.333 | 0.500 | 0.600 | 0.711 | 0.875 |
| 19 | 0.000 | 0.044 | 0.333 | 0.500 | 0.600 | 0.616 | 0.383 |
| 20 | 0.000 | 0.042 | 0.333 | 0.500 | 0.600 | 0.710 | 0.827 |

Table 2.1. Population values produced by the logistic map over 20 generations with 7 different values of the growth rate parameter $r$.

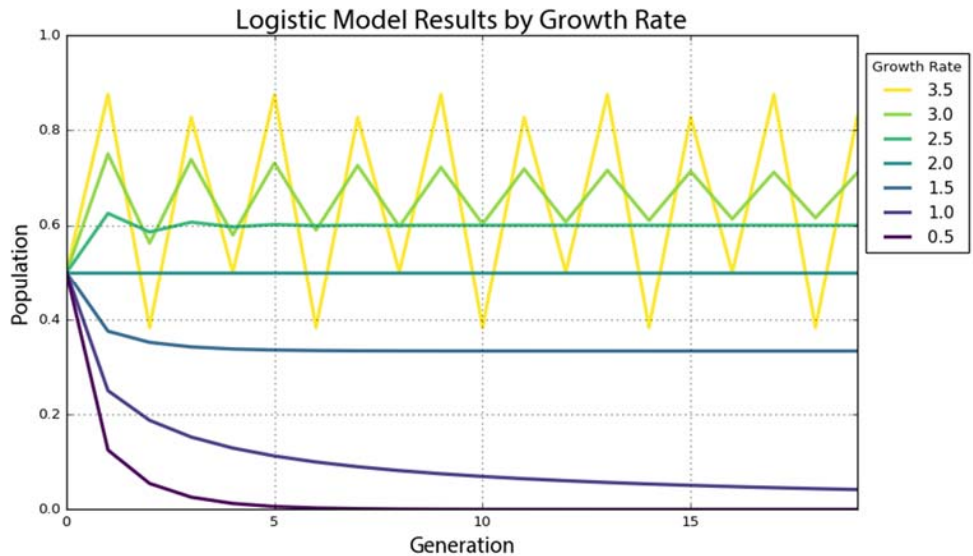

Figure 2.1. Time series graph of the logistic map with 7 growth rate parameter values over 20 generations.

Chaos can manifest itself in both continuous (i.e., with dynamics defined by differential equations) and discrete (i.e., with dynamics defined by an iterated map) nonlinear dynamical systems. The logistic map is a simple, one-dimensional, discrete equation that produces chaos at certain growth rates. We will explore this in depth momentarily, but first, we use Pynamical to run the logistic model for 20 time steps (we will henceforth call these recursive iterations of the equation *generations*) for growth rate parameter values of 0.5, 1, 1.5, 2, 2.5, 3, and 3.5. Table 2.1 presents the results: the columns represent growth rates and the rows represent generations. The model always starts with a population level of 0.5 and represents population as a ratio between 0 (extinction) and 1 (the maximum carrying capacity of our system). If we trace down the column in Table 2.1 under growth rate 1.5, we see that the population level eventually settles toward a final value of 0.333 after several generations. In the column for growth rate 2, we see an unchanging population level of 0.5 across every generation. This makes sense in the real world – if two parents produce two children, the overall population will neither grow nor shrink. Thus, a growth rate parameter value of 2 represents the replacement rate.

Figure 2.1 visualizes the time series in Table 2.1 as a chart with time on the *x*-axis and the system state (i.e., population) on the *y*-axis. This graph illustrates how the population changes over time at different growth rates. For instance, the violet line for growth rate 0.5 quickly drops to zero: the population dies out. The teal line that represents a growth rate of 2 (the replacement rate) stays steady at a population level of 0.5. The growth rates of 3 and 3.5 are more interesting. While the green line for growth rate 3 seems to slowly converge toward a stable value, the yellow line for growth rate 3.5 just seems to repeatedly bounce around four different values.

An *attractor* is the value, or set of values, that a system settles toward over time. When the growth rate parameter is set to 0.5, the system has a *fixed-point attractor* at population level 0, as depicted by the violet line dropping to 0. In other words, the population value is drawn toward a stable equilibrium of 0 over time as the model iterates: the logistic equation maps the value of a fixed-point attractor to itself. When the growth rate parameter is set to 3.5, the system oscillates between four values as depicted by the yellow line. This oscillating attractor is called a *limit cycle*. But when we adjust the growth rate parameter in this model beyond 3.57, we witness the onset of chaos. A chaotic system has a *strange attractor*, around which the system oscillates forever without ever repeating itself or settling into a steady state of behavior (Ruelle and Takens 1971; Shilnikov 2002). It never produces the same value twice and its structure is fractal, meaning the same patterns exist at every scale no matter how much we zoom into it (Grebogi et al. 1987).

## 2.4. System Bifurcations

To show this more clearly, we run the logistic model again, this time for 200 generations across 1,000 growth rate values between 0 and 4. When we produced the plot in Figure 2.1, we had only 7 growth rates. This time we have 1,000 so we need to visualize the results in a different way to make them comprehensible, using a *bifurcation diagram* that visualizes a system's attractors as a function of some parameter (May 1976; Gershenson 2004; Wu and Baleanu 2014). The bifurcation diagram in Figure 2.2 represents 1,000 discrete vertical slices, each corresponding to one of 1,000 growth rate parameter values evenly spaced between 0 and 4. To produce each of these visual slices, Pynamical ran the model 200 times then threw away the first 100 results, leaving just the final 100 generations for each growth rate. Each vertical slice thus visualizes the population values that the logistic map settles toward over time (i.e., the attractor) for that parameter value.

In Figure 2.2 we can see that for growth rates less than 1, the system always eventually collapses to zero (extinction). For growth rates between 1 and 3, the system always settles into an exact, stable population level. For instance, in the vertical slice above growth rate 2.5, there is only one population value represented (0.6) and it corresponds precisely to where the line for growth rate 2.5 settles in Figure 2.1's time series graph. At this parameter value, the system's attractor is a fixed point at 0.6. But for some growth rates, such as 3.9, the plot in Figure 2.2 shows 100 different values – in other words, a different value for each of its 100 generations. Here the system never settles into a fixed point or a limit cycle.

Why is this visualization called a bifurcation diagram? If we zoom into the growth rates between 2.8 and 4 to see what is happening at a finer scale (Figure 2.3), the possible population values fork into two discrete paths at the vertical slice above growth rate 3. At growth rate 3.2, the system oscillates exclusively between two population values: one around 0.5 and the other around 0.8. Thus, at that growth rate, applying the logistic map to one of these two population values yields the other. Just beyond growth rate 3.4, the diagram bifurcates again into *four* paths. This corresponds to the yellow line in Figure 2.1: when the growth rate parameter is set to 3.5, the system oscillates over *four* population values.

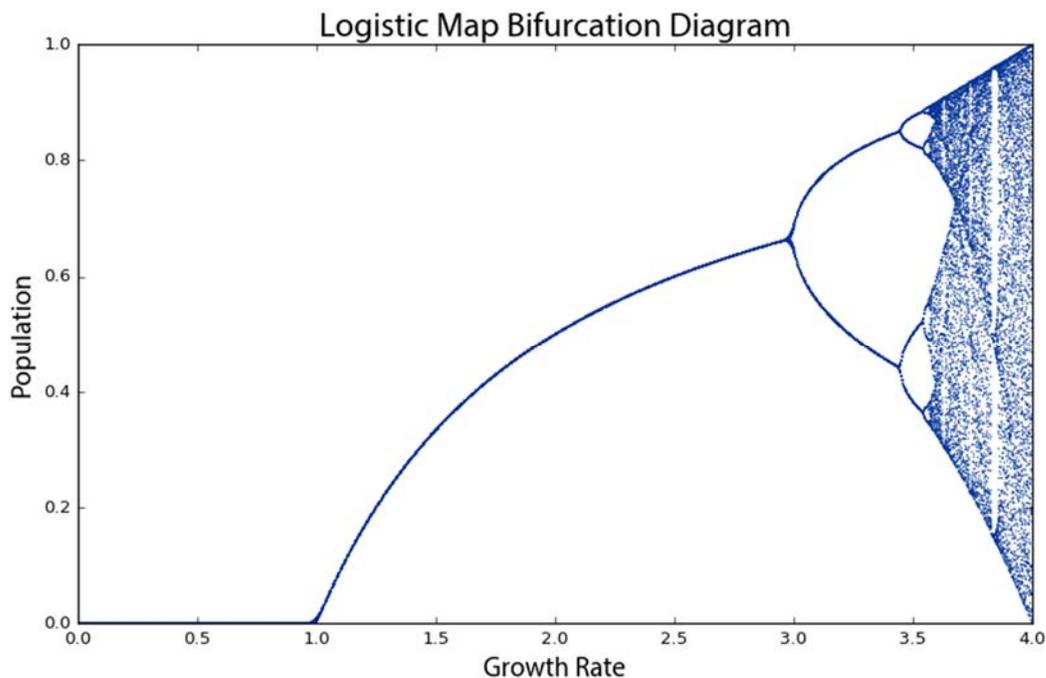

Figure 2.2. Bifurcation diagram of 100 generations of the logistic map for 1,000 growth rate parameter values between 0 and 4. The vertical slice above each growth rate depicts the system's attractor at that rate.

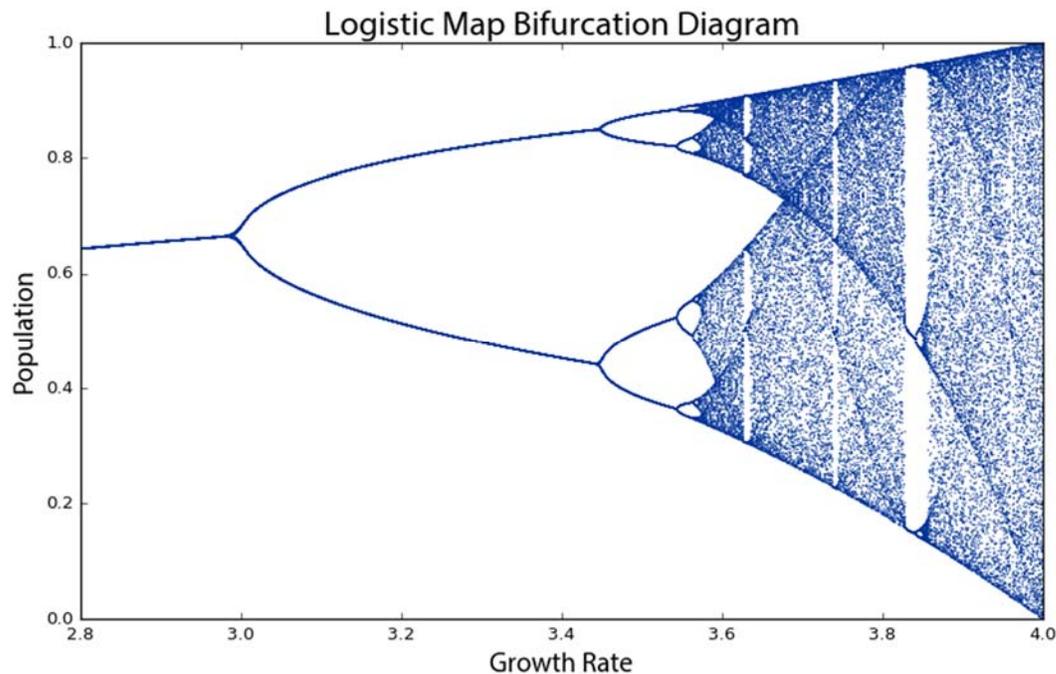

Figure 2.3. Bifurcation diagram of 100 generations of the logistic map for 1,000 growth rate parameter values between 2.8 and 4. The vertical slice above each growth rate depicts the system's attractor at that rate.

These are *periods*, just like the period of a pendulum. At growth rate 3.2, the system has a period-2 attractor. At growth rate 3.5, the system has a period-4 attractor. Just beyond growth rate 3.5, it bifurcates again into *eight* paths as the system oscillates over eight population values. These consecutive bifurcations are *phase transitions* from one behavior – such as a fixed-point attractor, to a qualitatively different type of behavior, such as a period-2 limit cycle attractor – as we vary this parameter value. Beyond a growth rate of 3.57, however, the bifurcations ramp up until the system is capable of eventually landing on any population value. This is known as the *period-doubling* path to chaos. As we adjust the growth rate parameter upwards, the logistic map will oscillate between two, then four, then eight, then 16, then 32 (and on and on to infinity) population values.

By the time we reach growth rate 3.99, it has bifurcated so many times that the system now jumps, seemingly randomly, between all population values. We only say *seemingly* randomly because it is definitely *not* truly random. Rather, this model follows simple deterministic rules yet produces apparent randomness due to its attractor having a period of infinite length. This is chaos: deterministic and aperiodic. If we zoom in again, to the narrow slice of growth rates between 3.7 and 3.9 (Figure 2.4), we begin to see the beauty of chaos. Out of the noise emerge swirling patterns and thresholds, on either side of which the system behaves very differently. For example, between the growth rates of 3.82 and 3.84, the system moves from chaos back into order, oscillating between just three population values: approximately 0.15, 0.55, and 0.95. But then at growth rates beyond 3.86 it bifurcates again and returns to chaos. Indeed, *any* one-dimensional system with a period-3 cycle such as this at some parameter value is capable of chaotic behavior at other parameter values (Li and Yorke 1975).

*Universality* refers to the phenomenon that very different systems can exhibit very similar behavior regardless of their underlying dynamics. It is commonly associated with Feigenbaum's discovery that all systems that undergo this period-doubling path to chaos obey a mathematical constant (Feigenbaum 1978; 1983). The distance between consecutive bifurcations along the horizontal axis shrinks by a factor that asymptotically approaches 4.669, now known as *Feigenbaum's constant* (Hofstadter 1985; Strogatz 2014). Regardless of the system's specific dynamics, the ratio of the bifurcations on its road to chaos always obeys this constant.

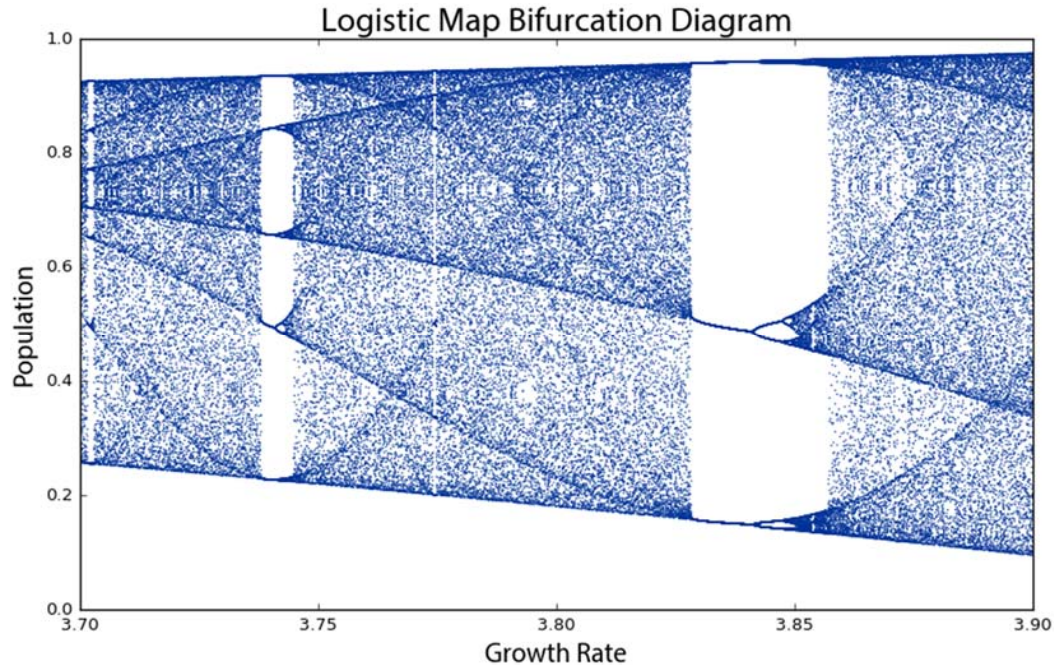

Figure 2.4. Bifurcation diagram of 100 generations of the logistic map for 1,000 growth rate parameter values between 3.7 and 3.9. The system moves from order to chaos and back again as the growth rate is adjusted.

## 2.5. Fractals and Strange Attractors

There is also a deep and universal connection between chaos and fractals (Tomida 2008). In Figure 2.4, the bifurcations around growth rate 3.85 may look familiar. If we zoom in on the center one (Figure 2.5), we see the same structure that we saw earlier at the macro-level. In fact, if we keep zooming infinitely in to this visualization, we will continue seeing the same structures and patterns at finer and finer scales, forever. How can this possibly be? We mentioned earlier that chaotic systems have *strange attractors* and that their structure can be characterized as *fractal* (Hénon 1976; Farmer 1983; Grassberger and Procaccia 1983). Fractals are shapes that are self-similar, meaning they have the same structure at every scale (Mandelbrot 1967; 1983; 1999). As we zoom in on them, we find smaller copies of the larger macro-structure. The bifurcation diagram (and thus the attractor) of the logistic map is a fractal: at the fine scale in Figure 2.5, we see a tiny reiteration of the same bifurcations, chaos, and limit cycles we saw in Figure 2.1's visualization of the full range of growth rates.

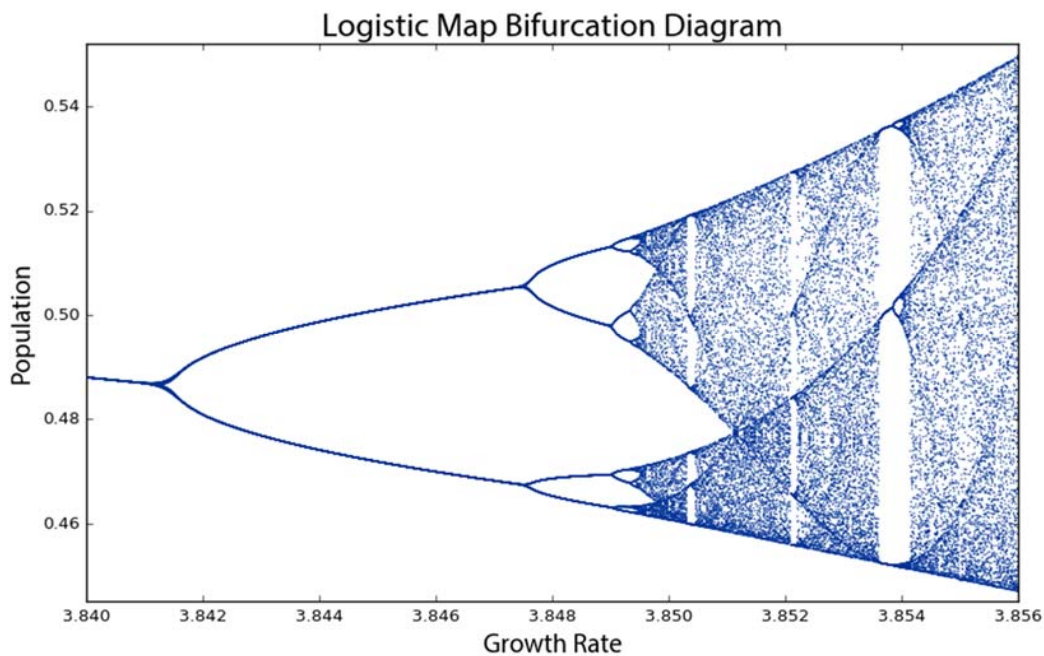

Figure 2.5. Bifurcation diagram of 100 generations of the logistic map for 1,000 growth rate parameter values between 3.84 and 3.856. This is the same structure that we saw earlier at the macro-level in Figure 2.3, because chaotic systems' strange attractors are fractal.

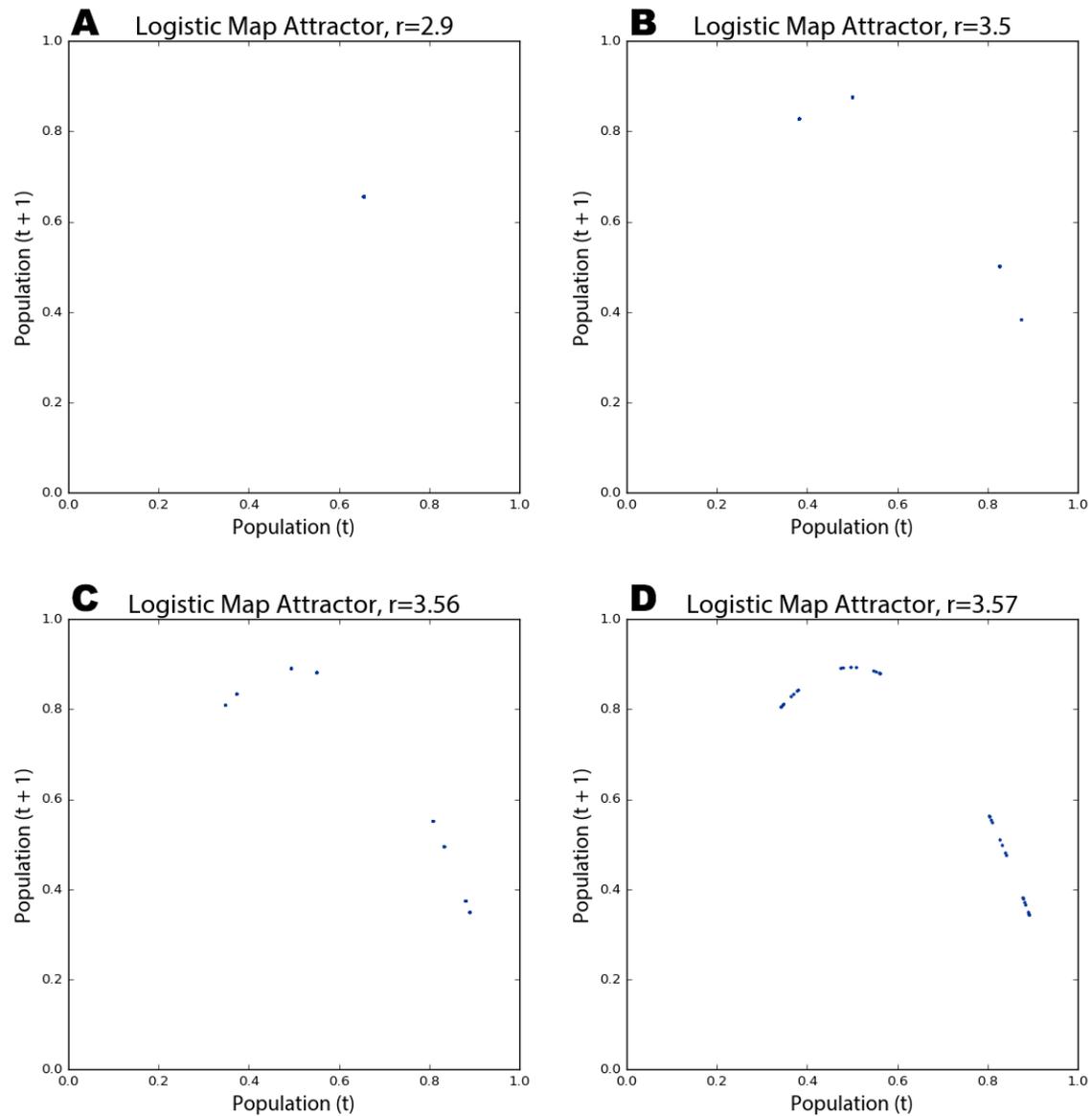

Figure 2.6. Phase diagrams of the logistic map over 200 generations for growth rate parameter values of 2.9 (A), 3.5 (B), 3.56 (C), and 3.57 (D). When the parameter is set to 2.9, the model converges at a single fixed-point. When the parameter is set to 3.5 or higher, the model oscillates over four points, then eight, and on and on as it bifurcates.

Another way to visualize this nonlinear time series is with a *phase diagram*, using a method called state-space reconstruction through delay-coordinate embedding (Bradley and Kantz 2015). Simply put, this plots the system's value at generation *t*+1 on the *y*-axis versus its value at *t* on the *x*-axis (Huikuri et al. 2000), giving us another visual window into the qualitative behavior of the system. The clever insight of this phase diagram is that it embeds one-dimensional time series data from our logistic map into two-dimensional *state space*: an imaginary space that uses system variables as its dimensions (Packard et al. 1980; Takens 1981; Theiler 1990). Each point in state space is a system *state*, or in other words, a set of variable values. While traditional systems analysis tends to focus on visualizing time series as in Figure 2.1, nonlinear dynamics tends to focus on visualizing these state spaces. Few real-world systems are fully observable, yet the dynamics in a properly reconstructed state space are identical to the true dynamics of the entire system (Bradley 2003).

In our case, the two variables are 1) the population value at generation $t$, and 2) the value at $t+1$. For example, with a growth rate of 3.5, the population value at generation 1 is 0.5, the value at generation 2 is 0.875, the value at generation 3 is 0.383, and so forth (see Table 2.1). Therefore, our two-dimensional phase diagram will have ($x$, $y$) points at (0.5, 0.875) and (0.875, 0.383) and so on (Figure 2.6b). Remember that our model follows a simple deterministic rule, so if we know a certain generation's population value, we can easily determine the next generation's value. Like earlier, to produce these phase diagrams Pynamical runs the logistic model for 200 generations and then discards the first 100 rows, to visualize only those values that the system settles toward over time.

In Figure 2.6a, the phase diagram shows that the logistic map homes in on a fixed-point attractor at 0.655 (on both axes) when the growth rate parameter is set to 2.9. This corresponds to the vertical slice above the $x$-axis value of 2.9 in the bifurcation diagram in Figure 2.2. Figure 2.6b depicts a period-4 limit cycle attractor: when the growth rate is set to 3.5, the logistic map oscillates over four points, as shown in this phase diagram (and in Figures 2.1 and 2.2). If we adjust the growth rate parameter up to 3.56, we witness a period-doubling bifurcation: Figure 2.6c shows the system now oscillating over eight points. As we approach the *chaotic regime* – the range of parameter values in which our system behaves chaotically – the period-doubling bifurcations start to come more quickly. Figure 2.6d shows that several additional bifurcations occurred between the growth rates of 3.56 and 3.57.

A kind of structure is slowly being revealed across Figure 2.6, but we can see it much more clearly as we push the growth rate parameter value deep into the chaotic regime. The phase diagram in Figure 2.7a reveals the system's attractor at a growth rate of 3.9. Figure 2.7b visualizes 50 different growth rate parameter values between 3.6 and 4, each with its own color. Those rates that exhibit chaos form parabolas due to the quadratic form of the logistic map's equation, but some gaps exist where the system occasionally settles down into periodic behavior (e.g., in the teal band when the growth rate is set to 3.83 – compare this band of periodicity with Figure 2.4).

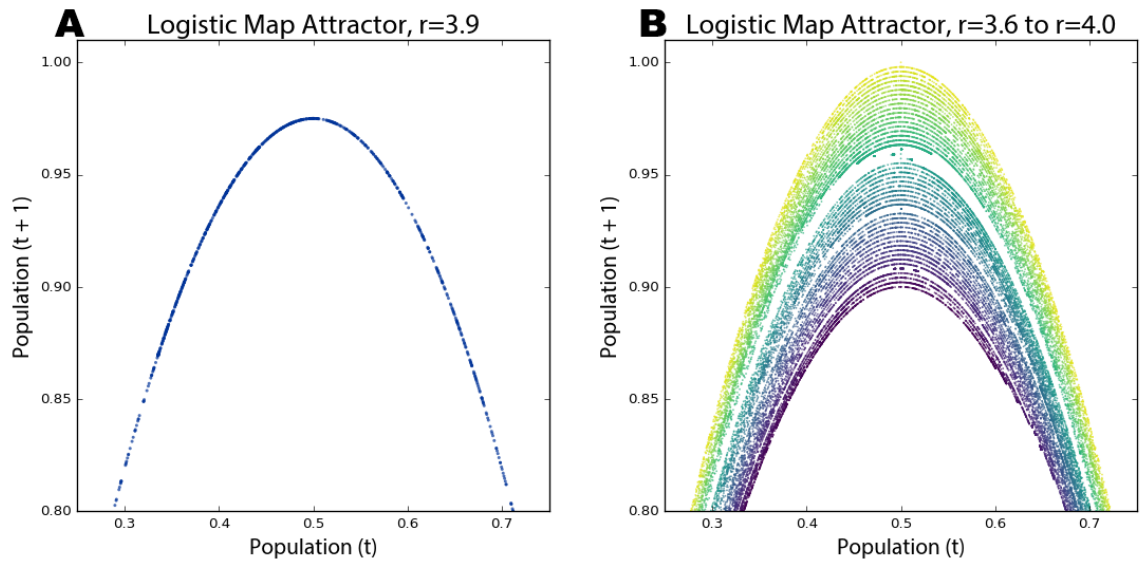

Figure 2.7. Cropped phase diagrams of the logistic map over 200 generations for (A) a growth rate parameter value of 3.9 and (B) 50 growth rate parameter values between 3.6 and 4 (the chaotic regime), each with its own colored line.

Strange attractors are revealed by these shapes as the system is somehow oddly constrained, yet never settles into a fixed point or limit cycle like it did in Figure 2.6. Instead it just bounces around different population values (i.e., points on the parabola) forever without ever repeating the same value twice. It is impossible to predict if any two consecutive observations appear near each other or far apart on the parabola. Further, the parabolas in Figure 2.7b never overlap due to their fractal geometry and the deterministic nature of the logistic map. Consider: if two different parameter values *could* ever land on the exact same point, their systems would have to evolve identically over time because the logistic map is deterministic.

We can see in these visualizations that this indeed never happens. While the dynamics of a chaotic system appear to have no pattern whatsoever, in reality they conform to a remarkable fractal pattern – a strange attractor – which confines the system to a limited slice of state space and ensures that no state will ever repeat (Kekre et al. 2014). Rather than having a whole-number dimension such as two or three, they are characterized by a fractional (hence, fractal) dimension (Grassberger and Procaccia 1983; Clarke 1986; Theiler 1990). The *fractal dimension* refers to the space-filling characteristics of a curve that, through self-similarity, becomes a bit more than a one-dimensional line yet a bit less than a two-dimensional plane.

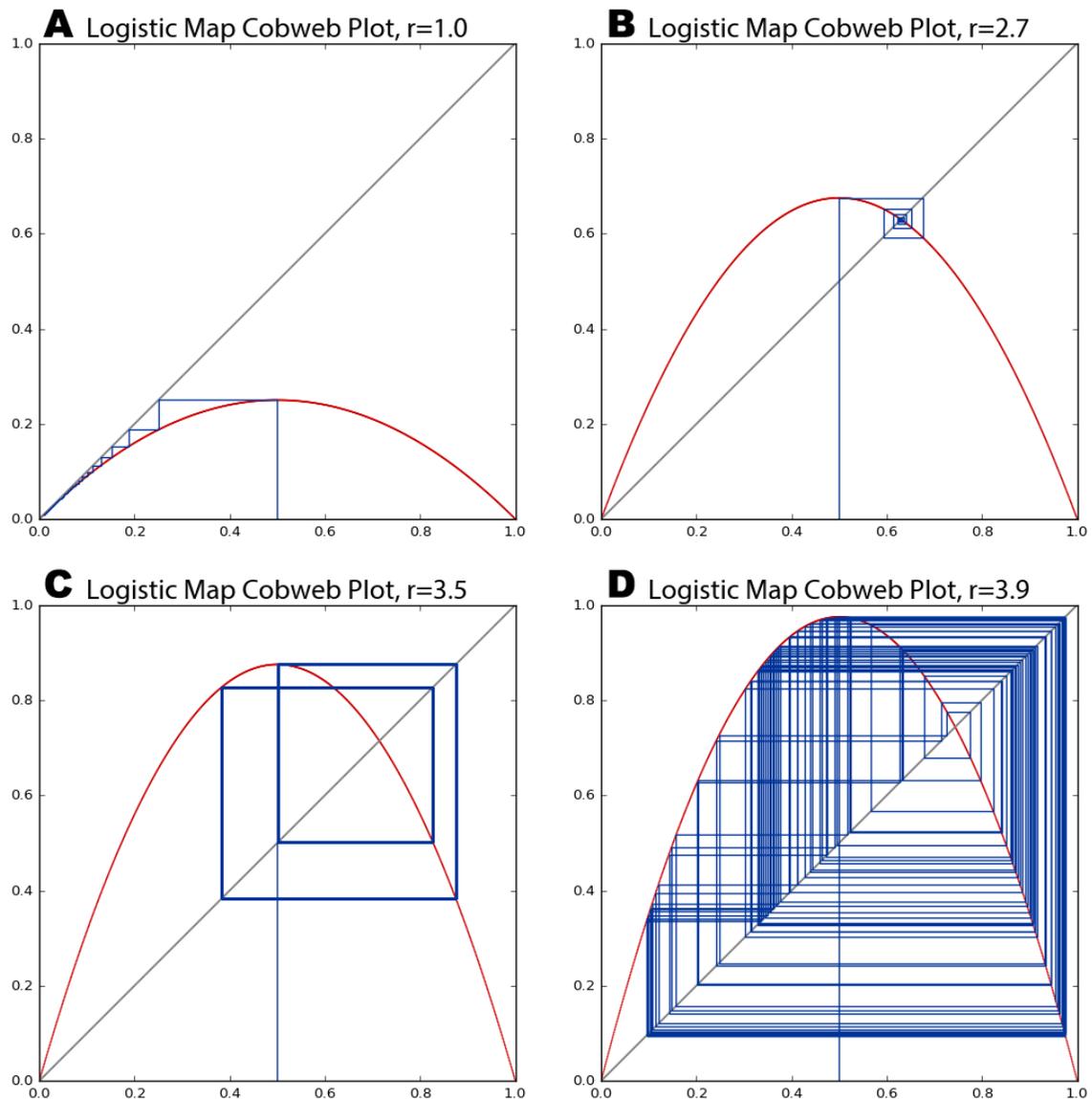

Figure 2.8. Cobweb plots of the logistic map for growth rate parameter values of (A) 1, (B) 2.7, (C) 3.5, (D) 3.9. The diagonal gray identity line represents *y=x*, the red curve represents the logistic map as *y=f(x)* for each of the four parameter values, and the blue line represents the system's trajectory over 100 generations.

These visualizations have all plotted quantitative data to better explain and understand the qualitative behavior of a nonlinear dynamical system. A *cobweb plot* is a visualization technique well-suited to revealing the qualitative behavior of one-dimensional maps, allowing us to analyze the long-term evolution of such systems under recursive iteration (Hofstadter 1985; Tomida 2008). The cobweb plots drawn by Pynamical in Figure 2.8 consist of three lines: a diagonal gray identity line representing $y=x$, a red curve representing the logistic map as $y=f(x)$ for a given parameter value, and a blue line tracing the path of the cobweb.

The blue lines intersect the red curve at those values our system lands on as it iterates from an initial population value of 0.5. In Figure 2.8a and b, the cobweb shows the system homing in on fixed-point attractors of 0 and 0.65, respectively. At a growth rate of 3.5 (Figure 2.8c) the system oscillates over four points in its limit cycle attractor, denoted by rectangular closed loops. The points where the blue lines intersect the red curve are the same as those revealed by the attractor in Figure 2.6b for the same parameter value. Finally, Figure 2.8d visualizes our system's behavior in the chaotic regime at a growth rate of 3.9. The chaotic orbit fills the plot with rectangles – an eventually infinite number of never-repeating trajectories that form a fractal cobweb throughout the diagram.

## 2.6. Determinism and Randomness

Phase diagrams are useful for visually revealing strange attractors in time series data, like that produced by the logistic map, because they embed this one-dimensional data into a two- or even three-dimensional state space. It can be difficult to ascertain if certain time series are deterministic or stochastic if we do not fully understand their underlying dynamics (Sander and Yorke 2015). Take the two series plotted by Pynamical in Figure 2.9 as an example. Both of the lines seem to jump around randomly. The red line *does* depict random data, but the blue line comes from our logistic model when the growth rate is set to 3.99. This is deterministic chaos, but it is difficult to differentiate from randomness. Instead in Figure 2.10 we visualize these same two data sets with phase diagrams rather than time graphs, giving us a clear window into the qualitative behavior of our systems. Now we can clearly see our chaotic system constrained by its strange attractor. By contrast, the random data set looks like the noise that it actually is.

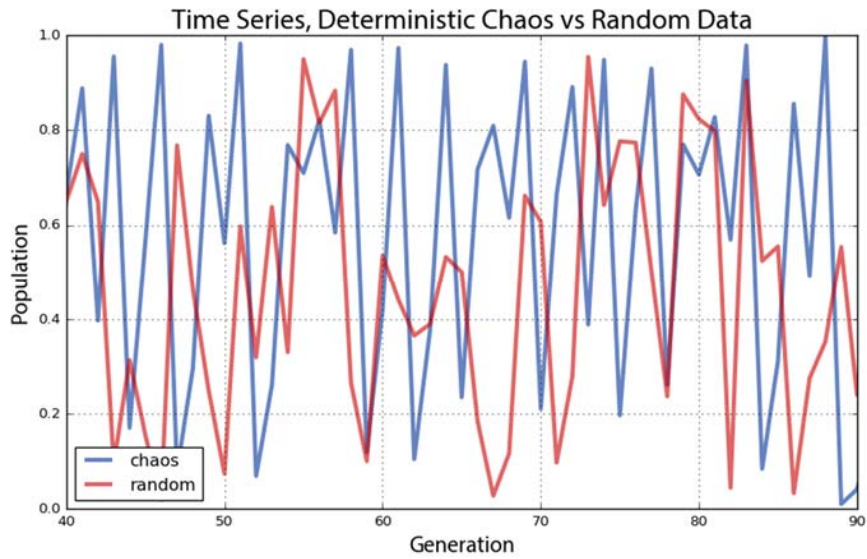

Figure 2.9. Plot of two time series, one deterministic/chaotic from the logistic map (blue), and one random (red).

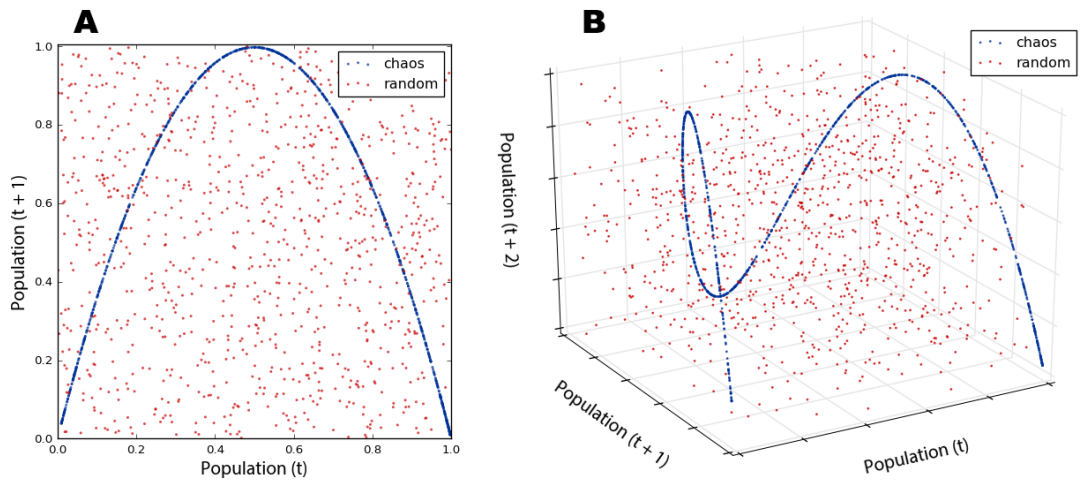

Figure 2.10. Phase diagrams of the time series in Figure 2.9. 10B is a three-dimensional state space version of the two-dimensional 10A.

This is particularly revealing in a three-dimensional phase diagram from Pynamical (Figure 2.10b) that embeds our time series into a three-dimensional state space by plotting the population value at generation *t*+2 versus the value at *t*+1 versus the value at *t*. This plot essentially extrudes our two-dimensional plot (Figure 2.10a), then pans and rotates the viewpoint. In fact, if we looked straight down at the x-y plane of the three-dimensional plot in Figure 2.10b, it would look identical to the two-dimensional plot in

Figure 2.10a. Strange attractors stretch and fold state space in higher dimensions, allowing their fractal forms to fill space without ever producing the same value twice.

To press this further, we can use Pynamical to visualize the rest of the logistic map's chaotic regime in three dimensions: the phase diagram in Figure 2.11 is a three-dimensional version of the two-dimensional state space we saw in Figure 2.7b. The color coding exposes the dynamical system's behavior across the chaotic regime – information virtually impenetrable without visualization. The structure of the strange attractor is revealed as it twists and curls around its three-dimensional state space. This structure again demonstrates that our *apparently* random time series data from the logistic model is not truly random at all. Instead, it is aperiodic deterministic chaos, constrained by a strange attractor. No matter how much we zoom in, the parabolas never overlap and no point ever repeats itself.

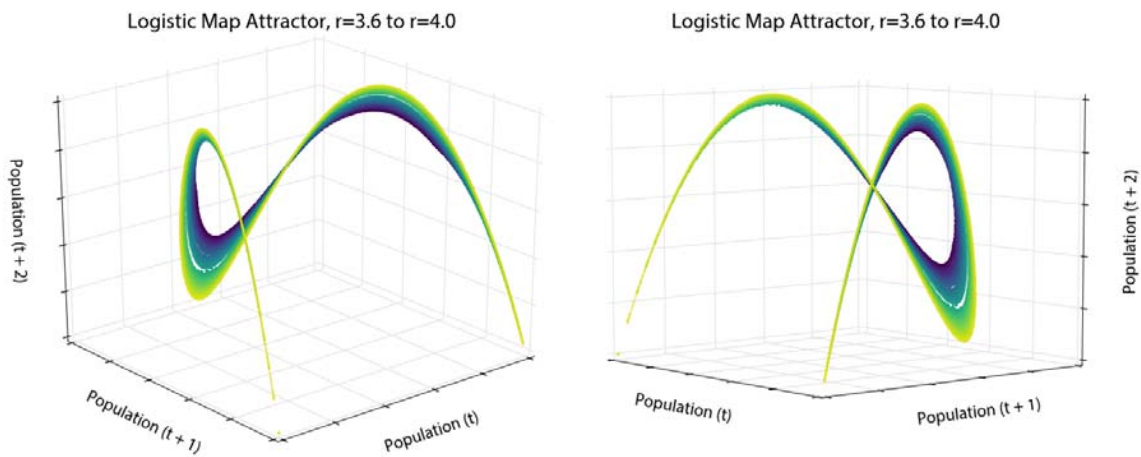

Figure 2.11. Two different viewing perspectives of a single three-dimensional phase diagram of the logistic map over 200 generations for 50 growth rate parameter values between 3.6 and 4, each with its own colored line.

## 2.7. The Limits of Prediction

Attractors have a *basin of attraction*: a set of points that the system's dynamics will pull into this attractor over time (Sprott and Xiong 2015). This is easily seen with a cobweb plot. Figure 2.12 shows how the logistic map's basin of attraction (when the growth rate is 2.7) pulls three different initial population values into the same fixed-point attractor. The initial state of the system will eventually become unknowable, because any one of many different possible points in the basin of attraction could have been the one pulled into the attractor.

By contrast, chaotic systems are characterized by their sensitive dependence on initial conditions (Zhang et al. 2016). Their strange attractors are globally stable yet locally unstable: they have basins of attraction, yet within a strange attractor infinitesimally close points diverge over time without ever leaving the attractor's confines. This divergence can be measured by *Lyapunov exponents* (Brown 1996), the calculation of which is described by Wolf et al. (1985). If the Lyapunov exponent's value is positive, then the two points move apart over time at an exponential rate. If the Lyapunov exponent is negative, then these points converge exponentially quickly, such as toward a fixed point or limit cycle. Finally, the Lyapunov exponent is zero when there is a bifurcation (Dingwell 2006).

For example, with our logistic model, the Lyapunov exponent is zero when the growth rate is set to 1 or 3 because they are bifurcation points; it is negative for most growth rates, such as $0 \leq r < 1$ and $1 < r < 3$, because they have fixed-point or limit cycle attractors; and it is positive for the chaotic regime (exclusive of those occasional windows when the system resumes brief periodicity, such as when the growth rate is 3.83). A positive Lyapunov exponent indicates that the system has a highly sensitive dependence on initial conditions, and is a common signature of chaos (Chan and Tong 2013; Kantz et al. 2013; Hunt and Ott 2015).

This nonlinear divergence of *very* similar values makes real-world modeling and prediction difficult, because we must measure the parameters and system state with infinite precision. Otherwise, tiny errors in measurement or rounding are compounded over time until the system eventually diverges drastically from the prediction. In the real world, infinite precision is impossible. It was through one such rounding error that Lorenz first discovered chaos. Recall his words at the beginning of this chapter: "the present determines the future, but the approximate present does not approximately determine the future."

As a demonstration of this, we run the logistic model with two *very* similar initial population values, shown in Figure 2.13. Both have the same growth rate parameter value of 3.9. The blue line represents an initial population value of 0.5 and the red line represents an initial population of 0.50001. These two initial conditions are extremely close to one another and accordingly their trajectories look essentially identical for the first 30 generations. After that, however, the minuscule difference in initial conditions compounds to the point that by the 40th generation the two lines show little in common. What began as nearly indistinguishable initial conditions produces completely different outcomes over time due to nonlinearity and exponential divergence.

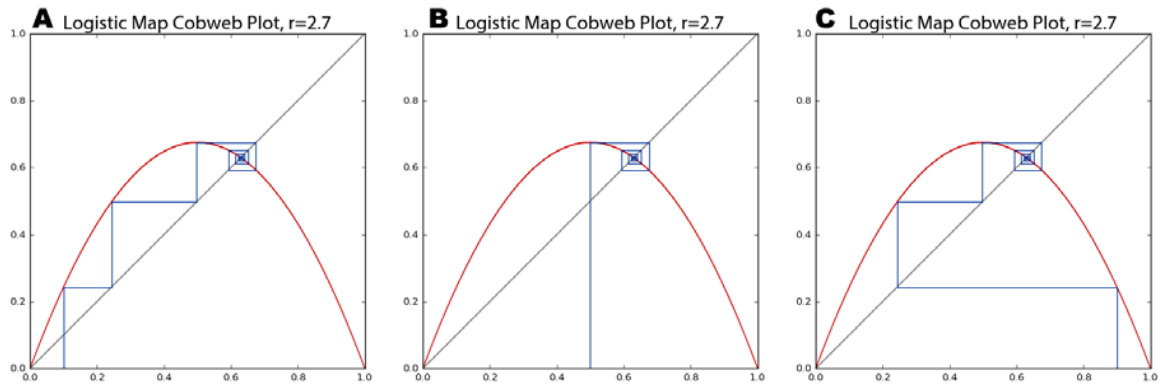

Figure 2.12. Cobweb plots of the logistic map pulling initial population values of 0.1 (A), 0.5 (B), and 0.9 (C) into the same fixed-point attractor over time. At this growth rate parameter value of 2.7, the Lyapunov exponent is negative.

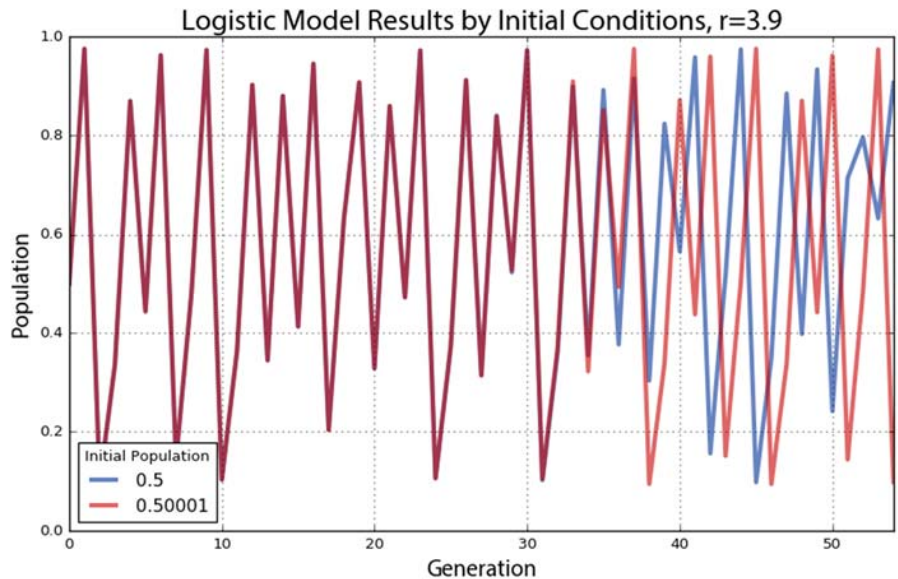

Figure 2.13. Plot of two time series with identical dynamics, one starting at an initial population value of 0.5 (blue) and the other starting at 0.50001 (red). At this growth rate parameter value of 3.9, the Lyapunov exponent is positive – thus the system is chaotic and we can see the nearby points diverge over time.

If our knowledge of these two systems began at generation 50, we would have no way of guessing that they were nearly identical in the beginning. With chaos, history is thus lost to time and prediction of the future is only as accurate as our measurements. Human measurements are never infinitely precise, so in real-world chaotic systems, errors compound and the future becomes entirely unknowable given long enough time horizons. This phenomenon is popularly known as the *butterfly effect*: a butterfly flaps its wings in China and sets off a tornado in Texas. Small events compound and irreversibly alter the future of the universe. In Figure 2.13, a tiny fluctuation of 0.00001 makes an enormous difference in the behavior and state of the system 40 generations later. Although this system's future cannot be predicted, we *can* characterize its dynamics geometrically with phase diagrams, bifurcation plots, and cobweb plots – and statistically with Lyapunov exponents and fractal dimensions.

## 2.8. Discussion

Pynamical and all the code used to develop these models and produce these visualizations are available in a public repository (see Appendix). Pynamical is built on top of Python's pandas, numpy, and matplotlib code libraries:

- numpy is a numerical library that handles the underlying numerical vectors
- pandas handles the higher-level data structures and analysis
- matplotlib is the engine used to produce the visualizations and graphics

Pynamical defines extensible functions to express the discrete map's equation and encapsulate the model that runs the equation iteratively. The logistic map, the Singer map, and the cubic map are built-in by default but any other iterated map can be defined and added. Pynamical also defines a function to convert model output into *x-y* points, as well as functions to plot these points as a bifurcation diagram, a cobweb plot, an animated cobweb plot, a two-dimensional phase diagram, a three-dimensional phase diagram, and an animated three-dimensional phase diagram. Animated cobweb plots of the entire parameter space and animated three-dimensional phase diagrams, extending those presented in this study, are also available in the repository. They shed particular light on the fractal nature of strange attractors as they stretch and fold state space, thus serving as a tool for pedagogy and visual information presentation.

Pynamical is easy to use and serves as a tool for introducing nonlinear dynamics and chaos. Sample code to produce some of the visualizations in the chapter demonstrates this simplicity. One merely imports Pynamical into the Python environment then runs the following code to produce the visualization:

Figure 2.2: `bifurcation_plot(simulate(num_rates=1000))`

Figure 2.4: `bifurcation_plot(simulate(min=3.7, max=3.9, num_rates=1000))`

Figure 2.6d: `phase_diagram(simulate(num_gens=100, min=3.57))`

Figure 2.8d: `cobweb_plot(r=3.9, x0=0.5)`

Figure 2.11: `phase_diagram_3d(simulate(num_gens=4000, min=3.6, num_rates=50))`

This chapter had two primary aims. First, it introduced the foundational concepts of nonlinear dynamics, chaos, fractals, self-similarity, and the limits of prediction through several visualization methods to analyze and understand the behavior of nonlinear dynamical systems. Second, it presented Pynamical, a software package for visualizing the behavior of discrete nonlinear dynamical systems. This package provides a free, fast, simple, extensible tool to introduce and analyze nonlinear dynamical systems' behavior visually – useful for research and pedagogy. Nonlinear systems are extremely difficult to solve analytically because they cannot be broken down into constituent parts. Instead, we used Pynamical to reveal hidden structure and patterns in time series whose underlying dynamics may not be well known. In particular, it revealed the qualitative behavior of nonlinear dynamical systems over time and in response to parameter variations.

This chapter used the logistic map to define such a set of nonlinear dynamics. As simple as this model was, at different growth rate parameter values it produced stability, periodic oscillations, or chaos. We used Pynamical to create bifurcation diagrams and cobweb plots to visualize this behavior across different parameter values. In the chaotic regime, the system jumped seemingly randomly between all population values. Accordingly, we used Pynamical to embed the data into higher-dimensional state space to create phase diagrams to visualize the system's strange attractor and understand its constrained, deterministic dynamics. Finally, we explored the butterfly effect's implications of nonlinearity on system sensitivity, as infinitesimal differences in initial conditions compounded over time until nearly identical systems had diverged drastically. Thus, in many nonlinear systems, there are fundamental limits to knowledge and prediction.

The modern study of complex systems evolved in the second half of the twentieth century from explorations in nonlinear dynamical systems, cybernetics, computation theory, systems theory, biology, and ecosystems study (Manson 2001; Manson and O'Sullivan 2006; Gershenson et al. 2016; Krivý 2016). During the 1990s, complexity theories largely supplanted chaos as an analytical frame for social systems. Although complexity draws on similar nonlinear principles, it emerges as a different beast. Instead of looking at simple, closed, deterministic systems, complexity examines large open systems made of many interacting parts. Unlike chaotic systems, complex systems retain some trace of their initial conditions and previous states, through path dependence. They are unpredictable, but in a different way than chaos is: complex systems have the ability to surprise through novelty and emergence.

The following chapter builds on this foundation of nonlinearity to unpack the theory of complex systems, examine how it applies to the study of cities, and introduce complex spatial network analysis.

# Chapter 3: Complexity and Cities

## 3.1. Abstract

This chapter presents the theoretical framework of complex systems and cities, culminating in network theory and analysis. A complex system is one characterized by the many nonlinear interactions among its component parts, resulting in unpredictable self-organization and emergent phenomena at different scales. Discussions of complexity and complex systems have appeared throughout the planning literature for years. These principles have been applied everywhere from the communicative turn and collaborative rationality, to cellular automata and agent-based urban models, to the design of resilient, livable neighborhoods. However, the interdisciplinary appeal and trendiness of complexity in the social sciences has resulted in a morass of ambiguous terminology, internal inconsistencies, and overloaded concepts open to multiple interpretations. This chapter unpacks the key foundational concepts of complex systems and network science in a brief, straightforward manner. In particular, it builds on the concepts presented in chapter 2 to introduce complexity in terms of systems and nonlinear dynamics before turning its attention to different types and measures of complexity. Next it discusses some of the key features of complex systems and how they apply to cities: equilibrium, stability, emergence, self-organization, and resilience. Finally, it presents complex networks – the primary lens this study uses in all subsequent chapters.

## 3.2. Introduction

Complexity theories have become a popular frame for conceptualizing and analyzing cities. There is no single complexity *theory* but rather a wide array of concepts and tools that can be applied to the study of complex systems across numerous disciplines (Manson and O'Sullivan 2006; Haken 2012). The term *complexity theory* generally refers to a nebulous union of these theories (see also Phelan [2001], Byrne [2001] and O'Sullivan [2004] for how this compares with complexity *science*). Such theories propose that certain large systems are characterized by the decentralized, nonlinear, dynamic interactions of their many constituent parts. These systems then behave in novel and unpredictable ways that cannot be divined by simply examining the components of the system, because the components' collective behavior is not a simple linear combination of their individual behaviors (Newman 2003; Yerra and Levinson 2005). Complexity problematizes traditional reductionist, linear methods of scientifically analyzing and predicting nonlinear systems like cities (Anderson 1972; Waldrop 1992; Cilliers 1998). It also opens up a new world of scholarship to researchers keen to formulate new kinds of sciences that take complexity into account (e.g., Wolfram 1994; 2002; see also recent pushes for and critiques of "urban science," e.g., Batty 2013c; Solecki et al. 2013; Alberti 2017; Kitchin 2017; Mattern 2017). These attempts usually follow Kuhn's (1962) theory of paradigm shifts: new evidence and modes of thinking undermine an established science, and a new science emerges to replace it.

Complexity theories have become a popular framework for scholarly enquiries into planning and urban studies over the past 30 years (McAdams 2008). Although it entails a fundamental shift away from the belief that predictive certainty is possible with complex systems, it *can* serve as a useful new lens for explaining urban phenomena, studying city form, and considering planning interventions. Further, complexity provides a comprehensive framework for assessing system behavior that could build stronger connections between quantitative and qualitative urban disciplines (Portugali 2006). However, complexity theories have sometimes been adopted into the social science literature in obscure or contradictory ways.

In particular, complexity theory in the planning literature has suffered from three notable problems. First, physical scientists often apply it atheoretically, either unconstructively problematizing planning methods, or naïvely (but mathematically) "proving" long-recognized urban phenomena (O'Sullivan and Manson 2015; cf. Stauffer 2004). Second, planning scholars sometimes cherry-pick concepts from complexity then use them abstractly or vaguely. For example, Portugali (2012) criticizes planning theorists like Manuel Castells and Patsy Healey for borrowing from complexity theory, then using the idea of complexity merely vernacularly – thereby losing all of its formalism and implications. Chettiparamb (2006) critiques Byrne (1997) for relying on undefined jargon and decontextualized assertions that render complexity theory vague, mystical, and New Age-y (see also Thrift [1999] and Auerbach [2016]). The mainstream and cross-disciplinary appeal of complexity has resulted in ambiguous terminology and overloaded concepts in the social sciences. Third, because of these first two problems, some urban scholars have dismissed "complexity" as merely fashionable nonsense.

This chapter wades into this morass to unpack the essential shared, foundational concepts of complexity theories – particularly as they might apply to cities – in a brief, straightforward manner. It is organized as follows. First, it draws the nonlinear foundations from chapter 2 out of the realm of simple closed systems and into the world of complex open systems. Then it discusses measures of complexity, providing a framework that will be fleshed out in applied detail in chapter 4. Next it reviews the concepts and ramifications of equilibrium, stability, emergence, self-organization, and resilience, drawing these concepts from the natural sciences into the study of cities, particularly urban form. This leads to the chapter's final section, which presents the science of networks and network analysis. Street network analysis has been central to network science since its nascence: its mathematical foundation, graph theory, was born in the eighteenth century when Leonhard Euler presented his famous Seven Bridges of Königsberg problem (Devlin 2000; Bonchev and Buck 2005; Derrible and Kennedy 2009). This discussion of networks lays the theoretical foundation for the empirical second half of this dissertation in chapters 5, 6, and 7.

## 3.3. Systems and Dynamics

A *system* is a set of interacting components that together form a whole. In the context of complexity theory, to say that a system is *complex* is to say that we cannot understand its behavior simply by examining its constituent parts (Newman 2003; Mitchell 2009). Complex systems comprise many interacting, hierarchical subcomponents whose recurrent interactions cause nonlinear feedback, collective behavior, and unpredictable emergent phenomena at multiple spatial and temporal scales (Simon 1962; Rickles et al. 2007). In contrast, *complicated* refers merely to being made up of many interrelated parts.

Examples are useful to disambiguate these types of systems. A wind-up clock is an example of a simple system with few interrelated parts. An automobile is an example of a complicated system with many interrelated parts. In contrast, stock markets, the climate, and cities are examples of complex systems. Complex systems are defined more by their internal relationships than they are by their constituent parts and it is argued that it is this networked structure and organization that makes them interesting (Manson 2001; Wilson 2006). The term *complexity* itself refers to the rich, dynamic system behavior arising from individual interactions between many heterogeneous subcomponents (Cilliers 1998). In particular, a *complex adaptive system* is one whose collective behavior exhibits adaptation and learning aimed at perpetuating itself in the face of a changing environment or conditions (Holland 2006; Miller and Page 2007).

As discussed in chapter 2, a *dynamical system* changes over time as its state evolves according to its initial conditions and the processes that describe its subcomponents' behavior. A system's *state* is the essential information about the system for an observer and is defined by the values of the relevant variables. A *variable* could be a system feature or a calculated indicator that an observer has decided to use to describe the system. *Process* is more difficult to define, but generally refers to some sequence of actions that changes the system state (O'Sullivan and Perry 2013). Related to process, *dynamics* can be interchangeably thought of as the system's "rules" or, thus, the paths the system state traces through time (for discussion in an urban context, see Albeverio [2008]). These paths of the variables through time are visualized with phase space diagrams, as discussed in chapter 2. *Phase space* is an abstract space that contains all possible system states, with each possible state represented by a single point (Figure 3.1).

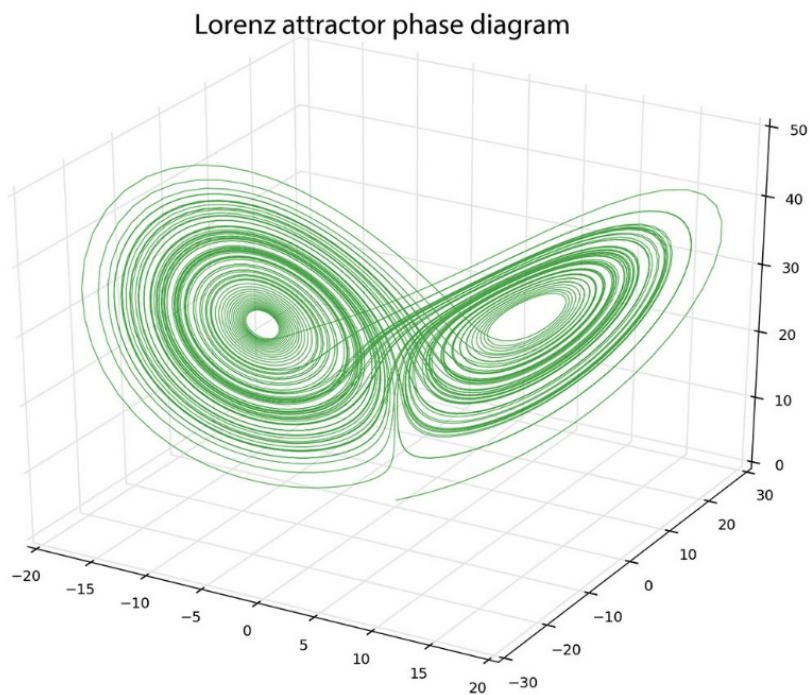

Figure 3.1. An example phase space diagram of the evolution of the Lorenz system over time.

Real-world complex systems are sensitive to outside influences because they are open systems. An *open system* is one that cannot be screened off from its environment, so researchers cannot safely ignore outside influences – or in other words, exogenous variables. Most real-world systems are open and pose problems for modeling because these exogenous influences must be taken into account (Batty and Marshall 2012; see also Schuster 2015). A *model* is simply an abstract representation of something, and could refer to conceptual models, mathematical models, statistical models, or even physical models (O'Sullivan and Perry 2013). Although real world systems tend to be open, all models are closed, to be tractable.

Complex systems and chaotic systems are both subtypes of nonlinear dynamical systems. Complexity theory deals with complex open systems that self-organize into emergent forms that could not have been predicted simply by understanding the constituent parts (Mitchell 2009). Chaos theory deals with simple, deterministic, nonlinear, closed systems resulting in a chaotic response to different initial conditions or perturbations (Reitsma 2003). As discussed in chapter 2, chaotic systems are unpredictable beyond limited time horizons because of their sensitivity to initial conditions, and these initial conditions may become unknowable later. Complex systems, in contrast, are unpredictable because of their capacity for novelty via emergence, discussed in detail below. They are sensitive to initial conditions in the sense that early historical accidents create path dependence that maintains their legacy over long time horizons. Chaos theory examines apparent disorder arising from simple order, while complexity theory examines large-scale order emerging from disorder at the local scale (ibid.).

## 3.4. Types and Measures of Complexity

Complexity lacks a single definition and measure, and often relies on what makes a phenomenon, behavior, or pattern complex in context. Shiner et al. (1999) provide an overview of three such formulations, discussed in detail in chapter 4. The first, based on entropy, is highest when objects are scrambled up with the greatest variety and diversity. The second balances variety and structure, and conforms to traditional definitions of complex adaptive systems. The third preferences order, self-organization, and emergence in which structure emerges from previous disorder. There are in turn several measures of complexity. Mitchell (2009) points out that no one measure is ideal (or could possibly capture the myriad denotations and connotations of complexity), but highlights several prominent ones: information entropy from information theory; statistical complexity, or the degree of structure and pattern in a system; the Lyapunov exponent, which mathematically defines a system's sensitivity to initial conditions; and the fractal dimension, which defines the irregularity of an object's form.

The former two are discussed in detail with regards to urban form and street networks in chapter 4. The latter two were already explored in chapter 2 with regards to simple nonlinear systems, but their implications for complex systems are worth considering briefly here. Sensitivity to initial conditions makes *prediction* of a nonlinear system difficult, as the initial state must be described with perfect accuracy (Rickles et al. 2007). Unfortunately, measurement of the real world always requires some amount of rounding and thus entails some amount of uncertainty. These tiny inaccuracies compound over time as the system evolves, making prediction difficult or even impossible. Theorists from Friedrich Hayek (1944; 1974) to Ilya Prigogine (1997) have thus questioned whether it is even possible to make accurate predictions of complex systems, given the requirements of data-gathering and precision *and* because of such systems' capability to surprise via emergence, a concept discussed in section 3.6.

Complex systems such as cities are sensitive to initial conditions in the sense of historical accidents, but their path dependence continues to reveal these conditions over long time horizons (Arthur 1988). *Path dependence* simply refers to the idea that history matters: complex systems' past states are remembered and play a role in future states – i.e., they are non-Markovian systems (Arthur 1989; Liebowitz and Margolis 1995). Further, it is possible for single events to alter a complex system in a way that persists for a long time (Allen and Sanglier 1981). In cities, historical accidents/natural subsidies (i.e., sensitivity to initial conditions) or exogenous perturbations (e.g., wars, new technology, or economic shocks) may significantly affect long-term system behavior. Some echo of a complex system's initial conditions remains apparent far into the future, whereas a simple chaotic system's initial conditions are eventually lost to time and become unknowable.

Finally, as introduced in chapter 2, the fractal dimension refers to the non-integer dimension of an object with an irregular form – e.g., a line so kinked that it can be characterized as something between a one-dimensional line and a two-dimensional plane.

Complex systems such as cities produce fractal self-similar forms that can be seen in urban peripheries and street networks (White and Engelen 1993; Batty and Longley 1994; Benguigui et al. 2000; Shen 2002) – and at the scale of urban design, as will be explored in chapter 4.

## 3.5. Equilibrium and Stability

*Equilibrium* is used in different ways in the social sciences literature. In urban economics, it typically refers to a point at which supply and demand are balanced, resulting in – to use location choice as an example – no incentive for anyone to move (Ogawa and Fujita 1980; Waddell 2000; Waddell et al. 2003; O'Sullivan 2008; Batty 2013c). However, complexity scholars typically borrow from physics instead to define equilibrium as a steady state of constant, maximum entropy in which a system does not change, adapt to its environment, or evolve structure (Holling 1973; Barthélemy 2017).

A common illustrative example is a gas diffusing into a vacuum until it is evenly dispersed. In the 1970s, Ilya Prigogine discovered that certain far-from-equilibrium open systems can evolve structures that locally contradict the second law of thermodynamics, which states that systems move toward maximum entropy (Nicolis and Prigogine 1977). Allen and Sanglier (1981; cf. Berry 1964) extended Prigogine's findings to the urban studies literature through their reformulation of central place theory in terms of these dissipative structures and bifurcation.

Given these different definitions, there is some vagueness in how the term "equilibrium" is used, often unqualified, in the urban complexity literature. Sometimes it refers to thermodynamic equilibrium, as scholars invoke it to argue that cities are far-from-equilibrium complex systems in the Prigogine sense (e.g., Batty 2013c; 2017). This stream of literature argues that cities are open systems and thus matter and energy – such as food, electricity, immigrants, building materials, etc. – flow into them, while entropy (namely, negentropy) is exported out of the system (Butera 1998). Structure and order evolve, locally violating the second law of thermodynamics. In this sense, cities do not move toward equilibrium; rather they are far from it, ever evolving and structuring their matter (White and Engelen 1993).

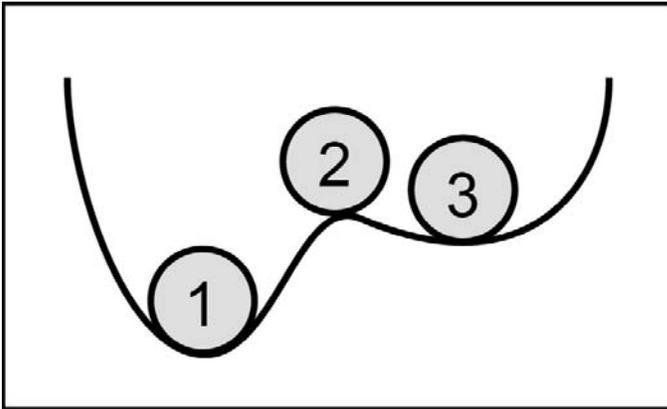

Figure 3.2. Three balls at rest on a slope, representing stable (ball 1), unstable (ball 2), and metastable (ball 3) states. Ball 1 will return to its current state after a large perturbation. Ball 2 will move away from its current state after even a slight perturbation. Ball 3 will remain in its current state after a small perturbation, but a large one can push it into a preferred stable state. Its semi-stable state is near a critical point of transition.

Other times, equilibrium is used to refer to an equilibrium of dynamics, where the system becomes limited and its state settles into an unchanging value or a set of values that it oscillates over (Phillips 2004). This is known as a stable equilibrium, or for disambiguation's sake, a stable state. Equilibrium in this context means that a system is in balance despite multiple forces acting on it, dominated by negative feedback that damps perturbations and pushes the system back toward the equilibrium. At a macro-level, a real-world complex system might appear to be in stable state based on its system-level state variables, but at a micro-level components may be dynamic and in flux (Batty 2013c; 2017). Thus, time and spatial scales are essential for considering equilibrium and disequilibrium (Barthélemy 2017). Consider residents settling into locations in a city. Over time, stable and consistent patterns may emerge city-wide, but at the human scale, residents are always moving in, around, and out of the city. Stable states make for tractable models of complex systems: Schelling (1971) famously demonstrated a simple simulation in which fairly tolerant residents relocate based on their subtle preferences for similar neighbors, resulting in a surprisingly segregated static equilibrium.

Complex systems can settle for periods of time into stable, metastable, or unstable states – or even shift between alternative stable states via phase transitions (May 1972; 1974; Holling 1973; Folke et al. 2004). These are depicted by the ball-and-cup diagram in Figure 3.2. A *stable state* (ball 1) is one in which the system is resilient to perturbation and its dynamics return it to this state after being perturbed. Stable states may include steady states – in which the system state remains at some fixed value – or limit cycles – in which the system oscillates over a consistent set of values. An *unstable state* (ball 2) is one which the system moves away from after even a slight perturbation: the system is precariously perched at a critical point between two possible states that it could settle into. A *metastable state* (ball 3) is one the system returns to after small but not large perturbations. The system may spend extended time in this semi-stable state, but a sufficient perturbation can push it into a preferred state.

*Alternative stable states* are also possible: ecosystems can exist in different stable states over long periods of time (Beisner et al. 2003). After a certain perturbation, they may transition from one to another via a phase transition, also known as a regime shift (Folke et al. 2004). Such behavior suggests a system with possible states that are separated by discontinuous thresholds rather than a smooth gradient, a common outcome of nonlinearity (May 1977). *Hysteresis* – the dependence of a system's behavior on both its present state and its past states – allows a system to exist in different states at different times but under the same conditions (Franz 1990; Beisner et al. 2003). This path dependence helps keep the system in the current state and suppresses transitions to other states it could otherwise be in.

Metastability, mentioned earlier, refers to a semi-stable state near a critical point. A *critical point* is the (typically unstable) point where transitions from one state to another occur (Downey 2012; Keane 2013). But unlike shifting from one fairly stable state to another, criticality itself connotes a system poised on the edge of catastrophe (Bak and Paczuski 1995). A system is critical if its behavior changes dramatically – for instance, transitioning from an ordered regime to a chaotic one – given some small input (Bak et al. 1988; Batty and Xie 1999). This critical state was popularly referred to in the past as the "edge of chaos" (Waldrop 1992; Oxley and George 2007).

When a parameter is adjusted to the critical point, the system undergoes a quick, radical change in its qualitative features, such as water freezing at 0° Celsius. A *parameter* is a factor that defines the system. Examples include the temperature of the water in the preceding example, the growth rate *r* in the logistic map in chapter 2, and the angle and depth of the slopes in Figure 3.2. A model parameter is similar to a variable, but either represents some universal value, or is directly controlled by the researcher rather than simply being observed. For the latter, consider the phase change of water at certain temperatures, from gas to liquid to solid. Here, temperature is the parameter being adjusted by the researcher to the critical point of phase transition. Alternatively, consider parameters such as the $CO_2$ carrying capacity of the Amazon rain forest, which may be some constant at any given time, but could change through global warming or pollution.

Finally, as discussed in chapter 2, *bifurcation* is the tendency for a system or one of its variables to jump suddenly from one attractor or stable state to another (Puu 2013). When this happens, a drastically different aspect of the system appears. Allen (2012) argues that bifurcation in complex human systems – such as cities – can be interpreted as an important historical juncture where the system could go one direction or another, with multiple possible future trajectories. With a sufficient understanding of the system and its dynamics, urban planners may adjust some parameter to steer the trajectory toward socially desirable outcomes (cf. Batty 2017; Kitchin 2017).

## 3.6. Emergence, Self-Organization, and Resilience

An *emergent* system property arises from interactions between subcomponents of a complex system (Miller and Page 2007). These subcomponents, however, do not themselves display this new system property and the property could not have been deduced merely by examining the subcomponents and their interactions (Aziz-Allaoui and Bertelle 2009). As Anderson (1972, p. 395) puts it, "the whole becomes not only more than but very different from the sum of its parts." Nonlinearity is the source of macro-scale (and often unpredictable) emergent system characteristics, as they result from many repeated (and possibly extremely simple) micro-scale interactions among subcomponents (Manson 2001). In other words, the researcher cannot just take the system apart, inspect the components to understand what the system does, and them put them back together again. Emergent phenomena are nonlinear characteristics of a system, such as catastrophes, thresholds, and self-organization.

*Self-organization* is an emergent phenomenon that occurs when a system orders itself into a "better" or more stable state without external control or a central overseer (Kauffman 1996; Cilliers 1998; Portugali 1999; Strogatz 2004). For instance, "tactical urbanism" initiatives have been theorized in terms of complexity and self-organization (Silva 2016).

Self-organization tends to be a *bottom-up* process by which one hierarchical level generates the features of the level above it (Allen 1998). The distinction between top-down and bottom-up processes should not be taken to be binary (for discussions of this in a planning context, see Adams and Tiesdell [2010]; Holcombe [2013]; Krivỳ [2016]). Rather, each simply refers to the general directionality of a process in terms of the system's hierarchy.

*Feedback* occurs when an output of the system "feeds back" into the system as an input (Hofstadter 1979). Negative feedback damps a variable's rate of change and pushes it toward a stable state. Positive feedback increases a variable's rate of change, as self-reinforcement. Furthermore, large-scale structures can emerge from small-scale subcomponent behavior and then influence future subcomponent behavior via *cross-scale feedback* (Allen 2012). Through *co-evolution*, subcomponents create their environment and are then in turn influenced by it. Culture, religion, and social norms – created by humans and in turn influencing humans – are examples of such emergent properties and their cross-scale feedback within cities and societies.

Resilience and robustness are complex adaptive system traits related to self-organization, feedback, and nonlinearity (Holling 1973; Miller and Page 2007). Walker et al. (2004, p. 6) define *resilience* as "the capacity of a system to absorb disturbance and reorganize while undergoing change so as to still retain essentially the same function, structure, identity, and feedbacks – in other words, stay in the same basin of attraction." Thus, a resilient system is able to return to its original stable state after a perturbation. *Robustness*, in contrast, tends to refer to the stability of a specific variable or system characteristic despite instability among some system components (Aligica 2014). In other words, resilience refers to returning to an original state after a perturbation, while robustness refers to perturbations having only a minimal effect in the first place.

*Self-organized criticality* describes a system that has a critical point as its attractor and continually evolves all by itself to this point of phase transition and catastrophe (Bak et al. 1987). At the critical point, system subcomponents are extremely connected and strongly influence one another. Here, even a small change to a single subcomponent is capable of producing vast effects that ripple through the entire system. The classic example is the sand pile model described by Bak et al. (1988). In this model, a sand pile has additional grains of sand continuously dropped on top of it. The pile evolves to a certain angle of

repose – the critical point – despite frequent trivially small avalanches. At this point, a single additional grain can suddenly cause a massive avalanche. After an avalanche, the system slowly evolves back to that critical angle and repeats. Batty and Xie (1999) argue that cities exhibit self-organized criticality as their urban forms evolve over time through discrete transitions. Forest fires serve as another example. Frequent small fires tend to prevent fuel build-up, but huge fires occasionally occur when no small fires have cleared the underbrush in a long time (Malamud et al. 1998).

Such systems accumulate energy over time and dissipate it through many small, and a few large, events. Thus, systems that exhibit self-organized criticality produce events – such as sand pile avalanches, forest fires, and earthquakes – that range from tiny to enormous (Turcotte 1999). In other words, these systems are *scale-free* – they have no characteristic size. Human beings, on the other hand, have a characteristic size since they tend to range between five and seven feet tall, with few outliers. Self-similarity, scale invariance, scale-free, and fractal geometry are equivalent concepts that indicate a lack of this characteristic scale (Rickles et al. 2007). Scale free systems follow a scaling law such as a *power law* with the form $p=b^{-a}$. Gaussian distributions result from processes that tend to sum to the center of the range, but with a power law distribution, the probability $p$ of an event is inversely proportional to its size $b$. Thus, there are very few massive events, some medium-sized ones, and lots of small ones.

Similar characteristics can be seen in numerous systems despite their different underlying dynamics. This phenomenon is called *universality* (cf. chapter 2, section 2.4). Accordingly, in the 1990s power laws became something of a popular signature for an underlying complex system. However, an over-reliance on power laws and universality has met controversy and criticism in recent years (Stumpf and Porter 2012; O'Sullivan and Manson 2015; Auerbach 2016). It can be challenging to differentiate between a power law distribution and other candidate distributions, particularly lognormal when it is difficult to observe tiny events to the left of the mode peak. Rarely does a real-world phenomenon follow a power law for all values of the independent variable, and spatial boundary definitions can further affect the phenomenon's distribution (Clauset et al. 2009; Veneri 2016). Moreover, there are innumerable ways to generate a power law distribution, so it alone cannot be an unambiguous indicator that a complex system underlies the observed phenomena (Mitzenmacher 2004).

This plays into a classic challenge of complex systems study: the *equifinality* problem is that different processes and models can result in the same outcome or pattern (Beven and Freer 2001). Many urban processes can be shown to produce similar patterns across spatial scales, but this does not help us understand what exactly is happening in each instance. For example, urban form may have a fractal spatial pattern, but this finding has yet to be connected convincingly to underlying social and economic processes (Manson and O'Sullivan 2006). Although much of the purpose of complexity studies lies in linking patterns to processes, there is risk in conflating pattern *with* process.

## 3.7. Networks

This chapter has thus far discussed complex systems theory, drawn from the foundations of nonlinear systems presented in chapter 2, including dynamics, stability, emergence, and self-organization. These characteristics appear in a system as a result of the many interactions between its connected, constituent parts. The interactions, connections, dynamics, and processes within a system are the subject of network science. The past 15 years have witnessed an explosion in the science of networks. Much of this research has been stimulated by recent advances in statistical physics and the study of complex systems (Blumenfeld-Lieberthal and Portugali 2010). In an urban context, the structural attributes of city networks can influence the way an urban system's physical links organize and influence complex human interactions, connections, and dynamics (Baynes 2009; Comunian 2011). As Glaeser (2011), among various others, argues, cities exist to connect people.

Network science is built upon the foundation of graph theory, a branch of discrete mathematics. A *graph* is an abstract representation of a set of elements and the connections between them (Tinkler 1979; Trudeau 1994). The elements are interchangeably called vertices or nodes, and the connections between them are called links or edges (Downey 2012). For consistency, we use the terms *nodes* and *edges* throughout this study. The following definitions are fundamental to graph theory and can be found in detail in e.g., Harary et al. (1965), Trudeau (1994), Albert and Barabási (2002), Dorogovtsev and Mendes (2002), Brandes and Erlebach (2005), Costa et al. (2007), Newman (2003; 2010), Barthélemy (2011), and Cranmer et al. (2017).

The number of nodes in the graph (i.e., the *degree* of the graph) is commonly represented as *n* and the number of edges as *m*. Two nodes are *adjacent* if an edge connects them, two edges are adjacent if they share the same node, and a node and an edge are *incident* if the edge connects the node to another node. A node's *degree* is the number of edges incident to the node, and its *neighbors* are all those nodes to which the node is connected by edges.

An *undirected* graph has undirected edges (i.e., each edge points mutually in both directions) but a *directed* graph, or digraph, has directed edges (i.e., edge *uv* points from node *u* to node *v*, but there is not necessarily an edge *vu*). A *self-loop* is an edge that connects a single node to itself. Graphs can also have parallel (i.e., multiple) edges between the same two nodes. Such graphs are called *multigraphs*, or *multidigraphs* if they are directed. An undirected graph is *connected* if each of its nodes can be reached from any other node. A directed graph is *weakly connected* if the undirected representation of the graph is connected, and *strongly connected* if each of its nodes can be reached from any other node. A *path* is an ordered sequence of edges that connects some ordered sequence of nodes. Two paths are *internally node-disjoint* if they have no nodes in common, besides end points. A *weighted* graph's edges have a weight attribute to quantify some value, such as importance or impedance, between connected nodes. The *distance* between two nodes is the number of edges in the path between them, while the *weighted distance* is the sum of the weights of the path's edges.

*Network science* is the study of typically real-world graphs – thus, networks inherit the terminology of graph theory. For a historical overview of the relationship between graph theory and network analysis, see Barnes and Harary (1983). Familiar networks include social networks (where the nodes are humans and the edges are their interpersonal relationships), the Internet (where the nodes are computers and the edges are the physical TCP/IP-based links that connect them), and the World Wide Web (where the nodes are web pages and the edges are hyperlinks that point from one page to another). A complex network is one with a nontrivial topology. Its *topology* is the configuration and structure of its nodes and edges, and by *nontrivial* we mean that this is neither fully regular nor fully random (Stewart 1995; Newman 2010). Compare this definition with category II complexity in the framework presented in chapter 4, section 4.4.1. Most large real-world networks are complex. Of particular interest to this study are *complex spatial networks* – that is, complex networks with nodes and/or edges embedded in space (Gastner and Newman 2006; O'Sullivan 2014). A street network is an example of a complex spatial network with both nodes and edges embedded in space, as are railways, power grids, and water and sewage networks (Barthélemy 2011).

A spatial network is *planar* if it can be represented in two dimensions with its edges intersecting only at nodes (Viana et al. 2013). A street network, for instance, *may* be planar (particularly at certain small scales), but most street networks are non-planar due to grade-separated expressways, overpasses, bridges, and tunnels. Despite this, most quantitative studies of urban street networks represent them as planar graphs (e.g., Strano et al. 2013) for tractability because bridges and tunnels are (in some places) reasonably uncommon, and thus the networks are *approximately* planar. However, this over-simplification to planarity for analytical tractability may be unnecessary and can cause analytical problems, as we shall discuss in chapter 5.

Complex networks have been studied extensively by urban scholars and planning researchers. From a qualitative perspective, Castells (e.g., 2009) argues that understanding flows and networks, rather than locations themselves, is critical for understanding cities. From a quantitative perspective, Batty (e.g., 2013b, 2013c) places urban modeling in the context of network evolution and flow. Law (2017) uses topological street network analysis for community detection to discover neighborhood boundaries. Most relevant to this study, however, is the rich body of transportation and urban form studies that use complex street networks for routing and characterizing the structure of cities. In particular, a typology of measures of the complexity of urban form – and particularly street networks – is developed in chapter 4, operationalized in chapter 5, and applied empirically in chapters 6 and 7.

## 3.8. Discussion

Complex systems are systems of interacting components that, through nonlinearity, can produce emergent phenomena and self-organized structure. The Polish mathematician Stanisław Ulam, an early pioneer of nonlinear studies, once observed that talking about "nonlinear science" is akin to "calling the bulk of zoology the study of non-elephants" (Campbell 2004, p. 455). Indeed, nearly all real-world systems are inherently nonlinear, and the emergent phenomena arising from complexity pervade the natural and human worlds. Human societies and cities are examples of large, complex systems.

Principles of complexity have been applied in urban planning from the communicative turn and collaborative rationality, to cellular automata and agent-based urban models, to the design of livable neighborhoods (Batty 1997; Innes and Booher 1999; 2000; 2004; 2010; O'Sullivan and Haklay 2000; Flyvbjerg and Richardson 2002; Healey 2003; 2007; White et al. 2015; Boeing 2015; 2017d). Complexity problematizes rationality and certainty in planning, as behavioral economics has similarly done in the wider social sciences in recent years (e.g., Tversky and Kahneman 1974; Kahneman and Tversky 1979; Johnson and Goldstein 2003; Thaler and Sunstein 2003; Glaeser 2005; Botti and Iyengar 2006; Bouchaud 2008; Kahneman 2011; Marsden et al. 2012; Chatman et al. 2013; Hoch et al. 2015). Complexity may be a lens through which planners can conceptualize and approach "wicked problems" and incrementalism (Lindblom 1959; Cartwright 1973; Rittel and Webber 1973; Christensen 1985; Salet et al. 2013; Altrock 2015; Yamu et al. 2016; for context see also Friedmann 1987; Flyvbjerg 2007; Acey 2016). The emergent features of stability, resilience, robustness, and connectivity are of particular interest to urban scholars. These form a bridge between qualitative theories of cities, such as Castells' spaces of flows, and quantitative studies of the cities – broadly, the study of urban form, design, and transportation.

The next chapter explores these theories of complexity within the discipline of urban design and builds a typology of measures of its complexity outcomes, emphasizing those relevant to street network analysis. It is worth noting that chapters 2 and 3 have emphasized the *process* and *dynamics* of complexity, before turning their attention to the (essentially static) *structure* of networks. The following chapters continue this shift in focus, but this transition from process to structure is not without its thorns. After all, we somewhat disregard the processual complexity of collective human behavior, self-organization, and emergence when we concentrate on the patterns, topology, and structure of networks rather than on the temporal dynamics that operate along them (cf. Barrat et al. 2012; Simmonds et al. 2013; Zhong et al. 2014; Gates and Rocha 2016; Barthélemy 2017).

However, the following chapters – particularly culminating in the discussion in chapter 7 – argue that street networks might be complex either directly through their form and topology (a complexity of pattern), or indirectly as that topology and its attributes influence how an urban system's physical links structure and sustain human interactions, connections, and behavior (a complexity of dynamics). In other words, this physical structure underlies the *concordia discors* of complex human interactions, returning us to the argument of Jacobs (1961), Alexander (1965), Glaeser (2011), and various others that cities exist to connect people. Urban design itself similarly intervenes in the density, connectedness, pattern, texture, façades, configuration, and grain of cities to likewise influence human dynamics and behavior (Whyte 1980; Willis et al. 2004; Rodriguez et al. 2006; Baran et al. 2008; Handy 2015). Moreover, through recursive co-evolution, physical structure influences dynamics, and dynamics in turn produce structure (Cilliers 1998). While such system dynamics are not the focus of this dissertation's empirics, this conceptual bridge links the theory of dynamics discussed in the early chapters with the empirical analysis of urban form and network structure presented in the following chapters.

# Chapter 4: Measuring the Complexity of Urban Form and Design

## 4.1. Abstract

This chapter develops a typology of methods and measures for assessing the complexity of the built form at the scale of urban design. In particular, it extends quantitative methods from network science, ecosystems studies, fractal geometry, and information theory to the outcomes of urban design and the analysis of its qualitative human experience. Metrics at multiple scales are scattered throughout these bodies of literature and have useful applications in analyzing the built form that results from local planning and design processes. This chapter unpacks the connections between neighborhood-scale built form and measures of its complexity, and the typology developed here applies to empirical research of multiple neighborhood types and design standards. Finally, the typology includes several street network-analytic measures of urban form – emphasizing complexity in terms of density, resilience, and connectedness – applied in the subsequent empirical chapters.

## 4.2. Introduction

This chapter examines measures of the complexity of urban form and design. Rich linkages between complexity theory and urban design have been underexplored by researchers at the neighborhood and street scales – the scales of daily human experience and of the *practice* of urban design. The urban design literature frequently cites the value of "complexity" in neighborhood design, but these arguments often lack the theoretical formalism found in complex systems science. Nevertheless, following from Jane Jacobs (1961) and Christopher Alexander (1964; 1965), this body of scholarship argues that neighborhood complexity is essential to the life of the city and the function of its neighborhoods. Prominent design paradigms today, such as Smart Growth and the New Urbanism, frequently speak both directly and indirectly to complexity and notions of complex systems (Duany, Plater-Zyberk & Co. 2001; Talen 2003; Sanders 2008; Congress for the New Urbanism 2015).

If complexity is important in the urban form, planners and designers require better tools to assess design outcomes and understand the built form. This chapter unpacks the connections between neighborhood-scale built form and measures of its complexity. The typology developed here applies to empirical research of multiple neighborhood types and design standards. Finally, this typology includes several street network-analytic measures of urban form, applied in the subsequent empirical chapters. This chapter is organized as follows. First, it briefly reviews the lineage and meaning of "complexity" in urban design theory. Then it explores potential measures of the complexity of urban form at the scale of urban design projects, in four categories: temporal, visual, spatial, and structural. The structural measures are sub-divided into fractal and network measures. Finally, this chapter organizes these various measures into a coherent typology.

## 4.3. Background: Complexity in Urban Design

Urban designers often discuss physical urban form and design projects in terms of "complexity" (e.g., Congress for the New Urbanism 2015). These discussions frequently borrow from the salient concepts of complex systems theory, but as discussed in chapter 3, they often do so loosely, making it difficult to assess claims and project outcomes. Nevertheless, various formulations of complexity have long been regarded as important in urban design, for several reasons. It can contribute to more lively, enjoyable, walkable, healthy, and vital neighborhoods (Jacobs 1961; Calthorpe et al. 1991; Congress for the New Urbanism 1996; Putnam 2001; Macdonald 2002; Carlson et al. 2012; Hamblin 2014; Marshall et al. 2015; Sung et al. 2015; McGreevy and Wilson 2016). It implies urban resilience, robustness, connectivity, and access, playing into wider debates about sustainability and resource efficiency (Peter and Swilling 2014; Pugh 2014; Wells 2014; cf. Deppisch and Schaerffer 2011). Complexity can be emancipatory (Byrne 2003) – improving social equity, spatial distributional justice, adaptiveness, and social contact and exchange (cf. Sennett 1992; Sandercock 2000; Pettigrew and Tropp 2006). This section summarizes this lineage of ideas about urban design and complexity – particularly through the notion that urban design influences the complexity of human habitats at the neighborhood scale and is closely tied to theories of livability. Then, in the subsequent section, we draw on complexity contextually and discuss several potential methods for measuring it.

Urban design is the physical shaping of the public realm and borrows from both architecture and city planning (Moudon 1992; Biddulph 2012). It includes deliberate top-down acts, informal bottom-up acts, and everything in between. The history of urban design reflects normative stances that have shifted through eras of classical formalism, romantic organicism, modernist simplifications, and post-Jane Jacobs gestures toward "organized complexity" (Barnett 2011). Jacobs's notions of complexity and bottom-up urbanism have been embraced by complex systems scholars studying cities – particularly scholars from the physical sciences (Batty 2005a; 2005b; 2008a; Bettencourt and West 2010; Lehrer 2010; Bettencourt 2013a; Bettencourt 2013b; Batty and Marshall 2016; Pollock 2016; cf. O'Sullivan and Manson 2015). However, urban scholars have also criticized her dichotomous theories as overly simplistic, given the role of capital and real estate markets underpinning her famous "sidewalk ballet" (Zukin 2011; Krivý 2016).

Urban design primarily interfaces with complexity through notions of diversity, connectivity, resilience, and livability. Livability has been defined in numerous ways and its meaning has evolved over time, but there is some common ground in the literature (Appleyard and Lintell 1972; Jacobs and Appleyard 1987; Bosselmann et al. 1999). Bosselmann (2008, p. 142) points out that "the original meaning of livability described conditions in neighborhoods where residents live relatively free from intrusions" but that the term has been progressively broadened to include sustainability, safety, comfort, available services, walkability, and transit. Macdonald (2005, p. 14) cites a modern vision of livable neighborhoods that create "lively, safe, and attractive streets, and [provide] public amenities such as parks, community centers, and schools."

Livability is in turn nested within broader debates around urban sustainability and justice, as it is inextricably dependent on the city's ability to meet its residents' ongoing needs into the future (Boeing et al. 2014; Boeing 2016b). Several planning models – some competing, some complimentary – have taken up the mantle of livability in the U.S. today, including smart growth, the new urbanism, traditional neighborhood development, and transit-oriented development. Each promotes a compact urban form, walkability, and improved access to transit. Finally, issues of social justice cannot be ignored in the theorization of livability, as uneven distributions of power, capital, and privilege inevitably cloud the question of livability for whom and at the expense of whom (Evans 2002; Harvey 2010; Boeing 2016a; Barajas et al. 2017).

These definitions imply the importance of physical form and design for various aspects of livability. Indeed, livability is perhaps the key way in which planners engage with neighborhood form to address complexity. Three subcomponents of livability that particularly rely on complexity emerge from the literature. The first is visual complexity: an interrelation of qualities related to perceptible variety that makes public space lively, attractive, and enjoyable. The second is neighborhood completeness: a diverse mixture of amenities in close proximity. The third is connectivity of the circulation network. This body of literature argues that walkability and, in turn, livability rely on completeness and connectivity to be feasible and on visual complexity to be desirable. This has become a key goal of modern urban design.

Urban design's interventions operate primarily at the scales of neighborhoods and blocks – usually with no more than a half mile radius (Boarnet and Crane 2001; Mehaffy et al. 2010; Porta et al. 2014). Metrics for measuring the outcomes of urban design thus must consider the neighborhood scale. Neighborhoods are related to the concept of community, but also have a specific geographic, spatial nature (Larice and Macdonald 2007; Drinan 2015; Talen et al. 2015; Law 2017). They are ubiquitous around the world and play an important role in complex urban systems. Clarence Perry (2007, p. 55) said "an urban neighborhood should be regarded both as a unit of a larger whole and as a distinct entity in itself." The concept of *neighborhood* has also been theorized by complexity scholars. For example, Portugali (2006; Portugali and Stolk 2014) argues that cognitive conceptions of neighborhoods arise out of complex human systems via information compression, an idea based on the reduction of information in synergetics (Haken 2012).

According to Smith (2010, p. 137), "The spatial division of cities into districts or neighborhoods is one of the few universals of urban life from the earliest cities to the present" (cf. Silver 1985; Peterman 2000; O'Sullivan 2009; Rohe 2009; Vanderbilt 2013; Madden 2014). Likewise, Mumford (1961) points out that since the earliest days of cities, natural neighborhoods would form organically around important points like temples. Most pre-twentieth century neighborhoods were "complete" because walking was the most common mode of travel (for modern formulations of the complete neighborhoods paradigm, see San Francisco Planning Department 2008; District of Columbia 2010; City of Portland 2012). Jackson (1985) describes such walking cities as dense and congested, with clear city/countryside distinctions and respectable locations nearest to the center of town, where accessibility was highest. Furthermore, "there were no neighborhoods exclusively given over to commercial, office, or residential functions... public buildings, hotels, churches, warehouses, shops, and homes were interspersed, or often located in the same structure" (ibid., p. 15). Jackson's vignette draws together themes of completeness, connectedness, and accessibility. From the days of the Greek agora to the birth of the automobile, much of urban life was spatially and socially centered on the public street and the public square – spaces symbolic of access, exchange, difference, and challenge (Holston 1989; Carmona 2015; Kimmelman 2016).

Before the revolution in transportation technologies that culminated in the automobile, proximity was paramount and people necessarily lived near employment and retail (Fishman 1987). However, Jacobs (1995) suggests that in response to the dreadful living conditions of industrial-era cities and new enabling technologies, two major manifestos emerged to dominate twentieth century neighborhood planning: Howard's garden cities and modernism's Charter of Athens. While the garden city movement largely respected the neighborhood, its legacy – suburban sprawl – did not (Mumford 1961; 2007). Nor did Le Corbusier and the modernist planners, setting the stage for twentieth century auto-dependency and single-use functional zoning (Hall 1996; Fishman 2003). This effectively became the age of sterile anti-complexity in urban design and land use. Its principles were embodied in utopian plans of urban dispersal, such as Frank Lloyd Wright's Broadacre City (Wright 1932; 1935; Grabow 1977; Nelson 1995; Meis 2014), the functionalist automobile-dominated urban designs of Corbusier (2007a; 2007b), and Robert Moses's "meat ax" carving its way through the disorderly urban fabric of mid-century New York (Caro 1974; cf. Kaufmann 1974; Boeing 2017e).

Scott (1998) critiques the modernist urban designs of cities like Chandigarh and Brasília by contrasting Corbusier's top-down simplified, rational, polished, utopian cities with bottom-up, organically built, messy everyday urbanisms dependent on localized tacit knowledge (cf. Jacobs 1961; Hayek 1974; Holston 1989). These modernist planners and architects confused geometric visual order for well-functioning sustainable social order in the built environment (Roy 2005; Sussman and Hollander 2015). In fact, informality itself may be a defining *sine qua non* of cities and urbanism (Sennett 1992; Sassen 2012; Greenfield 2013; cf. Kriv 2016). Scholars following in the wake of Jane Jacobs have argued that simplified single-purpose urban design destroys functional capacity and synergy. Rather, it is diversity, mixed uses, and complexity – grown naturally over time – that make a community livable. Over-simplified plans and interventions can cut into the living tissue of complex city systems, killing vital social processes. While healthy complex adaptive systems are resilient to perturbation, their resilience and adaptability may be destroyed through too many simplifying interventions (Marshall 2012a).

Yet every built environment has some deliberate design – especially in the public realm. Building façades are architected, roads are engineered, sidewalk widths are selected, and parks are laid out. Moroni (2010, p. 147) calls for rules that are simple, abstract, general, and purpose-independent to move away from a "flexible system of land-use planning" and toward "rules that enable society itself to be highly flexible." To this end, Moroni (2015) suggests urban codes based on principles rather than details, contain few simple rules that remain for long periods of time, give minimal discretion to public officials, and leave flexibility for individual creativity and experimentation. Such codes already exist to some extent in urban design as form-based codes that aim to balance bottom-up flexibility with top-down predictability (Talen 2009; 2011; 2016).

Generative design is a popular framework for conceptualizing such systems in action (Marshall 2004; 2005; Luca 2007; Alexander et al. 2008; Mehaffy 2008; Rakha and Reinhart 2012). Marshall (2012a, p. 203) similarly calls for a "system of planning" in which design and codes work together as a generative system that can give rise to a kind of emergent urbanism, with no guarantee that it will be optimal (cf. Eikelbloom et al. 2015). However, development control can then be exercised to nudge what emerges toward the public interest (ibid; cf. Allen 2012). This is a middle ground between attempts to plan everything and attempts to plan nothing. Marshall suggests such a system would enable urban design and planning to deliver true functional complexity for neighborhoods. Jane Jacobs (1961) similarly argued that the role of planners is to generate diversity and supply what is lacking in a neighborhood.

According to this mainstream of scholarship, urban design and planning can foster diversity, connectedness, complexity, resilience, and robustness – elements of a healthy complex adaptive system. Yet an open question remains. Beyond these qualitative formulations of complexity in urban form and design, how might it be defined quantitatively – especially at the neighborhood and block scale? A stream of planning literature has considered quantitative measures of the urban form, but without explicitly engaging with complexity (e.g., Cervero and Kockelman 1997; Song and Knaap 2004; Tsai 2005; Clifton et al. 2008; Ewing and Cervero 2010; Schwarz 2010; Song et al. 2013b). The next section defines complexity contextually and explores several potential methods for measuring it directly.

# 4.4. Measures of Complexity in Urban Form and Design

### 4.4.1. Overview

Various complexity metrics at multiple scales, from metropolitan to neighborhood to building, are scattered throughout different bodies of literature. Lloyd (2001) surveyed and categorized measures of complexity across numerous fields of inquiry. Bourdic et al. (2012) provide an overview of cross-scale spatial indicators and briefly touch on the neighborhood design assessment criteria of LEED-ND. Additional surveys and analyses of complexity indicators for ecosystems and cities have been produced by Parrott (2010) and Salat et al. (2010), mostly focusing on urban processes.

To take a step back, at a higher level, Shiner et al. (1999) characterize complexity from the perspective of order and disorder. They present three broad categories of complexity, depicted in Figure 4.1. Category I is positively correlated with disorder and includes algorithmic complexity and most measures of entropy. Here, complexity is highest when objects are scrambled-up with the greatest variety and diversity. Category II is a convex function of disorder, peaking at some midpoint between order and disorder. This balances between variety and structure and conforms to traditional definitions of complex adaptive systems (e.g., Gershenson and Fernández 2012). Category III takes complexity to be related more purely to order or structure alone, and includes notions of self-organization and emergence in which structure emerges from previous disorder.

This dissertation uses this framework of categories to consider measures of complexity in the built environment. While these categories are mutually incompatible (e.g., complexity cannot be simultaneously high *and* low when disorder is high), they do reflect different *aspects* of complexity that planners must consider and assess. The remainder of this study touches on all three categories as measures of complexity in urban design, depending on the context and character, but focuses on the second: the balance between structure and variety/messiness. We return to this framework again in the discussion that concludes this chapter, after presenting a typology of complexity measures for the urban form.

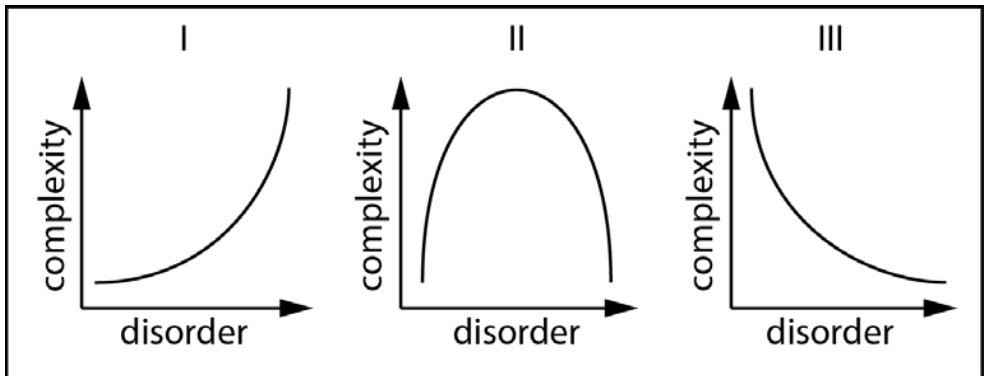

Figure 4.1. Three different categories of complexity. Category I increases monotonically with disorder. Category II is convex, peaking at a midpoint between order and disorder. Category III decreases monotonically as disorder increases. Adapted from Shiner et al. (1999).

The anthropologist Claude Lévi-Strauss called cities "the most complex of human inventions… at the confluence of nature and artifact" (1992, p. 137). If cities are complex systems, then indicators of their complexity – in all its many, varying facets – would be useful for grounding conversations and evaluating patterns and processes. Such an indicator of complexity could be a system-level state variable. However, just how to measure the complexity of a city system remains an open question. Further, how does this sense – or preference – vary from person to person and culture to culture? We may have some intuitive sense of the complexity of a place simply by observing it or moving through it, but how can this be formalized?

On one hand, a neighborhood can be examined as an urban ecosystem – a human habitat – that is a dynamical complex system. Thus, state variables such as population, density, employment, wealth, traffic volume, etc. can be (potentially) identified and (potentially) calculated at various scales to describe the state of the system as it changes over time. The system's dynamics can be explored and modeled with mathematical equations, statistical regressions, machine learning algorithms, cellular automata, or agent-based models to describe the processes occurring in the system. Forrester (1969) was an early pioneer of applying systems dynamics to cities, studying their stocks and flows. His use of differential equations and stock/flow modeling has been extended to cycles of urbanization and suburbanization (Orishimo 1987) and to the dynamics of parking (Cao and Menendez 2015). Such dynamic measures are less useful for the characterization and analysis of urban form.

On the other hand, a neighborhood can be examined as an output or product of human behavior and production. This focuses on the *physical form* of the neighborhood rather than its dynamical processes. Through co-evolution, as discussed in chapter 3, humans both shape their neighborhoods and are in turn shaped by them (Lynch 1954; Lynch and Rodwin 1958). The physical patterns that result constitute the urban form and can be examined in terms of network analyses, fractal structure, diversity (of various sorts), and information entropy. At a higher level of abstraction, neighborhood complex systems can be analyzed in terms of their resilience, robustness, and adaptiveness. How might the system's dynamics respond to perturbation given its spatial patterns, structure, connectedness, and efficiency?

The following framework borrows, adapts, and reformulates relevant metrics to measure complexity at the neighborhood scale, touching on temporal measures but focusing on spatial and structural measures such as those of urban morphology (Talen 2003; Marshall and Caliskan 2011). In particular, it provides a quantitative framework that accounts for both traditional urban planning/design measures as well as more abstract measures arising from the complexity sciences. It is worth noting that this framework is not aimed at quantifying all aspects of "good" neighborhood design. Rather it intends to formalize and measure the indistinct notion of *complexity* as it applies to urban design. Qualities related to vitality, sustainability, sense of place, and other prominent qualities may overlap in some ways with the complexity metrics in this typology, but they are otherwise not the focus of this work.

## 4.4.2. Temporal Measures of Urban Form

The first group of measures in this framework is temporal measures. Temporal measures describe time series data and in turn system dynamics. Such techniques include embedding the time series in state space, uncovering underlying attractors, estimating Lyapunov exponents (as discussed in chapter 2), and analyzing the system from an information theoretic perspective, such as Shannon entropy.

Nonlinear analysis techniques from the physical sciences, such as reconstructing attractors or estimating Lyapunov exponents, have not been found to be particularly effective in the ecology literature (Parrott 2010). Information theory, however, provides some measures of complexity that may be applied to urban design at the neighborhood scale. Shannon's (1948) original theory of information entropy concerns the average amount of information contained in the revelation of a message or event. *Shannon entropy* indicates that the more types of things there are and the more equal each type's proportional abundance is, the less predictable the type of any single object will be. This can be applied to abstract messages, time series, or spatial diversity, as discussed below. Entropy is lowest when the system is highly ordered and thus completely predictable. It is highest when the system's disorder is highest. Such a category I measure thus emphasizes disorder rather than peaking at some point between order and disorder (Batty 2005a, Yeh and Li 2001).

Derived from Shannon entropy, *mean information gain* assesses how much new information is gained from each subsequent datum in a time series (Proulx and Parrott 2008) and *fluctuation complexity* measures the amount of structure within a time series by evaluating the order of and relationship between values in the series. In other words, how likely is it we will observe some value *a* proximately after some other value *b*. Shannon entropy, mean information gain, and fluctuation complexity can be used to assess time series data arising from urban systems. However, more usefully, they might be abstracted and re-appropriated to evaluate the human experience of moving through the physical space that results from urban design (Kuper 2017).

### 4.4.3. Visual Complexity of Urban Form

In a simplified, low-information urban landscape, little new information is gained by a pedestrian through the visual revelations of each passing step. In a highly complex urban environment (in terms of a category I measure), however, an individual will be bombarded with enormous amounts of new information as he or she moves through space. In these cases, space is the medium and the unfolding visual tableau is the message. This message could be discretized into arbitrary units such as meters, or into units relative to the specific urban landscape, such as street blocks or land parcels.

Much of the research on human perceptions of the built environment follows in the wake of Gibson's (1979) ecological framework and Appleton's (1975; 1984) prospect-refuge theory (e.g., Tveit et al. 2006; Ode et al. 2010; Sang et al. 2015; Dosen and Ostwald 2016). Clifton et al. (2008) discuss qualities of the urban form and human perceptions at multiple scales. For neighborhood and street scale urban design, perceptions of human scale are related to building heights and signage, perceptions of coherence are related to consistency of building heights, and sense of enclosure is related to building/element spacing and tree canopy. "Good" visual complexity tends to reach an optimum at some balance point between order and disorder, with "unity in variety", implying a category II convex measure of complexity (Elsheshtawy 1997; Gunawardena et al. 2015).

Ewing and Clemente (2013) performed a literature review that yielded 51 perceptual qualities of urban environments, eight of which were selected for further study because of their importance across the literature: imageability, enclosure, human scale, transparency, coherence, legibility, linkage, and visual complexity (see also Ewing and Handy 2009). These researchers related visual complexity to the number of perceptible differences a person is exposed to while moving through the city. They found that humans prefer to experience information at a comfortable rate – too little deprives the senses and too much overloads them. Ewing and Clemente also found that good visual complexity depends on variety: types of buildings, design details, street furniture, signage, human activity, sunlight patterns, and the rich textural details of street trees and the urban forest (see also Schwab 2009; Pham et al. 2017). Complexity is lost when design becomes too top-down, controlled, and predictable in modern large-scale master plans. Poor complexity exists when urban design elements are too few, are too similar and predictable, or are too disordered to be comprehensible (ibid.). In this formulation, complexity follows a category II convex function with a maximum value at some midpoint between order and disorder.

Based on their literature review, the researchers develop a field manual for measuring visual complexity. It is operationalized in five steps. First, count the number of buildings within the study area. Second, count basic and accent building colors. Third, record the presence of outdoor dining on each block as a binary value, present/not. Fourth, count the individual number of pieces of public art within the study area. Fifth, count the number of pedestrians in the study area. These measures of complexity are part of a larger toolkit for measuring urban design according to the eight perceptual qualities cited earlier (ibid.). Cavalcante et al. (2014) provide an alternate, statistical image processing measure of urban visual complexity.

Fishman (2011) proposes that there exists a significant conflict between the two primary paradigms of modern urban design. The first paradigm, spearheaded by the modernists, seeks to *open up* the dense and messy urban fabric with towers-in-the-park, spacing, highways, and technology. The second, espoused by traditionalists and neotraditionalists, seeks instead to *enclose space* through human-scale architecture, walkability, and a dense, complex, organic urban fabric. Jacobs and Appleyard (1987) argue that buildings in varied arrangements (i.e., in accordance with Fishman's second paradigm) enhance visual complexity, but interminable wide buildings – a hallmark of modernist design – detract from it (see also Sussman and Hollander 2015). Jacobs (1995) argues that buildings need multiple varied surfaces for light to move constantly over to generate visual complexity. Macdonald (2005; see also Punter 2003) explores how Vancouver generates visual complexity to put proverbial eyes on the street, with many entryways and interesting ground-level design.

Slow-moving pedestrians need a high level of complexity to hold their interest, but fast-moving motorists find that same environment chaotic. Dumbaugh and Li (2011) find that urban designs that balance vehicle speeds, visual complexity, and traffic conflicts can increase motorist awareness, decrease collisions, and improve pedestrian safety. While streets obviously provide circulation functions, they also provide essential social and economic functions that must be considered in their design (Jones et al. 2008). Marshall (2012a) contends that urban environments with perceptual richness are more interesting and enjoyable for humans, possibly because our species evolved in natural environments with a high degree of visual complexity. Thus, appropriate visual complexity serves as a key component of livability because it creates rich, enjoyable, safe environments for humans.

### 4.4.4. Spatial Measures of Urban Form

The urban form that emerges from urban design is spatially embedded and can be characterized by various spatial measures of complexity. These measures assess the character of spatial patterns of the system at snapshots in time rather than looking at dynamics over time. Shannon entropy has been used to measure urban complexity (Batty 2005b) and mean information gain has been used to measure ecosystem spatial complexity (Proulx and Parrott 2008). Yeh and Li (2001) used entropy to monitor and measure urban sprawl. Applying these information theoretic metrics to space usually entails assessing raster data for predictability.

Diversity, however, is the most common spatial measure of complexity in the urban design and planning literature. Diversity is important for several reasons. Social diversity may enhance learning, adaptation, and unexpected social mixing. Jane Jacobs (1961) praised diverse land uses for their ability to create synergies from complementary functions. Boarnet and Crane's (2001) behavioral framework of the demand for travel fundamentally argues that urban design influences the (time) cost of travel by placing origins and destinations in closer or further proximity to one another (see also Cervero and Landis 1995; Giuliano 1995; Crane 2000; Stead and Marshall 2001; Handy et al. 2005; Cao et al. 2007; Glaeser et al. 2008; Chatman 2009; Greene et al. 2011). Cervero and Kockelman (1997) also argue for land use diversity as a key feature shaping human travel behavior in urban environments.

Salat et al. (2010) identify three types of urban spatial diversity related to complexity: diversity among similar objects, diversity in spatial distribution, and diversity of scale. Diversity among similar objects refers to different characteristics of the same type of thing – for example, the "thing" might be humans and the characteristics might be income, race, employment, education, and so forth. It does however imply that even distributions are optimal in that they score the highest. This is a questionable reflection of complexity and a risky goal for central planning. Measures of dispersion and physical shape are also useful in characterizing the uniformity, randomness, or spatial complexity of ecosystems and could be applied to the built environment as well.

Wissen Hayek et al. (2015) use UrbanSim (Waddell 2002; Krizek and Waddell 2002) and measures of land use mix and density to evaluate the quality of the neighborhood-scale urban environment. The Simpson diversity index measures the diversity of objects in total across space, and is a common measure of land use entropy (i.e., land use mix) in the urban planning literature. This index is often called the Herfindahl-Hirschmann index in economics and the Probability of Interspecies Encounter in the ecology literature. It is an *integral measure* that considers land use in a district as a whole, ignoring microscale structure and pattern (Song et al. 2013a).

In contrast, a *divisional measure* is sensitive to patterns *within* a district. This is a superior type of measure when considering questions of scale. The *dissimilarity index* measures how the land use mix within a district relates to the mix across the area as a whole – for two land use types, and for multiple (ibid.; cf. Decraene et al. 2013). Other measures of dissimilarity are explored by Bordoloi et al. (2013). These spatial distributions of objects concern how equitably some set of desirable or undesirable objects is spread across the city. For example, are all schools clustered in wealthy neighborhoods rather than being distributed evenly among all neighborhoods? Are waste treatment facilities clustered in poor neighborhoods rather than being distributed evenly among all neighborhoods? On the other hand, in a complex system, centers and clusters may form for inevitable or even "good" reasons. Agglomeration economies can cause job centers to cluster in certain areas (Jacobs 1969; Glaeser 2011; Sevtsuk 2014). Ecosystem services of urban forests are highest when green spaces are concentrated and clustered rather than evenly distributed throughout urban development (Krasny et al. 2014; Stott et al. 2015).

Diversity of scale addresses this specific issue further. Certain distributions within a complex system may be more efficient when they follow a power law (or more realistically – as we shall discuss in chapter 7 – a lognormal distribution) rather than an even distribution. For example, it is not likely ideal for a neighborhood to have the same number of arterial roads, collector streets, and local streets. Rather, there might be a small number of large arterial roads, a medium number of mid-sized collector streets, and a large number of capillary local streets. Murcio et al. (2015) similarly use urban transfer entropy to examine multi-scale urban patterns and flows.

### 4.4.5. Structural Measures of Urban Form: Fractal

Related to diversity, questions of scale and topological *structure* are addressed in this subsection. Measures of structure assess the internal physical configuration of a system. They have been applied to cities and are perhaps the most useful measures of the complexity outcomes of urban design because they characterize that which is most dependent upon the urban design process: physical structure and arrangement. Density itself might be a simple proxy for complexity as a greater number of things operating in the same area imply structure and connectivity. At the scale of urban design, these structural measures fall primarily into two categories: measures of fractal structure and network analysis.

Fractal structure refers to the "roughness" and self-similarity of some object, and how its detail relates to the scale at which it is observed. As discussed in chapter 2, fractals have a similar structure at every scale (Frame et al. 2015). But in the real world, fractals are not perfect and do not exist at all spatial scales – from the infinitesimal to the infinite – as abstract mathematical fractals do. However, self-similarity of patterns and structure over multiple scales exist throughout nature. Batty (e.g., 2005) has long demonstrated how city structure and urban peripheries are fractal.

Fractal structures tend to be distributed according to a power law. As mentioned earlier, in a power law distribution, there are few large items, a medium number of medium-sized items, and many small items. Consider the earlier example of an urban street network. At the largest scale, the city has a few major arterial roads and boulevards that serve as the key arteries for system-wide traffic circulation. But if we zoom into this picture, a larger number of mid-sized collector roads appear, branching off from these few large arteries. As we zoom in further to a fine scale, a denser mesh of local streets appears, branching off from these collector roads. Similar fractal analyses have been applied to the distribution and scale of other urban structures such as buildings as well as land uses. The fractal dimension, $D$, is a statistical measure of how a form's complexity changes with regard to the scale at which it is measured:

$$N \propto \varepsilon^{-D}$$

$$\log_\varepsilon N = -D = (\log N)/(\log \varepsilon)$$

In these formulae, $N$ represents the number of new objects generated as scale transitions and $\varepsilon$ is the scaling factor. This log-log ratio is similar to elasticities in economics. The fractal dimension of an object with one topological dimension refers to its space-filling characteristics that, through self-similarity, become a bit more than a one-dimensional line yet a bit less than a two-dimensional plane. Measures of fractal dimension include the Hausdorff dimension and the box-counting dimension (Shen 2002). For example, a Koch curve has a Hausdorff fractal dimension $D = -\log(4)/\log(1/3) = 1.26$.

The concept of fractal dimensions can also be applied to two dimensional surfaces, such as the surface of a city, the surface of a building, or the surface of elements of urban design (Cooper et al. 2013). The fractal dimension is closely related to the qualities of visual complexity in urban design and public architecture, discussed earlier. While modernist architecture sought to erase complexity with simplified, segregated, sterile forms, both traditional architecture and today's ideal tend to emphasize organic forms with rich detail at multiple scales (Marshall 2008). For instance, Salingaros (1998; 2000a;

2000b; 2001), argues (albeit somewhat abstrusely) that architecture and urban design must utilize fractal design to embrace the structure and organization of organic forms. The Eiffel Tower is an example of a built form that exhibits fractal structure. As Mandelbrot (1983, p. 131) puts it, "(well before Koch, Peano, and Sierpinski), the tower that Gustave Eiffel built in Paris deliberately incorporates the idea of a fractal curve full of branch points."

### 4.4.6. Structural Measures of Urban Form: Network

Beyond fractals, network science provides a second crucial lens with which to examine structure. Accessibility is a useful measure of urban design and is related to network analysis (Hansen 1959; Samaniego and Moses 2008; Levinson 2012). It concerns proximity, transportation mobility, and social interaction within the public sphere (Levine et al. 2012). Popular "walkability" tools – such as WalkScore and Walkonomics – and urban modeling tools such as pandana use street networks to determine accessibility (Foti 2014). Urban networks – considered here as primal, non-planar, weighted multidigraphs with self-loops – can be measured for their category II complexity based on their structure, particularly in terms of density, resilience, and connectedness. Such measures extend the toolkit commonly used by urban morphologists (Talen 2003). Extended definitions of and algorithms for the following measures can be found in Newman (2010) and Barthélemy (2011). The measures here are divided into metric measures and topological measures, but it is worth noting that in a planar graph, topological and metric properties are interrelated (Masucci et al. 2009).

*Metric structure* can be measured in terms of lengths (i.e., edge weights) and area and represents common transportation-design variables (e.g., Cervero and Kockelman 1997; Ewing and Cervero 2010). The *average street length*, the mean edge length in the undirected representation of the graph, serves as a linear proxy for block size and indicates how fine-grained or coarse-grained the network is (see Sevtsuk et al. [2016] for a discussion of block size). The *node density* is the number of nodes divided by the area covered by the network, and the *intersection density* is the node density of the set of nodes with more than one street emanating from them (thus excluding dead-ends). The *edge density* is the linear sum of all edge lengths divided by the area, and the *street density* is the linear sum of all edges in the undirected representation of the graph divided by the area. These four density measures all provide further indication of how fine-grained the network is. Finally, the average circuity divides the sum of all edge lengths by the sum of the great-circle distances between the nodes incident to each edge (cf. Levinson and El-Geneidy 2009; Barthélemy 2011; Strano et al. 2012; Giacomin and Levinson 2015). This circuity measure is the average ratio between an edge length and the straight-line distance between the two nodes it links.

Street connectivity *metrics* can behave inconsistently based on how study areas are drawn (Knight and Marshall 2015). Alternative *topological measures* of street network structure may more robustly indicate the connectedness and resilience of the network, and how these values are distributed. The *average node degree*, or mean number of edges incident to each node, quantifies how well the nodes are connected, are average (cf. Kansky's [1963] $\beta$ index). Similarly, but more concretely, the *average streets per node* measures the mean number of streets (i.e., edges in the undirected representation of the graph) that emanate from each node (i.e., intersections and dead-ends). This adapts the average node degree for physical form rather than circulation and flow. The distribution and proportion of number of streets per node characterizes the type, prevalence, and spatial distribution of intersection connectedness and dead-ends in the network.

The *eccentricity* of a node is the maximum of the shortest-path distances (weighted by length) between it and each other node in the network, and represents how far the node is from the node that is furthest from it (Urban and Keitt 2001). The *diameter* of a network is the maximum eccentricity of any node in the network and the *radius* of a network is the minimum eccentricity of any node in the network (Hage and Harary 1995). The *center* of a network is the node or set of nodes whose eccentricity equals the radius and the *periphery* of a network is the node or set of nodes whose eccentricity equals the diameter. These distances measure network complexity in terms of size, structure, and shape.

*Connectivity* measures the minimum number of nodes or edges that must be removed from a connected graph to disconnect the network (Urban and Keitt 2001; cf. Dill [2004] for a discussion of street connectivity in a less formal sense). This is a measure of resilience as complex networks with high connectivity provide more routing choices to agents and are more robust against failure. However, node and edge connectivity is less useful for approximately planar networks like street networks: *most* street networks will have a connectivity value of 1, because the presence of a single dead-end indicates that the removal of just one node or edge will disconnect the network. Instead, the *average node connectivity* of a network – the mean number of internally node-disjoint paths between each pair of nodes – more usefully represents the expected number of nodes that must be removed to disconnect a randomly selected pair of non-adjacent nodes (Beineke et al. 2002; Dankelmann and Oellermann 2003). This is a useful indicator of resilience.

Other measures of connectedness – such as intersection density, node degree distribution, and centrality/clustering (discussed below) – may capture the nature of a street network's resilience and connectedness better than standard node or edge connectivity can. Networks with low connectivity may have multiple single points of failure, leaving the system particularly vulnerable. This can be seen in urban design through permeability and choke points: if circulation is forced through single points of failure, traffic jams ensue and circulation networks can fail. Connectivity has also been linked to street network pedestrian volume (Hajrashouliha and Yin 2015; see also Jiang 2009; Jiang et al. 2009).

Network distances, degrees, and connectivity are significantly constrained by spatial embeddedness and approximate planarity (O'Sullivan 2014), so measures of clustering and centrality may better reveal topological structure and its distribution. The *clustering coefficient* of a node is the ratio of the number of edges between its neighbors to the maximum possible number of edges that could exist between these neighbors (Opsahl and Panzarasa 2009). The *weighted clustering coefficient* weights this ratio by edge length and the *average clustering coefficient* is the mean of the clustering coefficients of all the nodes in the network. These measure connectedness and complexity by how thoroughly the neighbors of some node are linked to each other. Jiang and Claramunt (2004) extend this coefficient to neighborhoods within an arbitrary distance, rather than just proximate, to make it more applicable to urban street networks.

Measures of centrality indicate the most important nodes in a network (Huang et al. 2016; Zhong et al. 2017). *Betweenness centrality* assesses the importance of a node by evaluating the number of shortest paths that pass through it (Freeman 1977; Barthélemy 2004; Ermagun and Levinson 2017). The average betweenness centrality is the mean of betweenness centralities of all the nodes in the network (Barthélemy 2011). In particular, the maximum betweenness centrality in a network specifies the proportion of shortest paths that pass through the most important node. This is an indicator of resilience:

networks with a high maximum betweenness centrality are more prone to failure or inefficiency should this single choke point fail. Betweenness centrality can also be calculated for weighted networks (Barrat et al. 2004). Barthélemy et al. (2013) uses betweenness centrality to identify top-down interventions versus bottom-up self-organization and evolution of the urban fabric in Paris.

*Closeness centrality* represents, for each node, the reciprocal of the sum of the distance from this node to all others in the network (optionally weighted by length): that is, nodes rank as more central if they are on average closer to all other nodes (Wang et al. 2011). The *average closeness centrality* is the mean of the closeness centralities of all the nodes in the network. *PageRank* – the algorithm Google uses to rank web pages – is a variant of network centrality, namely eigenvector centrality (Brin and Page 1998). PageRank ranks nodes based on the structure of incoming links and the rank of the source node, and may also be applied to street networks (Jiang 2008; Agryzkov et al. 2012; 2013; Chin and Wen 2015; Gleich 2015). Measures of centrality are typically used in combination to assess street networks. Porta et al. (2006a; 2006b; 2010) demonstrate a multiple centrality assessment methodology for analyzing urban street networks and identify signatures and differences between planned and self-organized cities. Crucitti et al. (2006a; 2006b) examine closeness, betweenness, and information as measures of urban network centrality. The Urban Network Analysis Toolbox (Sevtsuk and Mekonnen 2012) analyzes betweenness, closeness, and accessibility in urban street networks.

Finally, it is worth mentioning space syntax theory and dual graphs. The street networks discussed so far are *primal*: the graphs represent intersections as nodes and street segments as edges. In contrast, a *dual graph* inverts this network topology: a city's streets are represented as nodes and the intersections are represented as edges. Such a representation seems a bit odd, but provides certain advantages in analyzing the network topology (Crucitti et al. 2006a; 2006b). Dual graphs form the foundation of space syntax, another method of analyzing urban networks and configuration. Space syntax analyzes axial street lines and measures the depth from some network edge to others (Hillier et al. 1976; cf. Ratti 2004). Marcus and Legeby (2012) use space syntax to measure social capital in neighborhoods, through an explicit urban complexity lens. Jiang and Claramunt (2002) integrate an adapted space syntax – compensating for difficulties with axial lines – into computational GIS. Space syntax has formed the basis of many other adapted approaches to analytical urban design (e.g., Karimi 2012). This present study, however, focuses almost entirely on primal graphs because they retain all the geographic, spatial, metric information essential to urban form and design that space syntax disregards in its dual graphs (Ratti 2004; Ravulaparthy and Goulias 2014).

# 4.5. Typology of Complexity Measures

All of these methods of assessing the complexity of urban design, primarily at the neighborhood scale, can be fit together into a preliminary typology. Here the measures are grouped into five types: temporal, spatial, visual, fractal, and network. While temporal measures are ideal for assessing the complexity of dynamics and process, the spatial, visual, and structural (i.e., fractal and network) measures seem most promising for measuring *physical* complexity at the scale of urban design. In particular, the following three chapters explore the network measures in depth.

| Category | Measure of complexity | Description |
|---|---|---|
| Temporal | Embedding time series | Examine variables in state space to reveal possible deep structure and patterns in data |
| Temporal, Spatial | Shannon entropy | How unpredictable a sequence is, based on number of types and proportional abundance |
| Temporal, Spatial | Mean information gain | How much new information is gained from each subsequent datum |
| Temporal | Fluctuation complexity | Amount of structure within a time series |
| Temporal, Spatial | Urban Transfer Entropy | Analytic tool for examining multi-scale urban patterns and flows |
| Visual | Ewing and Clemente field guide | Set of methods for assessing the physical, visual complexity of the streetscape |
| Visual | Cavalcante streetscape measure | Image processing method to assess visual complexity on contrast and spatial frequency |
| Spatial | Simpson diversity index | Assesses land use mix: how homogeneous or heterogeneous is the area of analysis? |
| Spatial | Dissimilarity index | How does the land use mix within a subarea relate to the mix across the entire area? |
| Fractal | Hausdorff fractal dimension | How a form's complexity changes with regard to the scale at which it is measured |
| Fractal | Box-counting fractal dimension | How a form's complexity changes with regard to the scale at which it is measured |
| Spatial, Network | Destination accessibility | A function of land use entropy, amenity distribution, and network structure |
| Network | Average streets per node | How well connected and permeable the physical form of the street network is, on avg |
| Network | Proportion of streets per node | Characterizes the type, prevalence, and spatial distribution of intersection connectedness |
| Network | Average street length | How long the average block is between intersections; proxy for block size |
| Network | Node/intersection, edge/street density | How fine- or coarse-grained the street network is |
| Network | Average circuity | How similar network distances are to straight-line distances |
| Network | Diameter/periphery, radius/center | Measure network complexity in terms of max/min size, structure, and shape |
| Network | Node/edge connectivity | What is the minimum number of nodes/edges that must be removed to disconnect network? |
| Network | Average node connectivity | Average nodes that must be removed to disconnect some pair of non-adjacent nodes |
| Network | Clustering coefficient | Extent to which the neighbors of some node are linked to each other |
| Network | Average clustering coefficient | Mean of the clustering coefficients for all nodes |
| Network | Betweenness centrality | The importance of a node in in terms of how many shortest paths use that node |
| Network | Average betweenness centrality | Mean of the betweenness centralities for all nodes |
| Network | Closeness centrality | Nodes rank as more central if they are on average closer to all other nodes |
| Network | Average closeness centrality | Mean of the closeness centralities for all nodes |
| Network | PageRank | Ranking of node importance based on structure of incoming links |
| Network | Multiple centrality assessment | Uses primal, metric graphs to examine multiple indices of centrality |
| Network | Space syntax | Uses dual, topological graphs to examine closeness centrality of a named street |

Table 4.1. Typology of measures of the complexity of urban form/design drawn from the discussion in section 4.4.

# 4.6. Discussion

Practitioners and theorists have expounded on complexity's value long before and long after the days of Jane Jacobs. Complexity underlies urban resilience and sustainability planning (Jabareen 2013; Peter and Swilling 2014; Wells 2014; Mattsson and Jenelius 2015). Path dependence, hysteresis, and historical accidents all arise from complex systems and drastically affect the trajectory of urban form (Siodla 2015). These features are both products of urban design and constraints on urban design. More complex urban environments are more resilient, robust, and provide greater opportunities for social encounter, mixing, and adaptation through social learning. Complexity entails greater connectivity, diversity, variety, and sustainability. Today, prominent urban design movements such as the new urbanism and smart growth openly embrace the notion of complexity (Duany, Plater-Zyberk & Co. 2001; Talen 2003; Sanders 2008; Congress for the New Urbanism 2015).

This chapter sought to refine what "complexity" means in this context and in turn provide ways to measure it. Of the three categories of complexity in the framework presented in section 4.4.1, category I seems most appropriate for conceptualizing the complexity of *difference*: how scrambled-up land uses and socioeconomic traits are. In other words, planners must consider and track this category of complexity when assessing variety, be it land use entropy, amenity accessibility, or social mixing and encounter. Category II encapsulates the notion of *organized complexity*, where a balance between chaos and order is desirable, such as in visual complexity and structural complexity (Montgomery 1998).

Take the example of visual complexity discussed in section 4.4.3: too much disorder (overstimulation and bewilderment) is as undesirable as too much order (sterility and monotony). Category III is most useful for considering *ordered* elements of urban design, including when some have self-organized from an original disordered state. The gridded downtowns of Portland, New York, and other similar cities are high in category III complexity – as discussed in the following chapters. However, it was a similar lust for orderliness above all else that obsessed the modernist planners in their quest for perfect organization and rational geometric logic (Boyer 1983) – that is, a quest to erase the city's categories I and II complexity. This raises a question we return to later in this dissertation: is maximum order not the antithesis of complexity?

The typology of complexity measures presented in section 4.5 draws from different scientific disciplines to offer various measures of complexity that apply to urban form and particularly to urban design's scale of intervention. In particular, the measures of network structure characterize the complexity of the circulation network in terms of density, resilience, and connectedness – concepts leveraged throughout the second half of this dissertation. These attributes are rooted in urban planning and impact how an urban system's physical form influences and structures complex human interactions and connections – thus linking structure and dynamics.

*Density* refers to the concentration of elements per unit area, and in this study specifically refers to node, intersection, edge, and street density as discussed in Table 4.1 and section 4.4.6. *Connectedness* comprises a basket of attributes related to permeability, path routing, and node degrees. The average number of streets per node and its related proportion (Table 4.1) characterize the type, prevalence, and spatial distribution of intersection connectedness. Similarly, clustering coefficients, node connectivity, edge connectivity, and most usefully average node connectivity – i.e., the average number of nodes that must be removed to disconnect a randomly selected pair of non-adjacent nodes – can characterize how thoroughly linked and permeable a network is. In turn, as discussed in chapter 3, network *resilience* refers to the ability to recover or maintain similar functionality after a perturbation. Connectedness, along with centralities and measures of flow, can indicate how a street network can be resilient against floods, earthquakes, traffic collisions, or congestion and disruptions of other sorts (Batty 2013; Mariusz and Piotr 2014; Wang 2015).

The analytical framework developed here is generalizable to empirical research of multiple neighborhood types and design standards. In particular, network-analytic measures in this typology are applied empirically in the next two chapters. Chapter 5 presents a new toolkit for acquiring, constructing, analyzing, and visualizing urban street networks and demonstrates it with a case study. Chapters 6 and 7 conduct empirical studies of street networks at multiple scales using the metrics introduced here in chapter 4 and operationalized in the toolkit presented in chapter 5.

# Chapter 5: Acquiring, Analyzing, and Visualizing Street Networks

## 5.1. Abstract

Urban scholars have studied street networks in various ways, but there are data availability and consistency limitations to the current urban planning/street network analysis literature. To address these challenges, this chapter presents OSMnx, a new tool to make the collection of data and analysis of street networks simple, consistent, automatable and sound from the perspectives of graph theory, transportation, and urban design. OSMnx contributes five significant capabilities for researchers and practitioners: first, the automated downloading of political boundaries and building footprints; second, the tailored and automated downloading and construction of street network data from OpenStreetMap; third, the algorithmic correction of network topology; fourth, the ability to save street networks to disk as shapefiles, GraphML, or SVG files; and fifth, the ability to analyze street networks, including calculating routes, projecting and visualizing networks, and calculating metric and topological measures. These measures include those common in urban design and transportation studies, as well as advanced measures of the structure and topology of the network.

## 5.2. Introduction

As discussed in chapters 3 and 4, we can study the structure and interactions of complex systems through their networks. While physicists tend to look for simple models with few parameters to study complex systems (Barthélemy 2017), Cilliers proposes that "models of complex systems will have to be as complex as the systems themselves" (1998, p. 58). What might this mean for modeling complex systems like cities, using geospatial data? In his 1946 short story *On Exactitude in Science*, Jorge Luis Borges (1998, p. 325) wrote:

> …In that Empire, the Art of Cartography attained such Perfection that the map of a single Province occupied the entirety of a City, and the map of the Empire, the entirety of a Province. In time, those Unconscionable Maps no longer satisfied, and the Cartographers Guilds struck a Map of the Empire whose size was that of the Empire, and which coincided point for point with it.

In spite of – or possibly because of – its implausibility, Borges's 1:1 map of the empire has become a popular reference point among scientists and philosophers. Eco (1995) facetiously deconstructs its logistics and semiotics, before concluding that "every 1:1 map of the empire decrees the end of the empire as such and therefore is the map of a territory that is not an empire" (p. 106). McConchie (2016) similarly evokes Borges to consider possible futures of OpenStreetMap, a collaborative world mapping project that conceptually lies somewhere between Google Maps and Wikipedia. In one possible future, OpenStreetMap receives so many user contributions of trivial and unnotable geospatial objects (such as individual curbs, bushes, or paint markings on streets) that it becomes burdened with obsolete and unmaintainable data. It is eventually abandoned, mimicking the fate of Borges's map. But in another fanciful possibility, OpenStreetMap achieves some critical mass of users such that someday in the far future it approaches a well-maintained 1:1 representation of the world. Futuristic technological and geographical utopianism aside, OpenStreetMap today provides researchers with a massive geospatial data repository with overall good coverage and quality. This chapter examines this data source and presents new methods to study city form and urban street networks with it.

Street network analysis has held a prominent place in network science ever since Euler presented his famous Seven Bridges of Königsberg problem in 1736 (Devlin 2000; Bonchev and Buck 2005; Derrible and Kennedy 2009). Modern urban scholars and planners have studied street networks in numerous ways. Some studies have focused on the urban form (e.g., Southworth and Ben-Joseph 1997; Fecht 2012; Strano et al. 2013), others on transportation (e.g., Garrick and Marshall 2009; Marshall and Garrick 2010; Parthasarathi 2011; Parthasarathi et al. 2012; 2013; 2015), and others on the topology, complexity, and resilience of street networks (e.g., Jiang and Claramunt 2004; Porta et al. 2006a; Xie and Levinson 2007; Hu et al. 2008; Omer and Jiang 2010; Jiang et al. 2014; Brelsford et al. 2015; Barrington-Leigh and Millard-Ball 2015).

This chapter argues that current limitations of data availability, consistency, and technology have made researchers' work gratuitously difficult. In turn, the empirical literature often suffers from four shortcomings which this chapter examines: small sample sizes, excessive network simplification, difficult reproducibility, and the lack of consistent, easy-to-use research tools. These shortcomings are by no means fatal, but their presence limits the scalability, generalizability, and interpretability of empirical street network research.

To address these challenges, this study presents OSMnx, a new Python package developed by this author that downloads political boundary geometries and street networks from OpenStreetMap. OSMnx contributes five significant new capabilities for researchers and city planners: first, the automatic downloading of place boundaries and shapefiles; second, the tailored and automated downloading and construction of street networks from OpenStreetMap; third, the automated correction and simplification of network topology; fourth, the ability to save street networks to disk as shapefiles, GraphML, or SVG files; and fifth, the ability to analyze street networks, calculate routes, project and visualize the networks, and calculate network metrics and statistics. These metrics and statistics include both those common in urban design and transportation studies, as well as metrics that measure the structure and topology of the network.

This chapter is organized as follows. First it discusses the background of networks, street network analysis and representation, data sources such as census products and OpenStreetMap, and the current landscape of tools for this type of research – including their shortcomings. Next it presents OSMnx and discusses its functionality. Finally, it concludes with a discussion of this tool and its implications.

## 5.3. Background

### 5.3.1. Representation of Street Networks

Chapter 3 discussed the characteristics and analysis of street networks, including planarity – i.e., whether or not the network can be represented in two dimensions with its edges intersecting only at nodes. Most quantitative studies of urban street networks represent them as planar (e.g., Buhl et al. 2006; Cardillo et al. 2006; Barthélemy and Flammini 2008; Masucci et al. 2009; Strano et al. 2013) for tractability because bridges and tunnels are reasonably uncommon (in certain places) – thus the networks are approximately planar. However, this over-simplification to planarity for tractability may be unnecessary and can cause analytical problems, which we explore shortly.

The street networks discussed so far are primal: the graphs represent intersections as nodes and street segments as edges. In contrast, a dual graph (namely, the edge-to-node dual graph, also called the line graph) inverts this topology: it represents a city's streets as nodes and intersections as edges (Porta et al. 2006b). Such a representation seems a bit odd, but provides certain advantages in analyzing the network topology based on named streets (Crucitti et al. 2006a; 2006b). Dual graphs form the foundation of space syntax, a method of analyzing urban networks and configuration via axial street lines and the depth from one edge to others (Hillier et al. 1976; Hillier 1989; cf. Ratti 2004). Space syntax has formed the basis of many adapted approaches to analytical urban design (e.g., Karimi 2012).

This present study, however, focuses on primal graphs because they retain all the geographic, spatial, metric information essential to urban form and design that dual representations discard: all the geographic, experiential traits of the street (such as its length, shape, circuity, width, etc.) are lost in a dual graph. A primal graph, by contrast, can faithfully represent all the spatial characteristics of a street. Primal may be a better approach for analyzing spatial networks when geography matters, because the physical space underlying the network contains relevant information that cannot exist in the network's topology alone (Ratti 2004; Batty 2005c).

### 5.3.2. Current Tool Landscape

Several tools exist to study street networks. ESRI provides an ArcGIS Network Analyst extension, for which Sevtsuk and Mekonnen (2012) developed the Urban Network Analysis Toolkit plug-in. QGIS, an open-source alternative, also provides limited capabilities through built-in plug-ins. GIS tools generally provide very few network analysis capabilities, such as shortest path calculations (Fischer et al. 2004; Longley et al. 2005; Maantay and Ziegler 2006). In contrast, network analysis software – such as Gephi, igraph, and graph-tool – does not provide the GIS functionality essential to study spatial networks. Pandana is a Python package that does enable accessibility queries over a spatial network, but does not support other graph-theoretic network analyses (Foti 2014). NetworkX is a Python package for general network analysis, developed by researchers at Los Alamos National Laboratory. It is free, open-source, and able to analyze networks with millions of nodes and edges (Hagberg et al. 2008; Hagberg and Conway 2010).

Street network data come from many sources, including city, state, and national data repositories, and typically in shapefile format. In the U.S., the census bureau provides free TIGER/Line (Topologically Integrated Geographic Encoding and Referencing) shapefiles of geographic data such as cities, census tracts, roads, buildings, and certain natural features. However, TIGER/Line roads shapefiles suffer from inaccuracies (Wu et al. 2005; Frizzelle et al. 2009), contain quite coarse-grained classifiers (e.g., classifying parking lots as alleys), and topologically depict bollarded intersections as through-streets, which problematizes routing. Furthermore, there is no central repository of worldwide street network data, which can be inconsistent, difficult, or impossible to obtain in many countries.

OpenStreetMap – a collaborative mapping project that provides a free and publicly editable map of the world – has emerged in recent years as a major player both for mapping and for acquiring spatial data (Corcoran et al. 2013; Jokar Arsanjani 2015). OpenStreetMap data represent a type of Volunteered Geographic Information (VGI) – data that is both user-generated and geolocated. VGI is one of the most important and fastest-growing sources of geospatial big data (Goodchild 2007; Elwood et al. 2012; Jiang and Thill 2015; Dunkel 2015; Boeing and Waddell 2016). "Big data," though it varies in interpretation – and is often leaned on as a platitude – *is* a type of data that is meaningfully different from traditional and necessarily smaller-scale data (Mayer-Schönberger and Cukier 2013; Kitchin and McArdle 2016). These massive datasets can represent very large samples at incredibly fine spatial and temporal scales, and have significant implications for urban planning and research (Zook et al. 2010; Townsend 2013; Ching and Ferreira 2015).

However, others have critiqued the glorification of big data – and the smart cities and "urban science" paradigms it empowers – as a reformulation of unsophisticated positivist urban cybernetics and control (Sassen 2012; Sennett 2012; Greenfield 2013; Barnes and Wilson 2014; Wyly 2014a; 2014b; Goodspeed 2015; Kitchin 2016; Krivỳ 2016; Mattern 2017; Schweitzer and Afzalan 2017; cf. less-critical perspectives in Lazer et al. 2009; Liu et al. 2011; Argote-Cabanero et al. 2015). According to Kitchin (2017, p. 6), "scientific approaches to cities have been critiqued as being rather naïve and narrow in perspective, producing overly-simplified explanations and models, and a limited and limiting understanding of how cities work (foreclosing what kinds of questions can be asked and how they can be answered) and how urban issues can be tackled."

Nevertheless, VGI such as OpenStreetMap data provides a new lens with which to examine cities and human systems. Inspired by Wikipedia's mass-collaboration model, the OpenStreetMap project started in 2004 and has grown to over two million users today. Its data quality is generally quite high (Haklay 2010; Over et al. 2010; Zielstra and Hochmair 2011) – for example, Garmin GPS devices can now use OpenStreetMap data for navigation. Although data coverage varies worldwide, it is generally good when compared to corresponding estimates from the CIA World Factbook (Maron 2015). In the U.S., OpenStreetMap imported the 2005 TIGER/Line roads in 2007 as a foundational data source (Willis 2008). Since then, numerous corrections and improvements have been made. But more importantly, many additions have been made beyond what TIGER/Line captures, including pedestrian paths through parks, passageways between buildings, bike lanes and routes, and richer attribute data describing the characteristics of features, such as finer-grained codes for classifying arterial roads, collector streets, residential streets, alleys, parking lots, etc.

Several imperfect methods currently exist to acquire street network data from OpenStreetMap. First, OpenStreetMap provides an API, called Overpass, which can be queried programmatically to retrieve data from its database: streets or otherwise. However, its usage and syntax are notoriously challenging and several services have sprung up to simplify the process. Mapzen extracts chunks of OpenStreetMap data constrained to bounding boxes around 200 metropolitan areas worldwide. They also provide custom extracts, which can take up to an hour to run. Mapzen works well for simple bounding boxes around popular cities, but otherwise does not provide an easily scalable or customizable solution. Geofabrik similarly provides data extracts, generally at national or sub-national scales, but provides shapefiles as a paid service.

Finally, GISF2E is a tool (compatible with ArcGIS and an outdated version of Python) that can convert shapefiles such as Mapzen or Geofabrik extracts into graph-theoretic network data sets (Karduni et al. 2016). Its creators provide processed shapefiles for several cities online, but with some limitations. While GISF2E shapefiles' roads have a flag denoting one-way streets, it discards to and from nodes, thus making it unclear in which direction the one-way goes. It also treats nodes inconsistently due to arbitrary break points between OpenStreetMap IDs or line digitization. OpenStreetMap IDs sometimes map 1-to-1 with a named street, but other times a named street might comprise multiple OpenStreetMap IDs. Further, some streets have arbitrary "nodes" in the middle of a segment because the OpenStreetMap ID is different on either side.

### 5.3.3. Current Shortcomings

Due to the aforementioned limitations of street network data availability, consistency, and technology, the empirical literature often suffers from four shortcomings. First, the sample sizes in cross-sectional studies tend to be quite small due to the challenges of acquiring large data sets. Most cross-sectional studies tend to analyze somewhere between 10 and 50 or so networks for tractability at the city or neighborhood scale (e.g., Buhl et al. 2006; Cardillo et al. 2006; Jiang 2007; Marshall and Garrick 2010; Strano et al. 2013; Giacomin and Levinson 2015). Acquiring and assembling large numbers of street networks consistently from data sources spread across various governmental entities can be extremely difficult and time-consuming. However, small sample sizes can limit the representativeness and reliability of findings.

Second, studies usually simplify the representation of the street network to a planar or undirected graph for tractability (e.g., Buhl et al. 2006; Cardillo et al. 2006; Barthélemy and Flammini 2008; Masucci et al. 2009). Typically, researchers assemble street networks into some sort of graph-theoretic object from GIS data, for instance by splitting the centerlines of all the streets in a study area wherever they cross in two dimensions. These split lines become edges and the splitting points become nodes. However, this method presumes a planar graph: bridges and tunnels become splitting points (and thus nodes) even if the streets do not actually intersect in three dimensions. Unless the street network is truly planar, planar simplification produces a less-than-ideal representation that could yield inaccurate metrics, underestimate the lengths of edges, and overestimate the number of nodes. It may reasonably model a street network in a European medieval city center, but poorly models the street network in a city like Los Angeles with numerous grade-separated expressways, bridges, and tunnels in a truly non-planar network. Karduni et al. (2016) suggest the importance of using GIS attribute data to identify such non-planar features to create a correct topology (cf. Mandloi and Thill 2010).

The third problem is replicability. The dozens of decisions that go into analysis – such as spatial extents, topological simplification and correction, definitions of nodes and edges, etc. – are often ad hoc or only partially reported, making reproducibility challenging. Some studies gloss over the precise details of how their street networks were constructed (perhaps due to some combination of methodological complexity and journal word limits), yet numerous unreported decisions had to be made in the process. For example, various studies examine cities out to the urban fringe, but do not explain precisely how and where this periphery was defined (e.g., Strano et al. 2013). Some studies do not report if their networks are directed or undirected (e.g., Porta et al. 2006a; Strano et al. 2013). Directedness may matter little for pedestrian studies, but it substantially impacts the interpretation and values of various measures when directedness *does* matter.

Further, what are edges in the street network? Drivable streets? Pedestrian paths? What is a node in the street network? Is it where at least two different named streets come together? Does it denote any junction of routes? What about dead-ends? Different studies make different but perfectly valid decisions with these various questions (e.g., Frizzelle et al. 2009; Marshall and Garrick 2010; Sevtsuk and Mekonnen 2012; Foti 2014). Their definitions impact how we interpret various calculated features like degrees or intersection densities, and any research design decisions that go unreported can problematize replicability, interpretation, and generalizability.

Fourth, as discussed, the current landscape of tools and methods offers no ideal technique that balances usability, customizability, reproducibility, and scalability in acquiring, constructing, and analyzing network data. Taken together, these limitations make street network researchers' work difficult and can circumscribe the conclusions that may be drawn from the effort.

## 5.4. OSMnx: Methodology and Functionality

To address these challenges, the primary methodological contribution of this dissertation is the creation of a new tool to make the collection of data and creation and analysis of street networks simple, consistent, automatable, and sound from the perspectives of graph theory, transportation, and urban design. OSMnx is a free, open-source Python package developed by this author that downloads political boundaries and street networks from OpenStreetMap. It allows users to easily construct, project, visualize, and analyze non-planar complex street networks consistently. Users can construct a city's or neighborhood's walking, driving, or biking network with a single line of Python code – including node elevations and street grades. OSMnx is built on top of Python's NetworkX, matplotlib, and geopandas libraries for rich network analytic capabilities, beautiful and simple visualizations, and fast spatial queries with R-tree indexing.

OSMnx contributes five significant new capabilities for urban researchers and practitioners, which the following subsections discuss in order. First, OSMnx allows automated and on-demand downloading of political boundary geometries, building footprints, and elevations. Second, it can automate and customize the downloading of street networks from OpenStreetMap, and construct them into NetworkX multidigraphs. Third, it can correct and simplify network topology. Fourth, it can save street networks to disk in various file formats. Fifth and finally, OSMnx has built-in functions to analyze street networks, calculate routes, project and visualize networks, and quickly and consistently calculate various metric and topological measures.

### 5.4.1. Political Boundaries and Building Footprints

To acquire political boundary GIS data, one typically must track down shapefiles online and download them. However, bulk or automated acquisition and analysis (such as that required to analyze hundreds or thousands of separate geographies) requires clicking through numerous web pages to download shapefiles one at a time. With OSMnx, one can download place shapes from OpenStreetMap in a single line of Python code, and project them to UTM in one more line of code (all of the built-in projection in OSMnx calculates UTM zones automatically based on the centroid of the geometry).

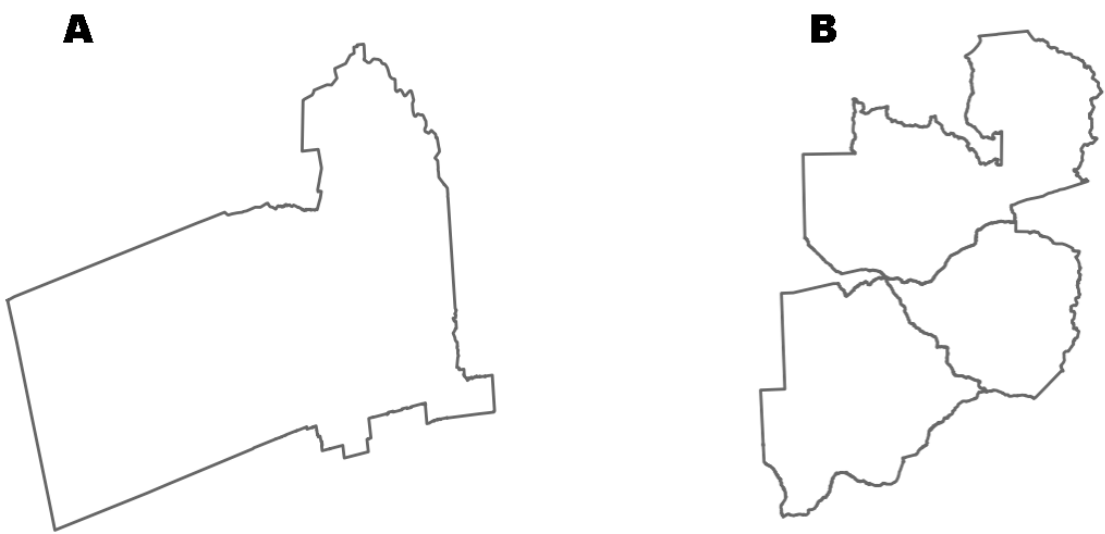

Figure 5.1. Political boundary vector geometries retrieved by OSMnx for A) the city of Berkeley, California, and B) the nations of Zambia, Zimbabwe, and Botswana.

One can just as easily acquire polygons for other place types, such as neighborhoods, boroughs, counties, states, or nations – any place geometry available in OpenStreetMap. Or, one can pass multiple places into a single query to construct a single shapefile with multiple features from their geometries. This can also be done with cities, states, countries or any other geographic entities, and the results can be saved as a shapefile to a hard drive (Figure 5.1). Similarly, building footprints can be retrieved for anywhere that OpenStreetMap has such data (Speranza 2016), as discussed further in section 5.4.6.

### 5.4.2. Download and Construct Street Networks

The primary contribution of OSMnx is the downloading and construction of street networks. To acquire street network GIS data, one must typically track down TIGER/Line roads from the U.S. census bureau, or individual data sets from other countries or their cities. However, this becomes preventively burdensome for large numbers of separate street networks as it does not entail bulk, automated analysis. Further, it ignores informal paths and pedestrian circulation routes that TIGER/Line lacks. Finally, TIGER/Line provides no street network data for outside the United States. In contrast, OSMnx handles all of these use cases.

OSMnx lets one download street network data and build topologically-corrected street networks, project and plot the networks, and save the street network as SVGs, GraphML files, or shapefiles for later use. The street networks are represented as multidigraphs and preserve one-way directionality. Moreover, once the network has been downloaded, OSMnx provides built-in functions to download the elevation of each node (from the Google Maps Elevation API) and calculate street grades. One can download a street network by providing OSMnx any of the following queries:

- a bounding box
- a latitude-longitude point plus a distance in meters (either a distance along the network or a distance in each cardinal direction from the point)
- an address plus a distance in meters (either a distance along the network or a distance in each cardinal direction from the point)
- a polygon of the desired street network's boundaries
- a place name or list of place names

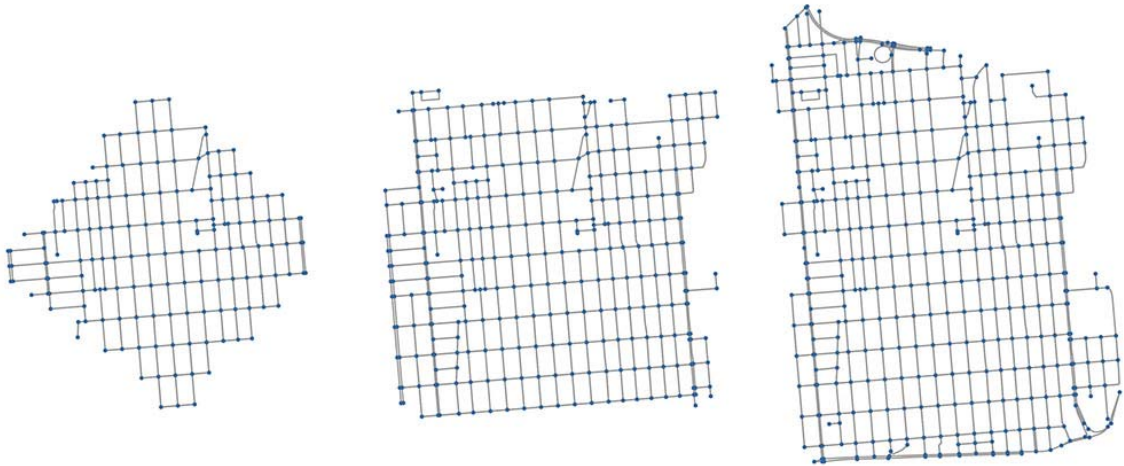

Figure 5.2. Three street networks at the same scale, created by address and network distance (left), bounding box (center), and neighborhood polygon (right).

One can also specify several different network types:

- *drive* - all drivable public streets (but not service roads)
- *drive_service* - all drivable public streets plus service roads
- *walk* - all streets and paths that pedestrians can use (this network type ignores one-way directionality by building reciprocal links in both directions between each pair of connected nodes)
- *bike* - all streets and paths that cyclists can use
- *all* - all (non-private) OpenStreetMap streets and paths
- *all_private* - all OpenStreetMap streets and paths, including those that are private-access only

The functionality to acquire street networks by place name or by polygon is particularly useful for researchers and planners. When passed a place name, OSMnx geocodes the name using OpenStreetMap's Nominatim API and constructs a polygon from its borders. It then buffers this polygon by 500 meters and downloads the street network data within its geometry from OpenStreetMap's Overpass API. Next it constructs a street network from this data, corrects the topology, calculates accurate degrees and intersection types per intersection (this ensures that intersections are not considered dead-ends simply because an incident edge connects to a node outside the desired polygon), then truncates the network to the original, desired polygon. One can just as easily request a street network within a borough, county, state, or other geographic entity. One can also pass a list of places (such as several neighboring cities) to create a unified street network within the union of their geometries.

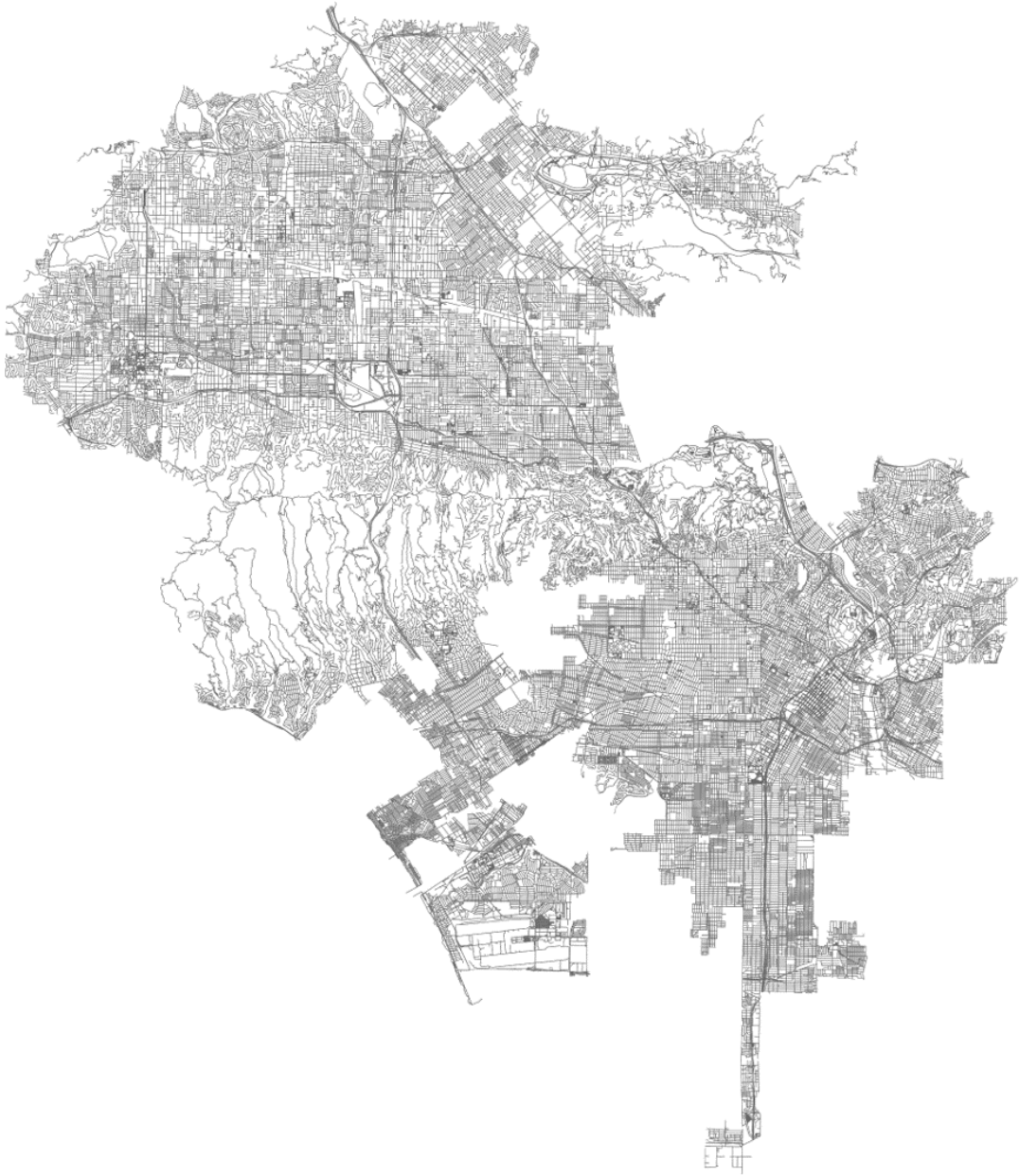

Figure 5.3. The drivable street network for municipal Los Angeles, created by simply passing the query phrase "Los Angeles, CA, USA" into OSMnx.

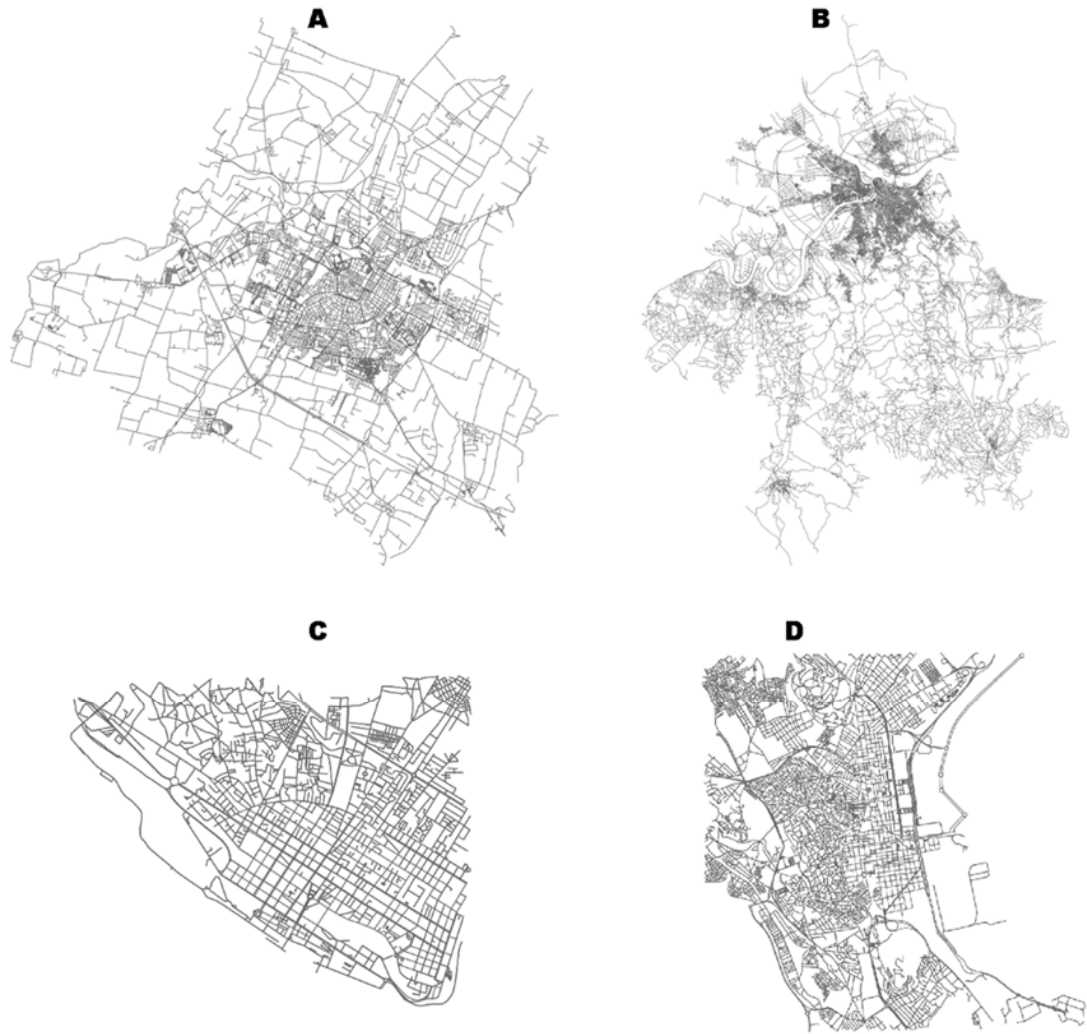

Figure 5.4. Street networks for A) Modena, Italy, B) Belgrade, Serbia, C) central Maputo, Mozambique, and D) central Tunis, Tunisia.

These geospatial operations take advantage of an R-tree spatial index to quickly identify nodes that lay within or outside of the polygons (Guttman 1984). An R-tree represents individual objects and their bounding boxes as the lowest level of the spatial index (the "R" is for "rectangle"). It then aggregates nearby objects and represents them with their aggregate bounding box in the next higher level of the index. At yet higher levels, the R-tree aggregates bounding boxes and represents them by their bounding box, iteratively, until everything is nested into one top-level bounding box. To search, the R-tree takes a query box and, starting at the top level, identifies which (if any) bounding boxes intersect it. It then expands each intersecting bounding box and sees which of the child bounding boxes inside it intersect the query box. This proceeds recursively until all intersecting boxes are searched down to the lowest level. Finally, it returns the matching objects from the lowest level.

However, an R-tree provides no speed-up when the features' bounding boxes are (approximately) identical, because it identifies (approximately) *every* point as a possible match: the bounding box of the polygon intersects every nested rectangle inside the index. This is a limitation of R-trees themselves. To work around this, OSMnx subdivides the polygons into smaller sub-polygons with concomitantly smaller minimum bounding boxes. It then iterates through these small sub-polygons to quickly identify which points lie within each, taking full advantage of the R-tree index's speed. This reduces the processing time of, for instance, metropolitan Los Angeles's street network (Figure 5.3) from approximately an hour to approximately a few seconds.

Of particular relevance to planning scholars and practitioners, OSMnx enables the acquisition of street networks around the world. In general, U.S. street network data sets are fairly easy to come by thanks to TIGER/Line shapefiles. OSMnx makes it *easier* by making them available with a single line of code, and *better* by supplementing them with all the additional data (both attributes and non-road routes) from OpenStreetMap. However, with OSMnx, one can just as easily acquire street networks from anywhere else in the world – places where such data might otherwise be inconsistent or difficult to come by (Figure 5.4).

### 5.4.3. Correct and Simplify Network Topology

Topological correction and simplification is performed by OSMnx automatically under the hood, but it is illuminating to break it out to see how it works. Simplification is essential for a correct topology because OpenStreetMap nodes can be inconsistent: they include intersections, but they also include all the points along a single street segment where the street curves. The latter are not nodes in the graph-theoretic sense, so we remove them algorithmically and consolidate the set of edges between "true" network nodes (i.e., intersections and dead-ends) into a single unified edge. These unified edges between intersections retain the full spatial geometry of the consolidated sub-edges and their relevant attributes, such as the full length of the street segment. OSMnx provides different simplification modes to provide researchers fine-grained control to define nodes rigorously. In *strict* simplification mode, a node is either:

1. where an edge dead-ends (i.e., the endpoint of a cul-de-sac), or
2. the endpoint from which an edge self-loops, or
3. the intersection between multiple streets where at least one of the streets continues through the intersection (i.e., if two streets dead-end at the same point, creating an elbow, the point is not considered a node)

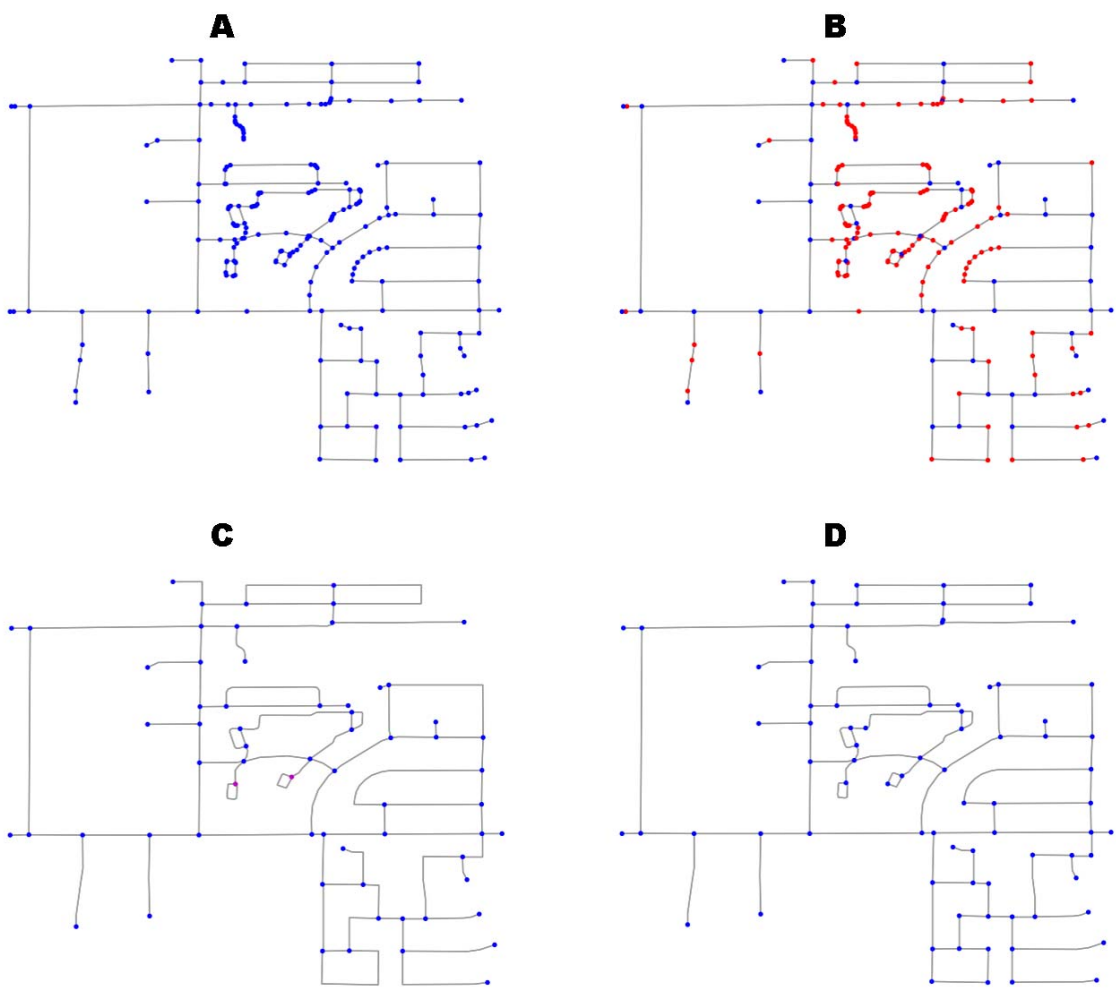

Figure 5.5. A) the original graph, B) non-graph-theoretic nodes highlighted in red and true intersections and dead-ends in blue, C) strictly simplified network, with self-loops noted in magenta, D) non-strictly simplified network.

In *non-strict* mode, conditions 1 and 2 remain the same, but 3 is relaxed to permit nodes at the intersection of two-streets, even if both streets dead-end there, as long as the streets have different OpenStreetMap IDs. In either mode, a node is always retained where there is a point at which a single street changes from one-way to two-way. The process of simplification is illustrated in Figure 5.5. When we first download and assemble the street network from OpenStreetMap, it appears as depicted in Figure 5.5a. For one-way streets, directed edges are added from the origin node to the destination node. For two-way streets, directed edges are added in both directions between nodes.

We want to simplify this network to only retain those nodes that represent dead-ends and the true junction of multiple streets. OSMnx does this automatically in strict mode, unless told to do otherwise. First, it identifies all non-intersection and non-dead-end nodes (i.e., all those that simplify form an expansion graph), as depicted in Figure 5.5b. Then it removes them, but faithfully maintains the spatial geometry and attributes of the street segment between the true intersection nodes. In Figure 5.5c, all the non-intersection nodes have been removed, all the true intersections and dead-ends remain in blue, and self-loop nodes are in purple. In strict mode, OSMnx considered two-way intersections to be topologically identical to a single street that bends around a curve. Conversely, if we wish to retain these intersections when the incident edges have different OpenStreetMap IDs, we may use non-strict mode, as depicted in Figure 5.5d.

### 5.4.4. Save Street Networks to Disk

OSMnx can save the street network to disk as a GraphML file (an open, standard file format for representing graphs on disk) to work with later in network analysis software like Gephi or NetworkX. Or, it can save the network as ESRI shapefiles of nodes and edges to work with later in any standard GIS software. When saving the street network as shapefiles, the graph is simplified to an undirected representation. However, one-way directionality and origin/destination nodes are preserved and saved as edge attributes for GIS routing applications. These shapefiles and GraphML files can be loaded back into OSMnx later for processing, analysis, or visualization. OSMnx can also save street networks as scalable vector graphics (SVG) files for design work in Adobe Illustrator (Figure 5.6).

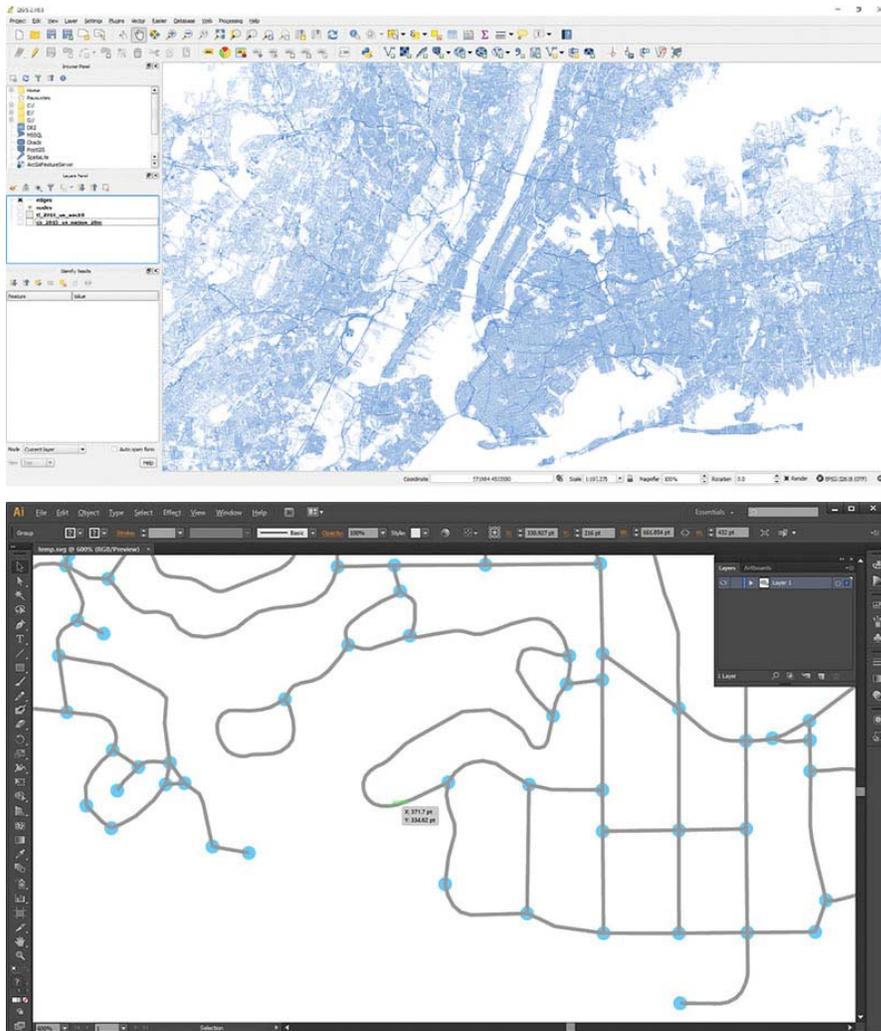

Figure 5.6. The street network for metropolitan New York from OSMnx saved as an ESRI shapefile and loaded in QGIS (above) and in Adobe Illustrator as SVG (below).

### 5.4.5. Analyze Street Networks

OSMnx easily analyzes networks and calculates network statistics, including spatial metrics based on geographic area or weighted by distance (Table 5.1). With a single command, OSMnx calculates the *nodes*' average neighborhood degrees (weighted and unweighted), betweenness centralities, closeness centralities, degree centralities, clustering coefficients (weighted and unweighted), PageRanks and the *network*'s, intersection count, intersection density, average betweenness centrality, average closeness centrality, average degree centrality, eccentricity, diameter, radius, center, periphery, node connectivity, average node connectivity, edge connectivity, average circuity, linear edge density per square kilometer, total edge length, average edge length, average degree, number of edges, number of nodes, node density per square kilometer, maximum and minimum PageRank values and corresponding nodes, the proportion of edges that self-loop, linear street density per square kilometer, total street length, average street length, number of street segments, average number of street segments emanating from each intersection, and the counts and proportions of node types.

The counts and proportions of streets per node improves on metrics used in other studies. Most studies simply use the node degree (either directed or undirected) to count how many edges are incident to the node. This has a couple of problems. First, it ignores the fact that nodes only have incident edges that connect to other nodes *within the spatial boundaries* requested. Thus, a 4-way intersection near the periphery may only have 3 incident edges in the network, because the fourth links a node outside the bounds and was therefore truncated. Second, bi-directional self-loops in a directed graph would count as *four* physical street segments connected to the intersection. This study instead created a new algorithm to correctly count physical streets per node, taking into account one-way and two-way streets, parallel edges, self-loops, and intersections with streets that do not appear in the graph because they link to a node outside the bounds.

| Measure | Definition |
| --- | --- |
| $n$ | number of nodes in the graph |
| $m$ | number of edges in the graph |
| average node degree | mean number of edges incident to the nodes |
| intersection count | number of intersections (non-dead-end nodes) in the graph |
| average streets per node | mean number of streets (edges in undirected representation of the graph) that emanate from each node (intersections and dead-ends) |
| counts of streets per node | a dictionary with keys = the number of streets emanating from the node, and values = the number of nodes with this number |
| proportions of streets per node | a dictionary, same as above, but represents a proportion of the total, rather than raw counts |
| total edge length | sum of all edge lengths in the graph, in meters |
| average edge length | mean edge length in the graph, in meters |
| total street length | sum of all edges in the undirected representation of the graph |
| average street length | mean edge length in undirected representation of the graph, meters |
| count of street segments | number of edges in the undirected representation of the graph |
| node density | $n$ divided by area in square kilometers |
| edge density | total edge length divided by area in square kilometers |
| street density | total street length divided by area in square kilometers |
| average circuity | total edge length divided by the sum of the great circle distances between the nodes incident to each edge |
| self-loop proportion | proportion of edges that have a single incident node (i.e., the edge links nodes $u$ and $v$, and $u=v$) |
| average neighborhood degree | mean degree of the nodes in the neighborhood of each node |
| average of the average neighborhood degree | mean of all the average neighborhood degrees in the graph |
| average weighted neighborhood degree | mean degree of the nodes in the neighborhood of each node, weighted by edge length |
| average of the average weighted neighborhood degree | mean of all the weighted average neighborhood degrees in the graph |
| degree centrality | the fraction of nodes that each node is connected to |
| average degree centrality | mean of all the degree centralities in the graph |
| clustering coefficient | extent to which node's neighborhood forms a complete graph |
| weighted clustering coefficient | extent to which node's neighborhood forms a complete graph, weighted by edge length |
| average weighted clustering coefficient | mean of the weighted clustering coefficients of all the nodes in the graph |
| PageRank | ranking of nodes based on structure of incoming edges (link analysis) |
| maximum PageRank | the highest PageRank value of any node in the graph |
| maximum PageRank node | the node with the maximum PageRank |
| minimum PageRank | the lowest PageRank value of any node in the graph |
| minimum PageRank node | the node with the minimum PageRank |
| node connectivity | the minimum number of nodes that must be removed to disconnect the graph |
| average node connectivity | the expected number of nodes that must be removed to disconnect a randomly selected pair of non-adjacent nodes |
| edge connectivity | the minimum number of edges that must be removed to disconnect the graph |
| eccentricity | for each node, the maximum distance from it to all other nodes, weighted by length |
| diameter | the maximum eccentricity of any node in the graph |
| radius | the minimum eccentricity of any node in the graph |
| center | the set of all nodes whose eccentricity equals the radius |
| periphery | the set of all nodes whose eccentricity equals the diameter |
| closeness centrality | for each node, the reciprocal of the sum of the distance from the node to all other nodes in the graph, weighted by length |
| average closeness centrality | mean of all the closeness centralities of all the nodes in the graph |
| betweenness centrality | for each node, the fraction of all shortest paths that pass through the node |
| average betweenness centrality | mean of all the betweenness centralities of all the nodes in the graph |

Table 5.1. Network measures calculated automatically by OSMnx.

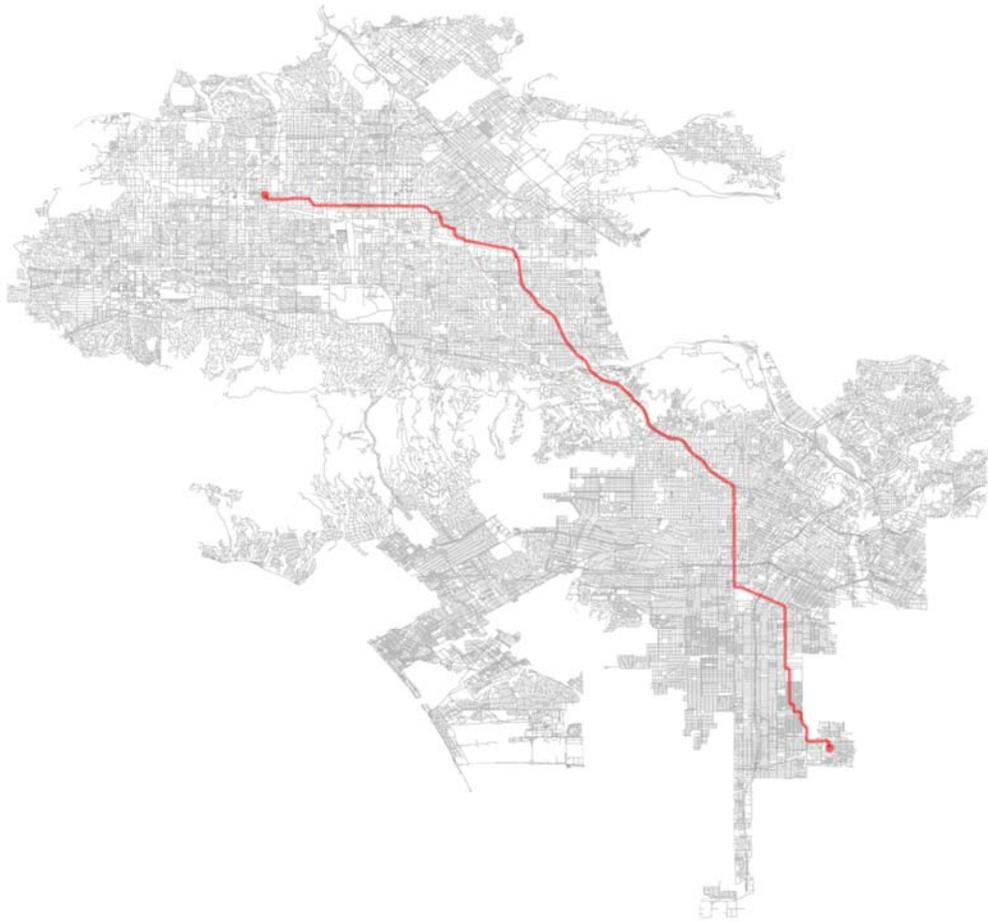

Figure 5.7. OSMnx calculates the shortest network path between two points in Los Angeles, accounting for one-way streets, and plots the route.

As demonstrated in Figure 5.7, we can also calculate and plot shortest-path routes between points – taking one-way streets into account – using Dijkstra's algorithm (Dijkstra 1959; Misa 2010). These shortest paths can be weighted by distance, travel time (assuming the availability of speed data), or any other impedance. For example, since OSMnx can automatically attach elevation data to each node and calculate street grades, a shortest path can be calculated that minimizes elevation change rather than trip distance.

### 5.4.6. Visualize Street Networks and Urban Form

OSMnx can visualize street networks and their metric and topological attributes in various ways. A few quick examples particularly relevant to planners, designers, and urban scholars help sketch this usage. For instance, planners can visualize a network's street segments by length to provide a sense of where a city's shortest and longest blocks are distributed. Planners can similarly visualize one-way versus two-way edges to provide a sense of where certain types of street circulation patterns are concentrated. Planners can also quickly visualize the spatial distribution of dead-ends (or, in fact, intersections of any type) in a city to analyze and visually communicate these points of low network connectivity (cf. Badger 2011; Barrington-Leigh and Millard-Ball 2015), as demonstrated in Figure 5.8.

OSMnx also produces figure-ground diagrams of street networks and building footprints, for urban design and the communication of planning decisions. The heart of Jacobs's (1995) classic book on street-level urban form and design, *Great Streets*, features dozens of hand-drawn figure-ground diagrams in the style of Nolli maps (cf. Hwang and Koile 2005; Verstegen and Ceen 2013). Each depicts one square mile of a city's street network. Drawing these cities at the same scale provides a revealing spatial objectivity in visually comparing their street networks and urban forms. We can re-create this visualization technique automatically and computationally with OSMnx, as shown in Figure 5.9. These Jacobsesque figure-ground diagrams are created completely with OSMnx and its figure-ground street network plotting function.

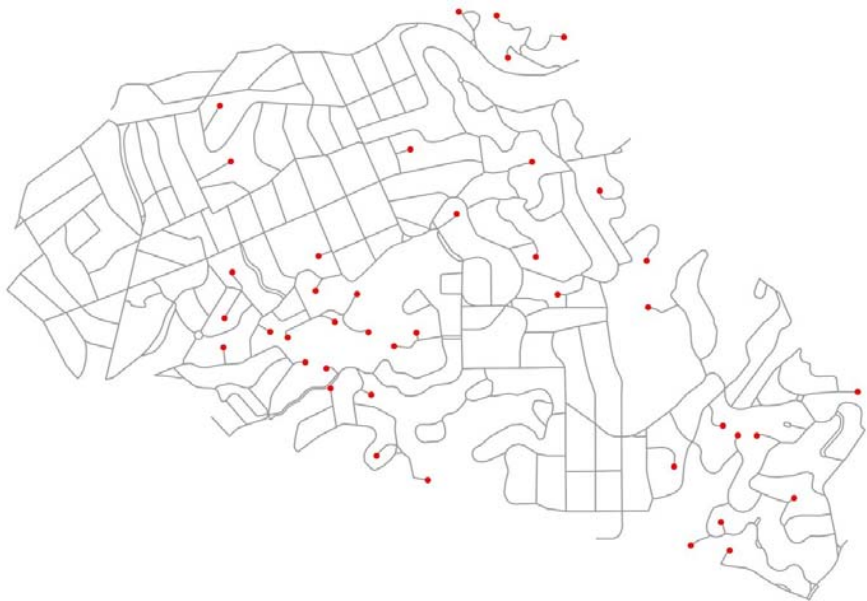

Figure 5.8. OSMnx visualizes the spatial distribution of dead-ends in the city of Piedmont, California.

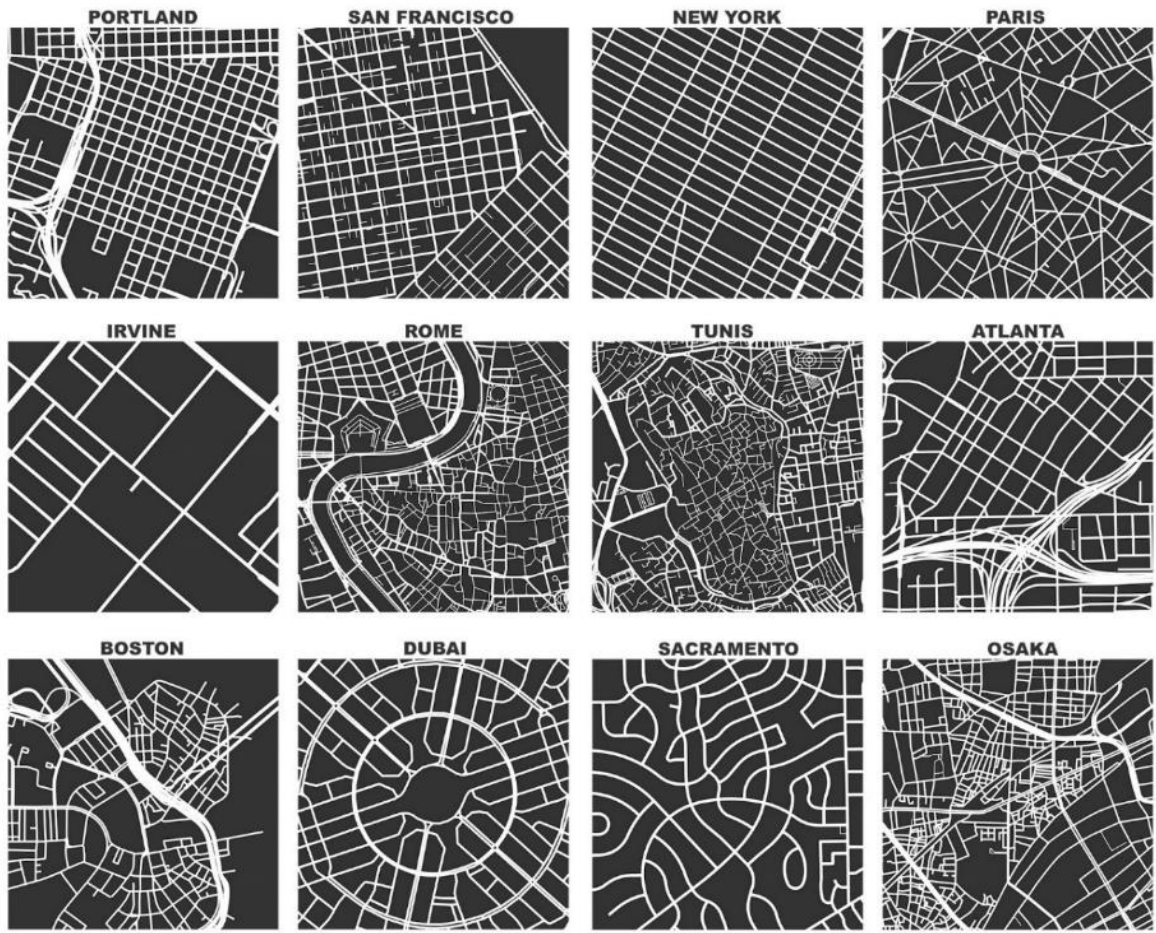

Figure 5.9. One square mile of each city's street network, created and plotted automatically by OSMnx. The consistent spatial scale allows us to easily compare different kinds of street networks and urban forms in different kinds of places.

At the top-left, Portland, Oregon and San Francisco, California typify the late nineteenth century orthogonal grid (Southworth and Ben-Joseph 1995; 1997; Cole 2014; Marshall et al. 2015). Portland's famously compact, walkable, 200-foot × 200-foot blocks are clearly visible but its grid is interrupted by the Interstate 405 freeway which tore through the central city in the 1960s (Speck 2012; Mesh 2014). In the middle-left, the business park in suburban Irvine, California demonstrates the coarse-grained, modernist, auto-centric form that characterized American urbanization in the latter half of the twentieth century (Jackson 1985; Jacobs 1995; Hayden 2004). In stark contrast, Rome has a fine-grained, complex, organic form evolved over millennia of self-organization and urban planning (Taylor et al. 2016). Because we represent all of these street networks here at the same

scale – one square mile – it is easy to compare the block sizes and intersection density in, say, Portland to those in Irvine. Contrast the order of the orthogonal grid in San Francisco and the functionalist simplifications of Irvine to the messy, complex mesh of pedestrian paths, passageways, and alleys constituting the circulation network in central Rome.

At the top- and middle-right, we see New York, Paris, Tunis, and Atlanta. Midtown Manhattan's rectangular grid originates from the New York Commissioners' Plan of 1811, which laid out its iconic 800-foot × 200-foot blocks (Ballon 2012; Koeppel 2015; Sevtsuk et al. 2016). Broadway weaves diagonally across it, revealing the path dependence of the old Wickquasgeck Trail's vestiges, which Native American residents used to traverse the length of the island long before the first Dutch settlers arrived (Shorto 2004; Holloway 2013). At the center of the Paris square mile lies the Arc de Triomphe, from which Baron Haussmann's streets radiate outward as remnants of his massive demolition and renovation of nineteenth century Paris (Hall 1996). The quantitative spatial signatures of Haussmann's project can clearly be seen via network analysis through the redistribution of betweenness centralities and block sizes (Barthélemy et al. 2013). At the center of the Tunis square mile lies its Medina, with a complex urban fabric that evolved over the middle ages (Micaud 1978; Kostof 1991). Finally, Atlanta is typical of many American downtowns: fairly coarse-grained, disconnected, and surrounded by freeways (Grable 1979; Jackson 1985; Allen 1996; Rose 2001; Kruse 2007).

The bottom row of Figure 5.9 shows square miles of Boston, Dubai, Sacramento, and Osaka. The central Boston square mile includes the city's old North End – beloved by Jane Jacobs (1961) for its lively streets, but previously cut-off from the rest of the city by the Interstate 93 freeway. This freeway has since been undergrounded as part of the "Big Dig" megaproject to alleviate traffic and re-knit the urban fabric (Flyvbjerg 2007; Robinson 2008). The Dubai square mile shows Jumeirah Village Circle, a master-planned residential suburb designed in the late 2000s by the Nakheel corporation, a major Dubai real estate developer (Boleat 2005; Kubat et al. 2009; Haine 2013). Its street network demonstrates a hybrid of the whimsical curvilinearity of the garden cities movement and the ordered geometry of modernism (cf. Kostof 1991). The Sacramento square mile depicts its northeastern residential suburb of Arden-Arcade and demonstrates Southworth and Ben-Joseph's (1997) "warped parallel" and "loops and lollipops" design patterns of late twentieth century American urban form.

Finally, the Osaka square mile portrays Fukushima-ku, a mixed-use but primarily residential neighborhood first urbanized during the late nineteenth century. Today, the freeway we see in the upper-right of this square mile infamously passes through the center of the high-rise Gate Tower Building's fifth through seventh floors (Yakunicheva 2014). This peculiar intermingling of street network and edifice arose when transportation planners were forced to compromise with private landowners seeking to redevelop their property, despite the prior designation of the freeway's alignment (Isaac 2014).

To compare urban form in different kinds of places, these visualizations have depicted modern central business districts, ancient historic quarters, twentieth century business parks, and suburban residential neighborhoods. The cities they represent are drawn from across the United States, Europe, North Africa, the Arabian Peninsula, and East Asia. Yet street network patterns also vary greatly *within* cities: Portland's suburban east and west sides look different than its downtown (as we will discuss in chapter 6), and Sacramento's compact, grid-like downtown looks different than its residential suburbs – a finding true of many American cities (as we will discuss in chapter 7). A single square mile diagram thus cannot be taken to be representative of broader scales or other locations within the municipality. These visualizations, rather, show us how different urbanization patterns and paradigms compare at the same scale. This can serve both as a tool for comprehending the physical outcomes of planning and informal urbanization, as well as a tool for communicating urban planning and design in a clear and immediate manner to laymen.

These uses can be seen perhaps even more clearly when we use OSMnx to visualize street networks along with building footprints, as shown in Figure 5.10. At the top-left, we see the densely-built form of midtown Manhattan, with large buildings filling most of the available space between streets. Within this square mile, there are 2,237 building footprints with a median area of 241 square meters (see Table 5.2). At the top-right, we see the medium-density perimeter blocks of San Francisco's Richmond district, just south of the Presidio. Here the building footprints line the streets while leaving the centers of each block as open space for residents. Within this square mile, there are 5,054 building footprints with a median area of 142 square meters. The bottom two images in Figure 5.10 reveal an entirely different mode of urbanization by visualizing the slums of Monrovia, Liberia and Port-au-Prince, Haiti. These informal settlements are much finer-grained, and are not structured according to the orderly logic of the American street grids in the top row. Monrovia's square mile contains 2,543 building footprints with a median area of 127 square meters. Port-au-Prince's square mile contains 14,037 building footprints with a median area of just 34 square meters.

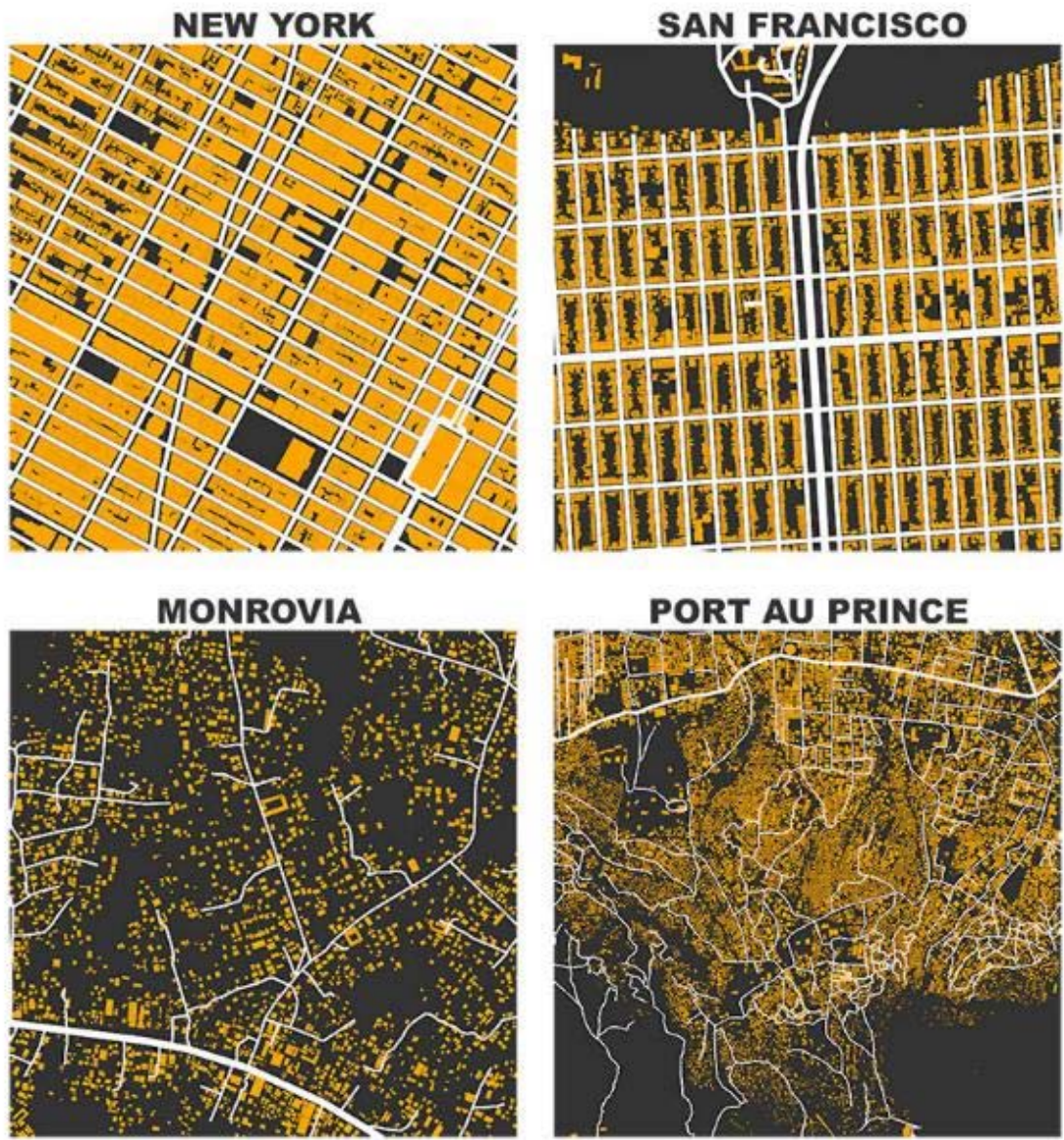

Figure 5.10. One square mile of each city's street network and building footprints, created and plotted automatically by OSMnx. The consistent spatial scale allows us to easily compare the urban form in different kinds of places, particularly the scale and pattern with which the street network interfaces with the rest of the built environment.

| Square mile | $n$ | Total area (m²) | Mean area (m²) | Median area (m²) | Coverage |
|---|---|---|---|---|---|
| New York | 2,237 | 1,551,235 | 693 | 241 | 60% |
| San Francisco | 5,054 | 806,057 | 159 | 142 | 31% |
| Monrovia | 2,543 | 398,637 | 157 | 127 | 15% |
| Port-au-Prince | 14,037 | 680,962 | 49 | 34 | 26% |

Table 5.2. Summary statistics for the building footprints in Figure 5.10: $n$ represents the number of footprints and *coverage* represents the percentage of the square mile covered by these footprints. Note that these coverage figures are gross, not net, and some of these square miles include parks and undeveloped peripheral areas.

To contrast this, the typical building footprint in the San Francisco square mile is 4.2 times larger than that in Port-au-Prince. In New York versus Port-au-Prince, the factor is 7.1 (and if we look at the means instead, it is 14.1). These visualizations provide researchers directly comparable illustrations of the pattern, texture, and grain of the urban form. In this way, OSMnx also provides planning practitioners an easy-to-use tool to visualize and examine street networks and building footprints as a planning and communication tool. Figure 5.10, for instance, could help planners and residents in Monrovia and Port-au-Prince collaboratively study how to percolate formal circulation networks into these informal settlements with minimal disruption to the existing urban fabric, homes, and livelihoods (Brelsford et al. 2015; cf. Zook et al. 2010; Gudmundsson and Mohajeri 2013; Masucci et al. 2013).

Holston (1989) suggests that built form figure-ground diagrams can reveal modernism's inversion of traditional urban spatial order. In pre-industrial cities, the figure dominates the ground as the diagram displays scattered open space between buildings, as seen in Figure 5.11. But in modernist cities, the ground dominates the figure as only a few scattered buildings are positioned as sculptural elements across the landscape's void. Recall Fishman's (2011) argument that we discussed in chapter 4, section 4.4.3: the modernist paradigm sought to open up the dense and messy urban fabric with towers-in-the-park, spacing, and highways. This phenomenon is clearly seen in Brasília, the modernist capital of Brazil, designed as a planned city in the 1950s by Lúcio Costa, Oscar Niemeyer, and Roberto Burle Marx (Figure 5.11). The structural order of the city also suggests "an ordering of social relations and practices in the city," as discussed in chapter 4 (Holston 1989, p. 125; see also Moudon 1997). These figure-ground diagrams provide a preliminary way to evoke and study the urban morphology and circulation networks that structure human activities and social relations.

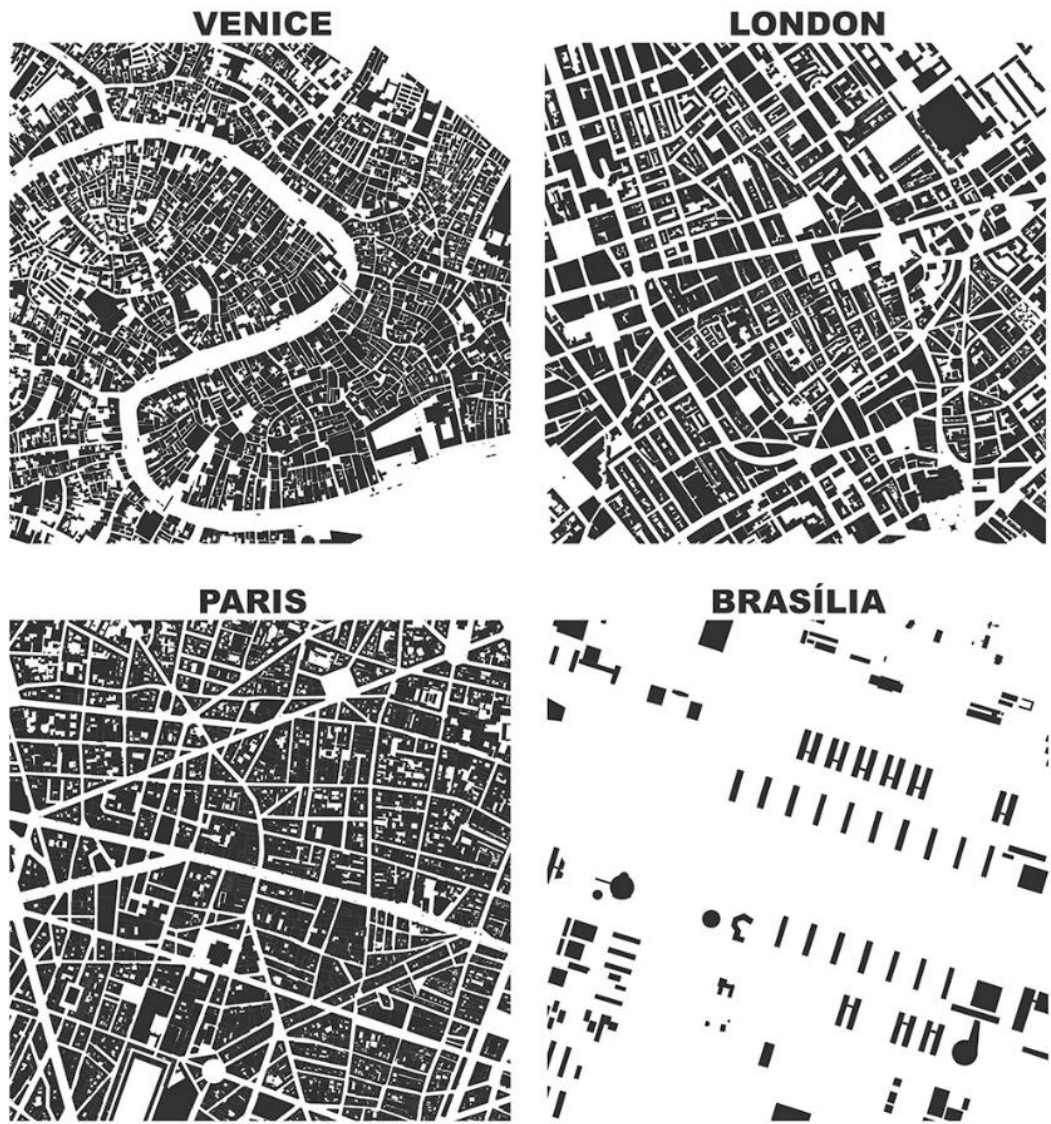

Figure 5.11. One square mile figure-ground diagrams of building footprints in the city centers of Venice, London, Paris, and Brasília reveal the modernist inversion of traditional urban spatial order. Downloaded and plotted automatically with OSMnx.

### 5.4.7. Summary

To briefly summarize this section's discussion of methods and functionality, OSMnx is a new research tool that simplifies and democratizes the process of collecting, constructing, correcting, analyzing, projecting, mapping, and visualizing complex urban street networks. It is built on top of NetworkX, matplotlib, and geopandas for rich network analytic capabilities, beautiful and simple visualizations, and fast spatial queries with R-tree indexing. This empowers research into urban form, street design and evolution, transportation, and resilience at multiple scales for any geography anywhere in the world.

## 5.5. Discussion

Street network analysis currently suffers from challenges of usability, planarity, reproducibility, and sample sizes. This chapter presented a tool, OSMnx, to make the collection of data and creation and analysis of street networks easy, consistent, scalable, and automatable for any study site in the world. It contributes five capabilities for researchers and practitioners: downloading place boundaries, building footprints, and elevation data; downloading and constructing street networks from OpenStreetMap; correcting network topology; saving street networks to disk as ESRI shapefiles, GraphML, or SVG files; and analyzing street networks, including calculating routes, visualizing networks, and calculating metric and topological measures of the network. In turn, it enables researchers and planners to ask new questions about network resilience, connectedness and sociodemographic segregation, accessibility, walkability, housing market responses to built form characteristics, neighborhood retrofitting (e.g., Dunham-Jones and Williamson 2011), and the comparative performance of alternative street layouts.

While these data describe various features of the built environment, they alone cannot tell us about the quality of the streetscape or pedestrian environment. OpenStreetMap is increasingly addressing this with richer attribute data about street width, lanes, speed limits, sidewalk presence, and street trees, but a general limitation of OSMnx is that it is dependent on what data exists in OpenStreetMap. While coverage is good across the United States and Western Europe, developing countries have less thorough, but still quite adequate (especially in cities), street network coverage. Moreover, any researcher or organization can digitize and add streets, building footprints, or other spatial data to OpenStreetMap at any time to serve as a public data repository for their own study, as well as anyone else's. In turn, OSMnx makes the acquisition, construction, and analysis of urban street networks easy, consistent, and reproducible while opening up a new world of public data to researchers and practitioners. OSMnx is open source and freely available to download from a public repository (see Appendix).

The following two chapters use this tool to conduct empirical studies of urban street networks at multiple scales. Chapter 6 presents a small case study of networks at the neighborhood scale in Portland, Oregon. Chapter 7 examines street networks at the metropolitan, municipal, and neighborhood scales across the United States.

# Chapter 6: Case Study: Portland, Oregon

## 6.1. Abstract

This chapter presents a small case study to demonstrate the usage of OSMnx for research. It collects the street networks for three small areas of Portland, Oregon to perform a cross-sectional analysis. First, it introduces these neighborhoods from a brief qualitative and historical perspective. Then it explores their comparative quantitative measures of network complexity and structure. Finally, it discusses these empirical findings and what planning and design insights may be drawn from them.

## 6.2. Introduction

Portland, Oregon is widely regarded as a model city for American urban planning, with compact blocks, a sustainable urban form, good walkability, and a robust transit network (e.g., Duany et al. 2010; Speck 2012). However, it was not always this way. In 1943, the city of Portland hired Robert Moses to design a loop-and-spoke freeway system to push the city into the era of the automobile – the "loop" remains to this day as the Interstate 405 (Mesh 2014). In the 1960s, the city planned to further expand this system with a network of several new freeways running throughout its neighborhoods, but was eventually defeated in a 1974 freeway revolt (ibid.). Furthermore, during the high modernist era of the late 1940s and early 1950s, Portland converted 40 miles of its streets to one-way in the name of improving traffic flow (Pryce 1950). As we shall see, unlike the freeway revolt, this massive conversion proceeded successfully and the city's downtown is presently composed almost entirely of one-way streets, despite a growing recognition today of the livability and economic benefits of two-way streets (e.g., Riggs and Gilderbloom 2015; Riggs and Appleyard 2016).

This chapter examines sections of Portland's street network to compare different urban designs from metric and topological perspectives. While this study does not model or optimize for traffic flow per se, it does consider the topological characteristics of these one-way street configurations. This is useful for planners given the aforementioned livability and economic benefits, as well as the substantial impacts on network resilience from a graph connectivity perspective. As discussed in chapter 5, this dissertation developed a new tool, OSMnx, for downloading, constructing, correcting, analyzing, projecting, and visualizing street networks from OpenStreetMap data. OSMnx can work with driving, walking, and biking circulation networks.

This chapter presents a small case study to demonstrate the functionality of OSMnx. It examines three small half-kilometer sections of the street network in different neighborhoods in Portland, Oregon. This scale of analysis and sample size are small, but they provide simple, comprehensible examples to illustrate the complex network concepts presented in chapter 3, the network complexity (metric and topological) measures presented in chapter 4, and the methodological tool presented in chapter 5. This chapter serves to tie these threads together empirically and present a visual demonstration before embarking on the large multi-scale analysis of 27,000 street networks in chapter 7.

This present chapter has three aims. First, as discussed, it demonstrates the functionality of OSMnx with a simple case study. Second, it presents empirical findings of three street network sections in Portland, Oregon and uses the quantitative measures to compare and contrast these network sections. Third, it offers some insights and suggestions arising from these findings. This chapter is organized as follows. First it lays out the methods by which it uses OSMnx to acquire and analyze these networks. Then it presents the findings from this analysis. Finally, it concludes with a discussion of these findings and their insights for urban planning and design.

## 6.3. Methods

To demonstrate OSMnx, we analyze three neighborhoods in Portland, Oregon. First, we define three square bounding boxes of half-a-square-kilometer each in the city's downtown, Laurelhurst, and Northwest Heights neighborhoods. The downtown study site is centered on the latitude-longitude coordinates (45.519, -122.68), the Laurelhurst study site is centered on the latitude-longitude coordinates (45.527, -122.625), and the Northwest Heights study site is centered on the latitude-longitude coordinates (45.54, -122.771). These half-a-square-kilometer study areas are small and do not conform to human definitions of the full local neighborhood extents, but instead are useful for consistent comparison across sites at a small spatial scale.

These three study sites were selected for their different urban forms and histories. Downtown Portland is geographically constrained between the Willamette River and the West Hills. Its street network, laid out in the late 1800s, is well-known for its dense, 200-foot, consistent, orthogonal blocks (Grammenos and Pollard 2009). This fine-grained, walkable network of streets has been heralded by many urban designers as something of an ideal American urban form (e.g., Duany et al. 2010; Speck 2012). Laurelhurst, in contrast, is a residential streetcar suburb in East Portland, developed primarily during the 1910s and 1920s (Snyder 1979). The neighborhood was originally platted by John Charles Olmstead – a nephew of Frederick Law Olmstead – and its homes and meandering streets exhibit the character of the American craftsman and garden suburb movements (Guzowski 1990; Works 2016). Finally, Northwest Heights is a sprawling neighborhood nestled in the rolling hills west of downtown and north of Beaverton, Oregon. Developed toward the end of the twentieth century, it features large, single-family homes and a winding, disconnected street network.

This study uses OSMnx to visualize these street networks then measure them quantitatively to assess their complexity, particularly through the lens of connectedness and resilience. OSMnx downloads the drivable street networks for each square study site, constructs the network, corrects the topology (as discussed in chapter 5, section 5.4.3), and then calculates full network measures (as described in chapter 5, section 5.4.5).

## 6.4. Findings

If we project these networks to UTM (zone 10 calculated automatically), and plot them as seen in Figure 6.1, we can get an initial qualitative sense of these sections of Portland's street network and how they compare to their disparate histories. OSMnx calculates the correct numbers of streets emanating from each intersection and dead-end (as discussed in section 5.4.5), even for peripheral intersections whose streets were cut off by the bounding box. Also notice in the network of Northwest Heights in Figure 6.1, at the far-right of the panel, that there are two nodes that appear to exist in the middle of a street segment, and thus should have been removed during simplification. However, these are actually multi-way intersections that OSMnx properly retained: they simply have a street that connects to a node outside the right edge of the bounding box.

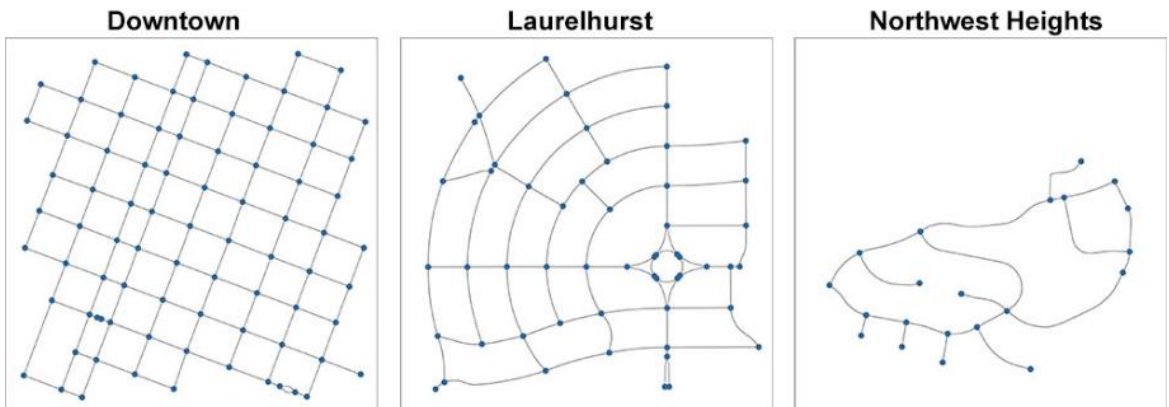

Figure 6.1. Three 0.5 km² sections (each centered on the coordinates presented in section 6.3) of the street network in Portland, Oregon projected and plotted automatically by OSMnx.

The different histories and designs of these street networks are reflected in Figure 6.1. They represent different historical eras, planning regimes, design paradigms, transportation technologies, and topographies (Snyder 1979; Guzowski 1990; Grammenos and Pollard 2009; Semuels 2016; Works 2016). Several quantitative measures can describe these differences as well (Table 6.1). In terms of density metrics, downtown has 164 intersections/km², Laurelhurst has 110, and Northwest Heights has 28. Given their respective histories, it is unsurprising that downtown has approximately 6 times the intersection density of Northwest Heights. Similarly, downtown has 21 linear km of physical street/km², Laurelhurst has 16, and Northwest Heights has 5. In the downtown network, the total street length equals the total edge length, because every edge is one-way. These values differ somewhat in the other two networks because of the presence of two-way streets. As a linear proxy for block size, the average street segment length is 76 meters in downtown, 92 meters in Laurelhurst, and 117 meters in Northwest Heights. These metric measures tell us quantitatively what we can see by visually inspecting the street networks: downtown's is fine-grained and dense, Northwest Heights's is coarse-grained and sparse, and Laurelhurst's is somewhere in between.

Topological measures can tell us more about complexity, connectivity, and resilience. On average, intersections in the downtown section have 3.9 streets connected to them, in Laurelhurst they have 3.6, and in Northwest Heights they have 2.4 (Table 6.1). Beyond the *average*, the statistical distribution of the number of streets per node reveals more about the type of network. Compare Figure 6.1 to Figure 6.2: the majority of the intersections downtown are 4-way intersections, whereas Laurelhurst features a fairly even mix of 3-way and 4-way intersections, and Northwest Heights has mostly 3-way intersections and dead-ends. In fact, one-third of its nodes are the latter.

|  | Downtown | Laurelhurst | NW Heights |
|---|---|---|---|
| Area (km$^2$) | 0.50 | 0.50 | 0.50 |
| Avg of the avg neighborhood degree | 1.64 | 2.98 | 2.75 |
| Avg of the avg weighted n'hood degree | 0.02 | 0.06 | 0.03 |
| Avg betweenness centrality | 0.07 | 0.08 | 0.14 |
| Avg circuity | 1.001 | 1.007 | 1.090 |
| Avg closeness centrality | 0.002 | 0.002 | 0.002 |
| Avg clustering coefficient | <0.001 | 0.108 | <0.001 |
| Avg weighted clustering coefficient | <0.001 | 0.023 | <0.001 |
| Intersection count | 82 | 55 | 14 |
| Avg degree centrality | 0.04 | 0.10 | 0.22 |
| Diameter (km) | 1.28 | 1.02 | 0.90 |
| Edge connectivity | 1 | 1 | 1 |
| Edge density (km/km$^2$) | 21.3 | 29.6 | 10.7 |
| Avg edge length (m) | 76.3 | 97.4 | 116.6 |
| Total edge length (km) | 10.7 | 14.8 | 5.4 |
| Intersection density (per km$^2$) | 163.7 | 109.8 | 28.0 |
| Average node degree | 3.42 | 5.53 | 4.38 |
| *m* | 140 | 152 | 46 |
| *n* | 82 | 55 | 21 |
| Node connectivity | 1 | 1 | 1 |
| Avg node connectivity | 1.33 | 2.11 | 1.44 |
| Avg node connectivity (undirected) | 2.87 | 2.50 | 1.44 |
| Node density (per km$^2$) | 163.7 | 109.8 | 41.9 |
| Max PageRank value | 0.030 | 0.029 | 0.106 |
| Min PageRank value | 0.002 | 0.004 | 0.017 |
| Radius (m) | 742.9 | 537.1 | 561.8 |
| Self-loop proportion | <0.001 | <0.001 | <0.001 |
| Street density (km/km$^2$) | 21.32 | 15.58 | 5.35 |
| Average street segment length (m) | 76.2 | 91.8 | 116.6 |
| Total street length (m) | 10.68 | 7.80 | 2.68 |
| Street segment count | 140 | 85 | 23 |
| Average streets per node | 3.93 | 3.58 | 2.38 |

Table 6.1. Descriptive statistics for three street network sections in the city of Portland, Oregon. For definitions and interpretation of these measures, see Table 5.1 in chapter 5 and section 4.4 in chapter 4.

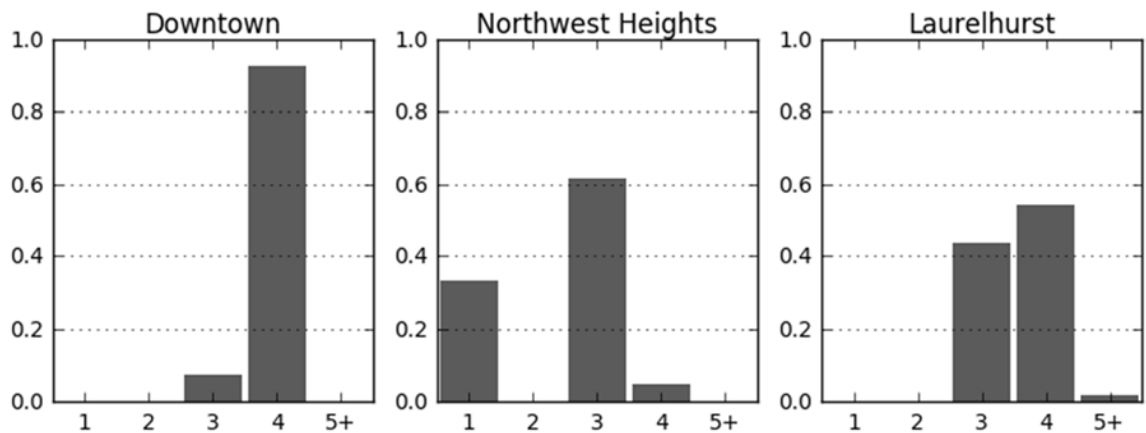

Figure 6.2. Distributions of node types in each street network: *x*-axis is the number of streets at the node and *y*-axis is the proportion of nodes (i.e., intersections and dead-ends) with that number of streets.

Unsurprisingly – for the reasons discussed in chapter 4 – the node and edge connectivity of each network is 1. More useful is the average node connectivity. Recall that this measure represents the average number of nodes that must be removed to disconnect a randomly selected pair of non-adjacent nodes. In other words, this is how many non-overlapping paths exist, on average, between two randomly selected nodes. Thus, on average 1.3 nodes must fail for two nodes to be disconnected in downtown, 2.1 in Laurelhurst, and 1.4 in Northwest Heights (Table 6.1). These values may initially seem surprising: by this measure, downtown has the *least* resilient network despite its density and fine grain. However, this is explained by the fact that *every* street in this downtown section is one-way, greatly circumscribing the number of paths between nodes. If we instead examine the *undirected* average node connectivity, it is 2.9 in downtown, 2.5 in Laurelhurst, and 1.4 in Northwest Heights. Thus, were all the edges in all three networks made bidirectional (i.e., two-way streets), downtown's average node connectivity would more than double and it would now have the *most* resilient network by this measure. This finding suggests that there could be considerable complexity, connectivity, and resilience gains in converting downtown's streets from one-way to two-way.

Finally, Figure 6.3 depicts the average betweenness centrality of the nodes. Note, however, that these neighborhood-scale analyses also show peripheral edge effects (Gil 2016) because they only consider flows originating from and traveling to nodes within the subset; that is, they ignore cross-city flows. In Table 6.1, we see that 7% of all shortest paths pass through an average node in downtown, 8% in Laurelhurst, and 14% in Northwest Heights. The spatial distribution of betweenness centralities in these three networks indicates the relative importance of each node (Figure 6.3): the lightest nodes have the most shortest-paths passing through them, and darkest nodes the fewest. In downtown, important nodes are concentrated at the center of the network due to its orthogonality (but due to the aforementioned edge effects and bounding box, the center is not inherently meaningful).

In Northwest Heights, the two most important nodes are also critical chokepoints connecting the west side of the network to the east. The most important node in Northwest Heights has 43% of all shortest paths running through it. In contrast, the most important node in downtown has only 15%. The street network in this section of Northwest Heights is thus more prone to disruption if its most important node fails (e.g., due to a traffic jam, flood, or earthquake) than downtown's is if its most important node fails.

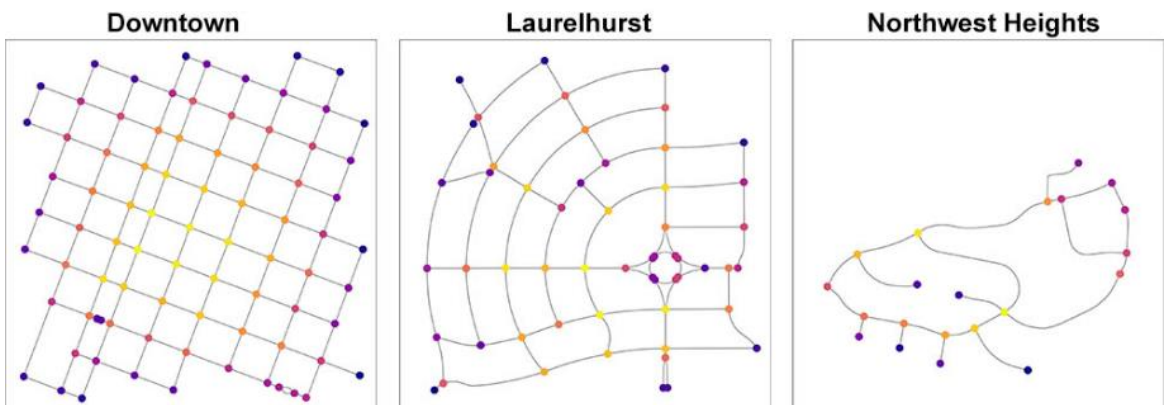

Figure 6.3. Three half-square-kilometer sections of the street network in Portland, Oregon. Nodes colored by betweenness centrality from lowest (dark) to highest (light) for flows originating from and traveling to nodes within the subset.

## 6.5. Discussion

This chapter presented a short case study demonstrating the abilities of OSMnx to download street network data from OpenStreetMap's APIs, construct the data into a graph-theoretic network object with NetworkX, analyze the network with various metric and topological measures of its complexity, and project and plot the network to visualize its characteristics and structure.

In this simple case study of Portland, Oregon, we saw how to assess the street network from both metric and topological perspectives using OSMnx. The quantitative analysis findings corresponded well with a qualitative assessment of the networks' visualizations. In particular, we found these networks differed substantially in density, connectedness, betweenness centrality, and resilience. Although it does not model traffic flow, this case study demonstrated how there could be substantial gains in network resilience from a graph connectivity perspective if one-way streets in the dense, orthogonal downtown were converted to two-way streets. It is important to note that this measure of average node connectivity examines directed flows. Therefore, its analysis of resilience gains is biased toward modes of travel that are constrained by graph directedness, such as automobility or bicycling, rather than toward those that are not, such as walking. However, pedestrian network resilience is strongly impacted by these other measures of connectivity, centrality, clustering, grain, permeability, and density. Overall, the measures of network structure in this case study characterized the complexity of the circulation network in terms of density, resilience, and connectedness – physical attributes that influence how an urban system structures its interactions, connections, and dynamics – as discussed in chapter 4.

As it is limited by its sample size, this small case study primarily serves illustrative purposes. Nevertheless, it demonstrates how to nearly instantaneously acquire, analyze, and visualize networks in just two or three lines of code with OSMnx. The small spatial scale of the analysis provides a succinct opportunity for clear visualization of the phenomena under discussion, as well as a neighborhood-scale interpretation of street network measures and their implications. However, these network subsets demonstrate peripheral edge effects as they only consider flows within the subset, ignoring the rest of the city. The next chapter addresses these limitations by applying OSMnx empirically in an urban form analysis of 27,000 street networks across the United States at multiple scales.

# Chapter 7: A Multi-Scale Analysis of Urban Street Networks

## 7.1. Abstract

This chapter presents wide-ranging empirical findings on U.S. urban form and street network characteristics, using OSMnx. It also demonstrates the scalability and flexibility of OSMnx as a new research tool. This chapter addresses the limitations identified in chapter 5 by conducting an analysis of street networks with large sample sizes, with clearly defined network definitions and extents for reproducibility, and using non-planar, directed graphs. In particular, it examines urban street networks through the framework of complexity developed in this dissertation, focusing on structure, connectedness, centrality, and resilience. In total it cross-sectionally analyzes 497 urbanized areas' street networks, 19,655 cities' and towns' street networks, and 6,857 neighborhoods' street networks. These sample sizes are larger than those in similar previous studies, and the preliminary empirical findings illustrate the use of OSMnx as a new research platform.

## 7.2. Introduction

On May 20, 1862, Abraham Lincoln signed the Homestead Act into law, making land across the United States Midwest and Great Plains available for free to all citizen applicants (Porterfield 2005). Under its auspices over the next 70 years, the federal government distributed 270 million acres of public land (10% of the entire U.S. landmass) to private owners in the form of 1.6 million homesteads (Lee 1979; Sherraden 2005). Towns with gridiron street networks sprang up rapidly across the Great Plains and Midwest, due to both the prevailing urban design paradigm of the day and the standardized rectilinear town plats used repeatedly to lay out instant new cities (Southworth and Ben-Joseph 1997). Through path dependence, the spatial signatures of these land use laws, design paradigms, and planning instruments can still be seen today in these cities' urban forms and street networks. A cross-sectional analysis of American urban form through its street networks at metropolitan, municipal, and neighborhood scales can reveal similar artifacts and histories across the nation.

Network analysis is a natural approach to the study of cities as growing, complex systems that self-organize to fill space with a capillary circulation network (Masucci et al. 2009). As discussed in chapter 5, the empirical literature on street network analysis is growing ever richer, but suffers from some limitations. First, sample sizes tend to be fairly small due to data availability, gathering, and processing constraints. Second, reproducibility is difficult when the dozens of decisions that go into analysis – such as spatial extents, topological simplification and correction, definitions of nodes and edges, etc. – are ad hoc or only partly reported. Third, and related to the first two, studies frequently oversimplify to planar or undirected primal graphs for tractability, or use dual graphs despite the loss of geographic and metric information. Fourth, the current landscape of tools and methods offers no ideal technique that balances usability, customizability, reproducibility, and scalability in acquiring, constructing, and analyzing network data.

The previous two chapters in this dissertation specifically addressed the fourth limitation above by introducing OSMnx and demonstrating its use in a small case study of Portland, Oregon. This chapter aims to address the first three limitations by conducting an analysis of street networks at multiple scales, with large sample sizes, with clearly defined network definitions and extents for reproducibility, and using non-planar, directed graphs. In particular, it examines urban street networks – represented here as primal, non-planar, weighted multidigraphs with possible self-loops – through the framework of complexity developed in this dissertation, focusing on structure, connectedness, centrality, and resilience.

Most studies in the street network literature that conduct topological and/or metric analysis tend to have sample sizes ranging around 5 to 50 networks. This chapter instead conducts a large analysis of 27,000 urban street networks at multiple overlapping scales across the United States. Namely, it examines the street networks of every U.S. incorporated city and town, urbanized area, and Zillow-defined neighborhood. In total, the study presented in this chapter uses OSMnx to download, construct, and analyze 497 urbanized areas' street networks, 19,655 cities' and towns' street networks, and 6,857 neighborhoods' street networks. It uses these street networks to conduct four analyses: at the metropolitan scale, at the municipal scale, at the neighborhood scale, and with a case study looking deeper at the neighborhood-scale street networks in the city of San Francisco.

This chapter is organized as follows. In the next section, it presents the data sources, tools, and methods used to collect, construct, and analyze these street networks. Then it presents findings of the analyses at three spatial scales: metropolitan, municipal, and neighborhood. Next it presents a case study at the neighborhood scale in San Francisco.

Finally, it concludes with a discussion of these findings and their implications for street network analysis, urban form, and city planning.

## 7.3. Methods

This study uses OSMnx – discussed in chapters 5 and 6 – to download, construct, correct, analyze, and visualize street networks at metropolitan, municipal, and neighborhood scales. To define the study sites and their spatial boundaries, this study uses three sets of geometries. The first set defines the metropolitan-scale study sites using the 2016 national TIGER/Line shapefile of U.S. census bureau urban areas, released 19 August 2016. Census-defined urban areas comprise a set of census tracts that meet a minimum density threshold (U.S. Census Bureau 2010). In particular, we retain only the *urbanized areas* subset in this data set (i.e., areas with greater than 50,000 population), discarding the smaller *urban clusters* subset. The second set of geometries defines the municipal-scale study sites using 51 TIGER/Line shapefiles (again, 2016) of U.S. census bureau places for all 50 states plus the District of Columbia. In particular, we discard the subset of census-designated places in this data set, instead retaining every incorporated city and town in the United States.

The third and final set of geometries defines the neighborhood-scale study sites using the 2016 Zillow neighborhood boundary shapefiles. These 42 shapefiles contain the geometries of neighborhoods in major cities in 41 states plus the District of Columbia. This is a fairly new data set comprising nearly 7,000 neighborhoods in large U.S. cities, but as Schernthanner et al. (2016) point out, Zillow does not publish the methodology it uses to construct these geometries. However, despite being new, it already has a track record in the academic literature. For instance, Besbris et al. (2015) used Zillow neighborhood boundaries to examine neighborhood stigma, and Albrecht and Abramowitz (2014) used these data to study neighborhood-level poverty in New York.

For all of these geometries, we use OSMnx's graph_from_polygon function to download the (drivable, public) street network contained within the geometry. First it buffers each geometry by 0.5 km, then downloads the street network within this buffered geometry. Next it constructs a street network from this data, corrects the topology, calculates street counts per node (this ensures that intersections are not considered dead-ends simply because an incident edge connects to a node outside the desired polygon), then truncates the network to the original, desired polygon – as discussed in chapter 5. OSMnx saves each of these networks to disk as GraphML and shapefiles. Finally, it automatically calculates metric and topological measures for each, as discussed in chapter 4. However, in a planar graph, topological and metric properties are somewhat interrelated (Masucci et al. 2009). For reference, these measures are recapitulated (briefly, to avoid excess repetition) in the following paragraphs, but they are discussed in detail in chapters 4 and 5 and summarized in Table 5.1.

*Average street length*, the mean edge length in the undirected representation of the graph, serves as a linear proxy for block size and indicates how fine-grained or coarse-grained the network is. *Node density* is the number of nodes divided by the area covered by the network. *Intersection density* is the node density of the set of nodes with more than one street emanating from them (thus excluding dead-ends). The *edge density* is the sum of all edge lengths divided by the area, and the physical *street density* is the sum of all edges in the undirected representation of the graph divided by the area. These density measures all provide an indication of how fine-grained the network is. Finally, the *average circuity* represents the average ratio between an edge length and the straight-line distance between the two nodes it links.

The *average node degree*, or mean number of edges incident to each node, quantifies how well the nodes are connected, on average. Similarly, but more concretely, the *average streets per node* measures the mean number of physical streets (i.e., edges in the undirected representation of the graph) that emanate from each intersection and dead-end. This adapts the average node degree for physical form rather than directed circulation. The statistical and spatial distributions of number of streets per node characterize the type, prevalence, and dispersion of intersection connectedness in the network. *Connectivity* measures the minimum number of nodes or edges that must be removed from a connected graph to disconnect it. The *average node connectivity* of a network – the mean number of internally node-disjoint paths between each pair of nodes in the graph – represents the expected number of nodes that must be removed to disconnect a randomly selected pair of non-adjacent nodes (Beineke et al. 2002).

The *clustering coefficient* of a node is the ratio of the number of edges between its neighbors to the maximum possible number of edges that could exist between these neighbors. The *weighted clustering coefficient* weights this ratio by edge length and the *average clustering coefficient* is the mean of the clustering coefficients of all the nodes in the network. Centrality indicates the importance of nodes in a network. *Betweenness centrality* evaluates the number of shortest paths that pass through each node (Barthélemy 2004). The maximum betweenness centrality in a network specifies the proportion of shortest paths that pass through the most important node. This is an indicator of resilience: networks with a high maximum betweenness centrality are more prone to failure or inefficiency should this single choke point fail. *Closeness centrality* ranks nodes as more central if they are on average closer to all other nodes. Finally, *PageRank* ranks nodes based on the structure of incoming links and the rank of the source node.

In total, this study cross-sectionally analyzes 27,009 places: 497 urbanized areas, 19,655 cities and towns, and 6,857 neighborhoods (note that places with no streets within their boundaries are not analyzed). These sample sizes are larger than those in any previous study. In the following sections, we present the findings of these analyses at the metropolitan scale, the municipal scale, the neighborhood scale, and through a case study looking deeper at the neighborhood-scale street networks in the city of San Francisco, California.

## 7.4. Analysis of Metropolitan-Scale Street Networks

There is substantial variation in street network characteristics across the entire data set of 497 urbanized areas (Table 7.1). This is unsurprising: the nation's urbanized areas span a wide spectrum of sizes, from the Delano, CA Urbanized Area's 26 km² to the New York--Newark, NY--NJ--CT Urbanized Area's 8,937 km² – thus, density and count-based measures demonstrate substantial variance. Further, these urbanized areas span a wide spectrum of terrains, development eras and paradigms, and cultures.

Nevertheless, looking across the data set provides a sense of the breadth of American metropolitan street networks. New York's urbanized area – America's largest – has 373,309 nodes and 79 million meters of linear street (or 417,570 and 83.4 million if we include service roads). Delano, California's urbanized area – America's smallest – has 874 nodes and 222,328 meters of linear street (or 964 and 231,000 meters if we include service roads). The typical (Table 7.1, *median*) American urbanized area is approximately 185 km$^2$ in land area, has 5,830 nodes, and 1.3 million linear meters of street. Its street network is about 7.4% more circuitous than straight-line, as-the-crow-flies edges between nodes would be. The most circuitous network is 14% more circuitous than straight-line would be, and least is only 2%.

| | **mean** | **σ** | **min** | **median** | **max** |
|---|---|---|---|---|---|
| Area (km$^2$) | 460.657 | 858.125 | 25.685 | 184.898 | 8937.429 |
| Avg of the avg neighborhood degree | 2.886 | 0.109 | 2.626 | 2.875 | 3.228 |
| Avg of the avg weighted n'hood degree | 0.032 | 0.018 | 0.021 | 0.03 | 0.321 |
| Avg circuity | 1.076 | 0.019 | 1.023 | 1.074 | 1.14 |
| Avg clustering coefficient | 0.042 | 0.009 | 0.015 | 0.042 | 0.071 |
| Avg weighted clustering coefficient | 0.002 | 0.001 | <0.001 | 0.001 | 0.006 |
| Intersection count | 12582 | 26054 | 751 | 4593 | 307848 |
| Avg degree centrality | 0.001 | 0.001 | <0.001 | 0.001 | 0.007 |
| Edge density (km/km$^2$) | 13.455 | 2.137 | 7.961 | 13.352 | 21.233 |
| Avg edge length (m) | 158.588 | 17.653 | 117.341 | 157.332 | 223.08 |
| Total edge length (km) | 6353 | 12625 | 427 | 2393 | 1.42e8 |
| Proportion of dead-ends | 0.213 | 0.055 | 0.077 | 0.207 | 0.416 |
| Proportion of 3-way intersections | 0.593 | 0.046 | 0.444 | 0.591 | 0.778 |
| Proportion of 4-way intersections | 0.187 | 0.063 | 0.054 | 0.178 | 0.422 |
| Intersection density (per km$^2$) | 26.469 | 6.256 | 12.469 | 26.029 | 49.423 |
| Average node degree | 5.153 | 0.302 | 4.307 | 5.143 | 6.056 |
| $m$ | 40890 | 83678 | 2516 | 14955 | 981646 |
| $n$ | 16032 | 32585 | 874 | 5830 | 373309 |
| Node density (per km$^2$) | 33.628 | 7.641 | 17.675 | 33.071 | 61.655 |
| Max PageRank value | 0.001 | 0.001 | <0.001 | 0.001 | 0.003 |
| Min PageRank value | <0.001 | <0.001 | <0.001 | <0.001 | <0.001 |
| Self-loop proportion | 0.008 | 0.008 | <0.001 | 0.006 | 0.071 |
| Street density (km/km$^2$) | 7.262 | 1.221 | 4.217 | 7.171 | 11.797 |
| Average street segment length (m) | 161.33 | 17.77 | 119.57 | 160.29 | 225.92 |
| Total street length (km) | 3480 | 7026 | 222 | 1269 | 79046 |
| Street segment count | 22011 | 45725 | 1281 | 7868 | 533757 |
| Average streets per node | 2.764 | 0.162 | 2.223 | 2.770 | 3.217 |

Table 7.1. Measures of central tendency and dispersion for selected measures of the 497 urbanized area street networks. For definitions and interpretation of these measures, see Table 5.1 in chapter 5 and section 4.4 in chapter 4.

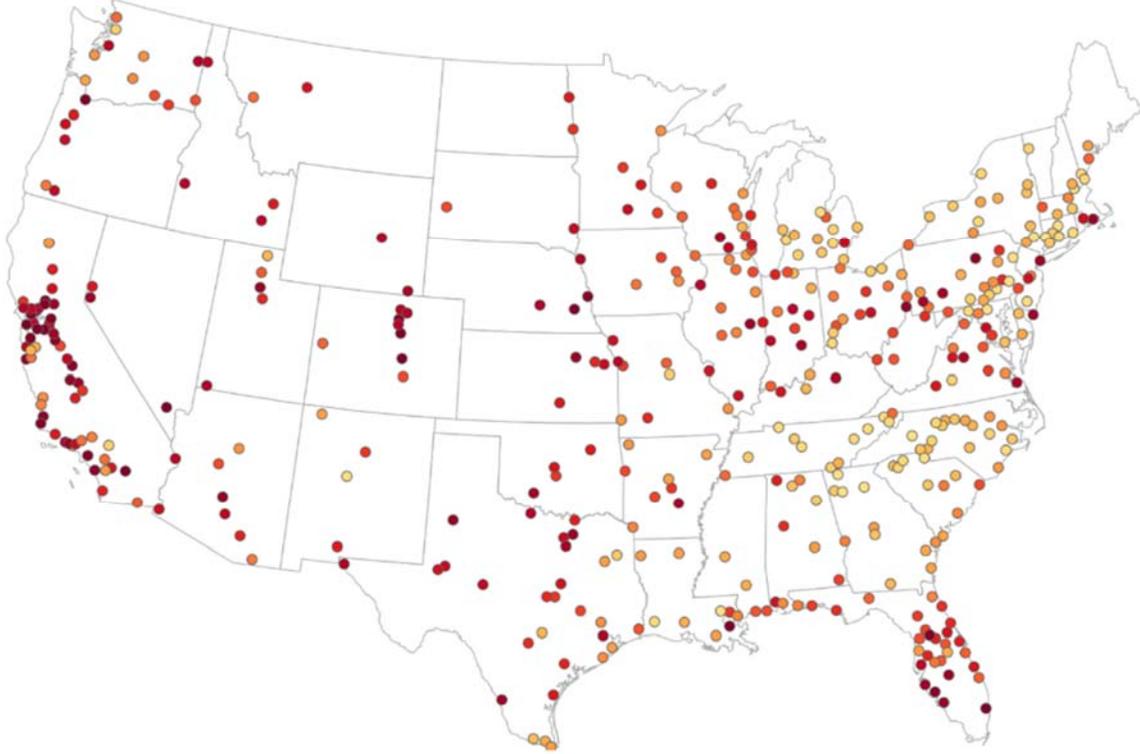

Figure 7.1. Intersection density per urbanized area, from lowest (pale yellow) to highest (dark red), in the contiguous U.S.

Looking at network complexity in terms of density and connectivity, in the typical urbanized area, the average street segment length (a proxy for block size) is 160 meters. The longest average street segment is the 226-meter average of urbanized Danbury, Connecticut. Puerto Rican cities hold the top four positions for shortest average street segment length, but among the 50 states plus Washington DC, the shortest average street segment is the 125.3-meter average of urbanized Tracy, California, indicating a much finer street network. The urbanized area of Portland, Oregon, with its famously compact walkable blocks, ranks second at 125.5 meters on average.

The typical urbanized area has 26 intersections per km². Both the densest and the sparsest are in the deep south: the sparsest is 12.5 (Gainesville, Georgia's urbanized area) and the densest is 49.4 (New Orleans, Louisiana's urbanized area). However, New Orleans is an anomaly in the deep south. Figure 7.1 depicts the intersection density in each American urbanized area, from lowest density in dark red to highest density in light yellow. The map makes clear how the highest intersection densities are concentrated to the west of the Mississippi River. The lowest intersection densities are concentrated in a belt running from Louisiana, up through the Carolinas and Appalachians, and into New England. In general, only the largest cities on the east coast (e.g., Boston, New York, Philadelphia, Washington) and Florida escape this trend.

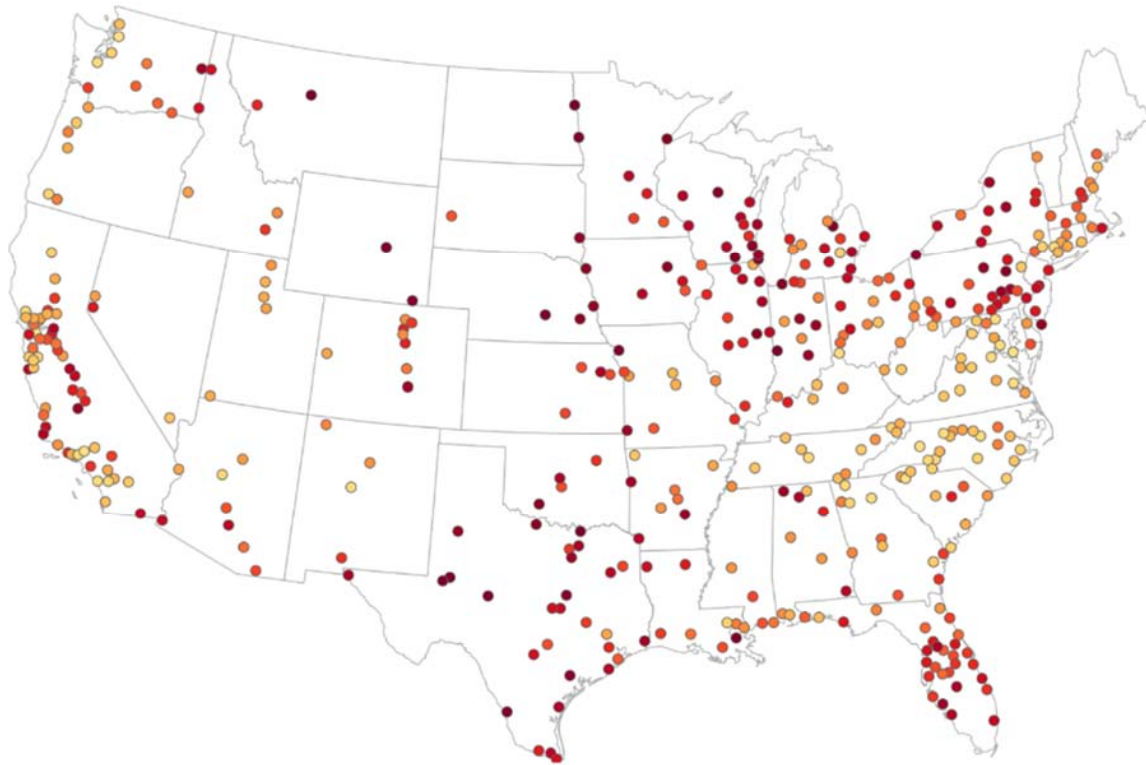

Figure 7.2. Average streets per node per urbanized area, from fewest (pale yellow) to most (dark red), in the contiguous U.S.

The distribution of node types (i.e., intersections and dead-ends) provides a clear indicator of network connectedness. The typical urbanized area has 2.8 streets per node on average: lots of 3-way intersections, fewer dead-ends, and even fewer 4-way intersections. The grid-like San Angelo, Texas urbanized area has the most streets per node (3.2) on average, and (outside of Puerto Rico, which contains the seven lowest urbanized areas) the sprawling, disconnected Lexington Park, Maryland urbanized area has the fewest (2.2). These two urban areas fit the trend seen in the spatial distribution across the U.S. in Figure 7.2: urbanized areas in the great plains and Midwest have particularly high numbers of streets per node on average, indicating more grid-like, connected networks. Cities in the southern and western U.S. tend to have fewer streets per node, reflecting more dead-ends and a disconnected network. This finding is discussed in more detail in the upcoming section.

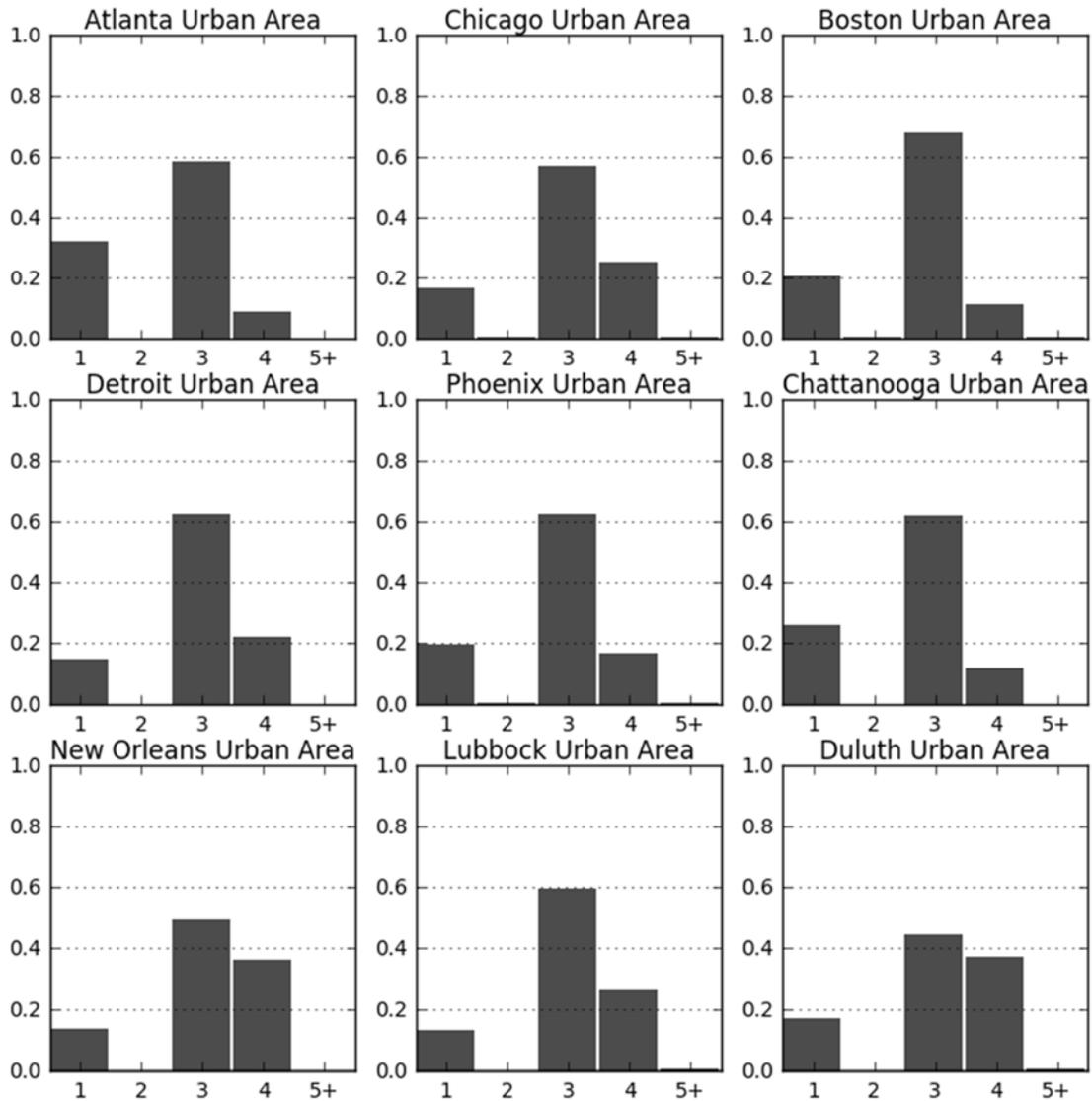

Figure 7.3. Distribution of node types in 9 urbanized areas, with number of streets emanating from the node on the *x*-axis and proportion of nodes of this type on the *y*-axis (cf. Figure 6.2).

In the typical urbanized area, 18% of nodes are 4-way intersections, 59% are 3-way intersections, and 21% are dead-ends. However, this distribution varies somewhat between urbanized areas. Examining a small sample of 9 urbanized areas, chosen to maximize variance, reveals this in clearer detail. In Figure 7.3, urban Atlanta and Chattanooga have high proportions of dead-ends, each over 30% of all nodes, and few 4-way intersections, indicating a disconnected street pattern. The urban areas of Phoenix, Boston, Detroit, and Chattanooga have particularly high proportions of 3-way intersections, each over 60%, indicating a prevalence of T-intersections. Conversely, Chicago, New Orleans, Duluth, and Lubbock have high proportions of 4-way intersections, indicating more grid-like connected networks. But what is perhaps most notable about Figure 7.3 is that these nine urbanized areas, despite being chosen to maximize variance, are overwhelmingly similar to each other. At the metropolitan scale, every large American urban agglomeration is characterized by a preponderance of 3-way intersections.

The relationship between fine-grained networks and connectedness/grid-ness is, however, not clear-cut. Intersection density has only a weak, positive linear relationship with the proportion of 4-way intersections in the urbanized area ($r^2$=0.17), as seen in Figure 7.4. But the relationship between network circuity and grid-ness is somewhat clearer. Average circuity has a negative linear relationship with the proportion of 4-way intersections in the urbanized area ($r^2$=0.43).

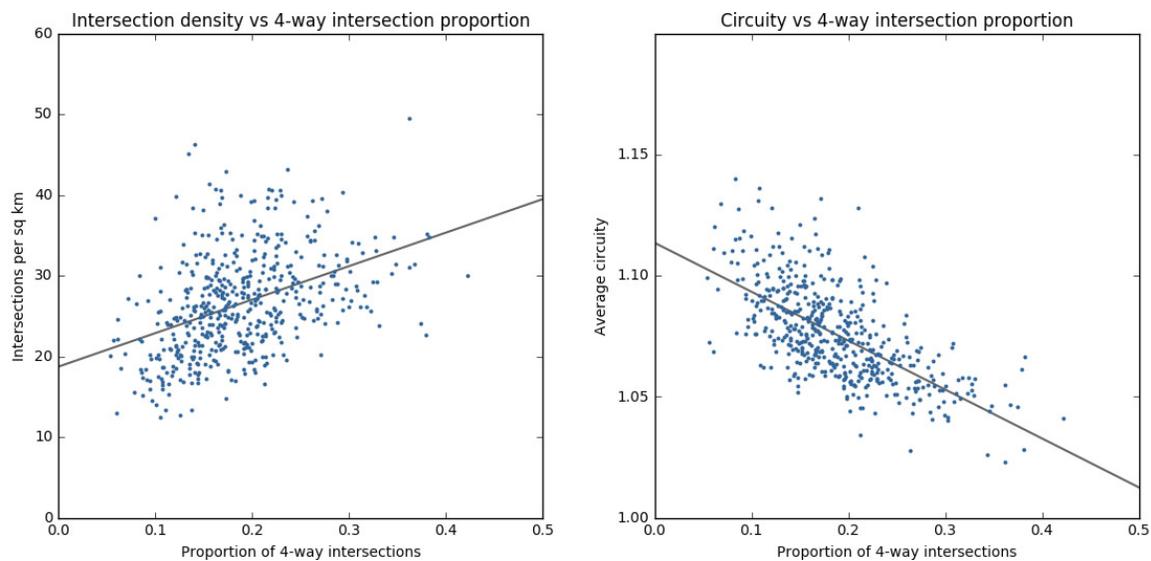

Figure 7.4. Scatterplots of intersection density versus 4-way intersection proportion (left) and average circuity versus 4-way intersection proportion (right), with simple regression lines to indicate the trend.

| Urban area core city | Land area km² | Avg circ­uity | Avg cluster coeff. | Dead-end ratio | 3-way ratio | 4-way ratio | Intersect density (/km²) | Street density (km/km²) | Avg street len (m) | Avg str/ node |
|---|---|---|---|---|---|---|---|---|---|---|
| New York | 8937 | 1.06 | 0.04 | 0.18 | 0.62 | 0.20 | 34.44 | 8.84 | 148 | 2.86 |
| Atlanta | 6850 | 1.10 | 0.04 | 0.32 | 0.58 | 0.09 | 18.39 | 6.16 | 186 | 2.45 |
| Chicago | 6325 | 1.07 | 0.04 | 0.17 | 0.57 | 0.25 | 27.05 | 7.77 | 163 | 2.92 |
| Philadelphia | 5132 | 1.08 | 0.05 | 0.17 | 0.63 | 0.20 | 26.65 | 7.30 | 159 | 2.87 |
| Boston | 4852 | 1.09 | 0.05 | 0.20 | 0.68 | 0.11 | 24.23 | 6.44 | 154 | 2.71 |
| Dallas | 4612 | 1.07 | 0.05 | 0.15 | 0.61 | 0.23 | 34.16 | 9.16 | 156 | 2.95 |
| Los Angeles | 4497 | 1.06 | 0.03 | 0.21 | 0.56 | 0.22 | 39.45 | 10.59 | 151 | 2.82 |
| Houston | 4303 | 1.08 | 0.04 | 0.20 | 0.57 | 0.22 | 33.49 | 8.62 | 145 | 2.83 |
| Detroit | 3461 | 1.07 | 0.04 | 0.15 | 0.63 | 0.22 | 31.10 | 8.56 | 159 | 2.93 |
| Washington | 3424 | 1.09 | 0.05 | 0.26 | 0.56 | 0.17 | 31.22 | 8.26 | 146 | 2.66 |
| Miami | 3204 | 1.10 | 0.05 | 0.17 | 0.59 | 0.23 | 40.54 | 10.61 | 149 | 2.89 |
| Phoenix | 2968 | 1.09 | 0.05 | 0.20 | 0.62 | 0.17 | 35.31 | 9.10 | 150 | 2.77 |
| Minneapolis | 2647 | 1.08 | 0.05 | 0.19 | 0.57 | 0.23 | 29.54 | 8.68 | 167 | 2.84 |
| Seattle | 2617 | 1.07 | 0.03 | 0.30 | 0.54 | 0.16 | 31.57 | 8.20 | 143 | 2.57 |
| Tampa | 2479 | 1.10 | 0.05 | 0.20 | 0.58 | 0.21 | 31.35 | 8.46 | 153 | 2.83 |
| St. Louis | 2392 | 1.10 | 0.04 | 0.22 | 0.62 | 0.15 | 29.68 | 8.16 | 154 | 2.73 |
| Pittsburgh | 2345 | 1.09 | 0.04 | 0.23 | 0.60 | 0.16 | 23.57 | 6.71 | 165 | 2.72 |
| San Juan | 2245 | 1.11 | 0.02 | 0.36 | 0.56 | 0.08 | 26.57 | 6.43 | 131 | 2.36 |
| Cincinnati | 2040 | 1.07 | 0.03 | 0.31 | 0.54 | 0.14 | 17.96 | 6.10 | 186 | 2.51 |
| Cleveland | 2004 | 1.07 | 0.04 | 0.19 | 0.66 | 0.14 | 19.13 | 6.51 | 198 | 2.76 |
| Charlotte | 1920 | 1.08 | 0.04 | 0.30 | 0.57 | 0.11 | 21.00 | 6.43 | 170 | 2.51 |
| San Diego | 1897 | 1.08 | 0.03 | 0.28 | 0.54 | 0.17 | 28.89 | 8.32 | 159 | 2.62 |
| Baltimore | 1857 | 1.09 | 0.04 | 0.23 | 0.59 | 0.17 | 27.72 | 7.56 | 152 | 2.72 |
| Indianapolis | 1828 | 1.08 | 0.05 | 0.23 | 0.59 | 0.17 | 27.62 | 7.63 | 157 | 2.70 |
| Kansas City | 1756 | 1.06 | 0.04 | 0.21 | 0.58 | 0.20 | 32.09 | 8.57 | 152 | 2.79 |
| Denver | 1729 | 1.07 | 0.05 | 0.20 | 0.57 | 0.22 | 40.60 | 9.84 | 138 | 2.84 |
| Orlando | 1548 | 1.11 | 0.06 | 0.20 | 0.61 | 0.18 | 26.30 | 7.44 | 163 | 2.79 |
| San Antonio | 1547 | 1.07 | 0.05 | 0.17 | 0.60 | 0.21 | 28.33 | 7.91 | 162 | 2.87 |
| Nashville | 1460 | 1.08 | 0.03 | 0.27 | 0.59 | 0.14 | 19.08 | 6.10 | 181 | 2.60 |
| Milwaukee | 1413 | 1.06 | 0.06 | 0.14 | 0.55 | 0.30 | 28.27 | 7.81 | 157 | 3.03 |

Table 7.2. Selected street network stats for the 30 largest urbanized areas (by land area). For definitions and interpretation of these measures, see Table 5.1 in chapter 5 and section 4.4 in chapter 4.

Due to the substantial variation in urbanized area size, from 25 to 9,000 km², the preceding analysis covers a wide swath of metropolitan place types. To better compare apples-to-apples, we can focus on the 30 largest urban areas cross-sectionally to examine how their metric and topological measures compare (Table 7.2). This provides more consistent spatial scales and extents, while offering a window into the similarities and differences in the built forms of America's largest urban agglomerations.

Among these urban areas, Milwaukee has the least circuitous network (6% more circuitous than straight-line edges would be), and Orlando has the most (12%). San Juan and Atlanta have the fewest streets per node on average (2.36 and 2.45, respectively), while Milwaukee has the most (3.03). Cincinnati has both the lowest intersection density (18/km$^2$) and street density (6.1 km/km$^2$) while Denver has the highest intersection density (40.6/km$^2$) and Miami and Los Angeles have the highest street density (10.6 km/km$^2$, apiece). In other words, Cincinnati has a particularly coarse-grained network with few connections and paths. This can also be seen in the average street segment length, a proxy for block size: Cincinnati has the second highest (186 m), bested only by Cleveland (198 m). In contrast, the two lowest are Denver's 138-meter average and San Juan's 131-meter average.

These metropolitan-scale analyses consider trends in the built form at the scale of broad self-organized human systems and urbanized regions. However, they aggregate multiple heterogeneous neighborhoods and municipalities – the scales of human life, urban design projects, and planning jurisdiction – into single units of analysis. To disaggregate and analyze finer characteristics, the following sections examine municipal- and neighborhood-scale street networks across the United States.

# 7.5. Analysis of Municipal-Scale Street Networks

There is similarly great variation in street network characteristics across the entire data set of 19,655 cities and towns (Table 7.3). Again, this data set comprises the street networks of every incorporated city and town in the United States. Following the recent work by Barthélemy and Flammini (2008) and Strano et al. (2013), we examine the relationship between the total street length $L$ and the number of nodes $n$. Barthélemy and Flammini proposed a model of cities in which $L$ and $n$ scale as $n^{1/2}$, and Strano et al. confirmed this finding empirically with a small sample of ten European cities' street networks. However, their small sample size may limit the generalizability and interpretability of their finding. To investigate this empirically, we examine the relationship between the total street length $L$ and the number of nodes $n$ for 19,655 U.S. cities and towns and find a strong linear relationship ($r^2$=0.98), contradicting Strano et al., as depicted in Figure 7.5. A nearly identical linear relationship is found at the metropolitan and neighborhood scales.

| | mean | $\sigma$ | min | median | max |
|---|---|---|---|---|---|
| Area (km²) | 16.703 | 107.499 | 0.039 | 3.918 | 7434.258 |
| Avg of the avg neighborhood degree | 2.940 | 0.297 | 0.400 | 2.953 | 3.735 |
| Avg of the avg weighted n'hood degree | 0.033 | 0.141 | <0.001 | 0.029 | 9.357 |
| Avg circuity | 1.067 | 0.159 | 1.000 | 1.055 | 20.452 |
| Avg clustering coefficient | 0.048 | 0.041 | <0.001 | 0.040 | 1.000 |
| Avg weighted clustering coefficient | 0.010 | 0.018 | <0.001 | 0.005 | 0.524 |
| Intersection count | 324 | 1266 | 0 | 83 | 62996 |
| Avg degree centrality | 0.093 | 0.136 | <0.001 | 0.052 | 2.667 |
| Edge density (km/km²) | 12.654 | 6.705 | 0.006 | 11.814 | 58.603 |
| Avg edge length (m) | 161.184 | 80.769 | 25.822 | 144.447 | 3036.957 |
| Total edge length (km) | 159.067 | 578.521 | 0.052 | 40.986 | 24728.326 |
| Proportion of dead-ends | 0.192 | 0.093 | <0.001 | 0.184 | 1.000 |
| Proportion of 3-way intersections | 0.572 | 0.110 | <0.001 | 0.579 | 1.000 |
| Proportion of 4-way intersections | 0.237 | 0.129 | <0.001 | 0.217 | 1.000 |
| Intersection density (per km²) | 29.363 | 21.607 | <0.001 | 24.719 | 259.647 |
| Average node degree | 5.251 | 0.668 | 0.800 | 5.268 | 7.166 |
| $m$ | 1046 | 3924 | 2 | 275 | 176161 |
| $n$ | 401 | 1516 | 2 | 103 | 71993 |
| Node density (per km²) | 35.449 | 24.409 | 0.047 | 30.718 | 296.740 |
| Max PageRank value | 0.034 | 0.046 | <0.001 | 0.021 | 0.870 |
| Min PageRank value | 0.005 | 0.018 | <0.001 | 0.002 | 0.500 |
| Self-loop proportion | 0.005 | 0.015 | <0.001 | <0.001 | 1.000 |
| Street density (km/km²) | 6.528 | 3.435 | 0.003 | 6.109 | 29.302 |
| Average street segment length (m) | 162.41 | 81.04 | 25.82 | 145.48 | 3036.96 |
| Total street length (km) | 86.096 | 331.048 | 0.026 | 21.005 | 15348.01 |
| Street segment count | 558 | 2208 | 1 | 140 | 107393 |
| Average streets per node | 2.851 | 0.282 | 1.000 | 2.852 | 4.000 |

Table 7.3. Selected summary stats for every incorporated city and town in the United States. For definitions and interpretation of these measures, see Table 5.1 in chapter 5 and section 4.4 in chapter 4.

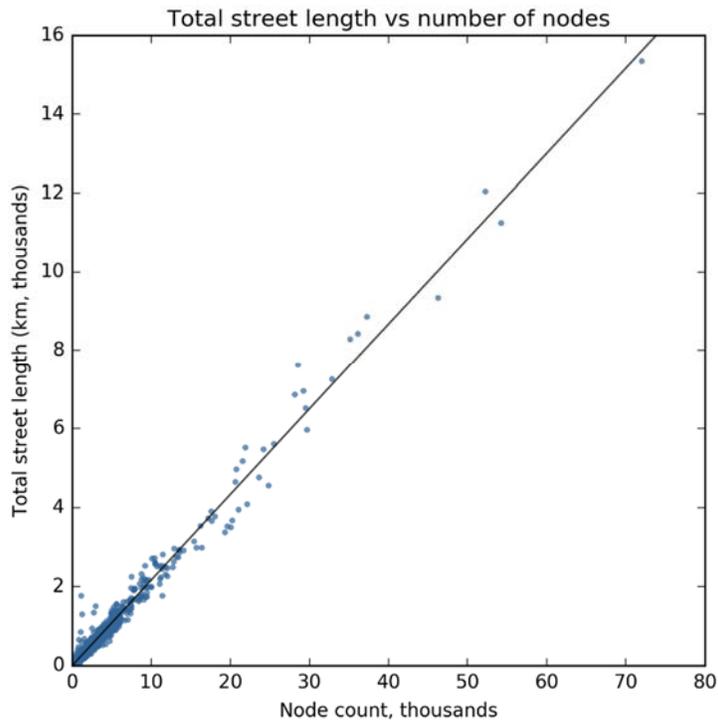

Figure 7.5. The linear relationship between total street length and number of nodes for 19,655 U.S. cities and towns.

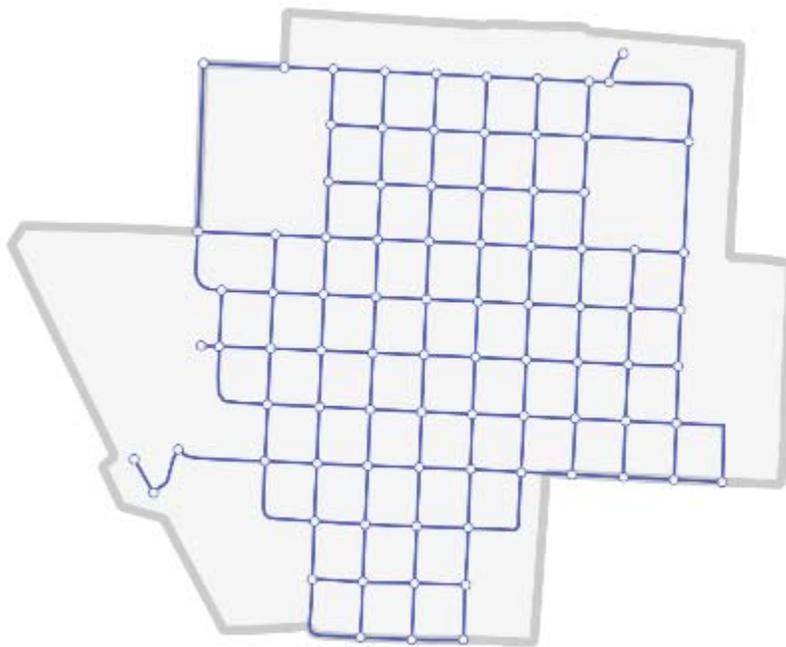

Figure 7.6. The town of Orleans, Nebraska exhibits a compact grid-like street network archetypal of towns across the Great Plains. Municipal extents are shown in gray.

Moreover, previous findings (e.g., Masucci et al. 2009) suggest the distribution of street segment lengths in an urban street network follows a power-law distribution. However, we find that these networks typically follow a lognormal distribution instead (see theoretical discussion in chapter 3). This finding makes sense: most street networks are *not* truly scale-free. While there are very few very long street segments (say, 1 km), more medium-length segments (say, 250 m), and many short segments (say, 80 m), there are very few very short segments (say, 10 m). This theoretical illustration suggests the lognormal distribution this analysis typically finds across municipal U.S. street networks. One exception, of course, lies in consistently sized, orthogonal grids filling a city's incorporated spatial extents. Such distributions are extremely peaked around a single value: the linear length of a grid block.

This analysis finds that such cities are not uncommon, particularly between the Mississippi River and the Rocky Mountains. These Great Plains states are characterized by a unique street network form that is both orthogonal and reasonably dense. The former is partly the result of topography (flat, plains terrain that allows idealized grids) and design history (platting and development during the late nineteenth century) that favor orthogonal grids, as discussed earlier. The latter seems to result from the fact that most towns across the Great Plains exhibit minimal suburban sprawl. Thus, the municipal boundaries snugly embrace the grid-like street network (e.g., Figure 7.6), without extending to accommodate the vast peripheral belt of twentieth century sprawl, circuity, and loops and lollipops that characterizes many cities in California that were settled in the same era but later subjected to substantial sprawl.

For example, if we measure connectedness in terms of *average number of streets per node* at the city-scale and then aggregate these cities by state (Table 7.4), we find that Nebraska, Kansas, South Dakota, Montana, North Dakota, Oklahoma, and Iowa have, in order, the highest medians (Figure 7.7). This indicates the most grid-like networks. If we measure density and connectedness in terms of *intersection density* at the city-scale and then aggregate these cities by state, we find that Rhode Island, Nebraska, New Jersey, Kansas, and Montana have, in order, the highest medians. We again see three Great Plains states near the top, alongside densely populated East Coast states. Nebraska also has the smallest block sizes (measured via the proxy of average street segment length) while the largest are concentrated in the deep South, upper New England, and Utah (Figure 7.8).

Figure 7.7. Contiguous U.S. states by median average number of streets per node in city and town street networks, colored from lowest/least-connected (pale yellow) to highest/most-connected (dark red).

Figure 7.8. Contiguous U.S. states by median average street segment length in city and town street networks, colored from longest/coarsest-grain (pale yellow) to shortest/finest-grain (dark red).

Municipal boundaries vary greatly in their extents around the built-up area. For example, while Rhode Island averages 56 intersections/km$^2$ in its cities and towns, Alaska averages only 1.3. This is an artifact of Alaska's municipal boundaries often extending thousands of square kilometers beyond the actual built-up area. In fact, Alaska has four *cities* (Anchorage, Juneau, Sitka, and Wrangell) with such large municipal extents that their land areas exceed that of the *state* of Rhode Island. These state-level aggregations of municipal-scale street network characteristics show clear variation across the country that reflect topography, economies, culture, planning paradigms, and settlement eras. But they also aggregate and thus obfuscate the variation within each state and within each city. To explore these smaller-scale differences, the following section examines street networks at the neighborhood scale.

| State | Intersect density (per km$^2$) | Avg streets per node | Avg circuity | Avg street segment length (m) |
|---|---|---|---|---|
| AK | 1.28 | 2.43 | 1.10 | 223.50 |
| AL | 9.70 | 2.64 | 1.07 | 190.81 |
| AR | 15.75 | 2.78 | 1.06 | 166.32 |
| AZ | 12.45 | 2.77 | 1.08 | 171.80 |
| CA | 32.58 | 2.74 | 1.07 | 143.79 |
| CO | 29.26 | 2.88 | 1.06 | 136.68 |
| CT | 28.05 | 2.70 | 1.07 | 165.87 |
| DC | 58.91 | 3.26 | 1.04 | 122.23 |
| DE | 25.30 | 2.80 | 1.06 | 127.80 |
| FL | 26.26 | 2.87 | 1.07 | 150.75 |
| GA | 15.25 | 2.78 | 1.07 | 177.50 |
| HI | 8.00 | 2.42 | 1.07 | 177.93 |
| IA | 24.08 | 3.02 | 1.04 | 129.36 |
| ID | 33.85 | 2.91 | 1.06 | 132.08 |
| IL | 29.02 | 2.93 | 1.05 | 137.77 |
| IN | 35.25 | 2.93 | 1.05 | 125.72 |
| KS | 43.94 | 3.14 | 1.04 | 124.39 |
| KY | 25.12 | 2.68 | 1.07 | 151.28 |
| LA | 17.14 | 2.79 | 1.06 | 162.62 |
| MA | 32.33 | 2.76 | 1.07 | 135.98 |
| MD | 28.67 | 2.79 | 1.07 | 133.69 |
| ME | 7.69 | 2.67 | 1.07 | 198.93 |
| MI | 20.93 | 2.90 | 1.05 | 153.50 |
| MN | 18.96 | 2.87 | 1.06 | 152.92 |

| | | | |
|---|---|---|---|
| MO | 29.87 | 2.89 | 1.06 | 138.29 |
| MS | 14.76 | 2.75 | 1.06 | 174.86 |
| MT | 38.94 | 3.11 | 1.04 | 126.89 |
| NC | 19.28 | 2.65 | 1.06 | 166.69 |
| ND | 34.28 | 3.07 | 1.04 | 123.93 |
| NE | 45.89 | 3.16 | 1.04 | 119.79 |
| NH | 12.22 | 2.69 | 1.10 | 175.88 |
| NJ | 44.98 | 2.88 | 1.04 | 130.79 |
| NM | 18.50 | 2.93 | 1.05 | 152.02 |
| NV | 13.86 | 2.77 | 1.07 | 147.35 |
| NY | 21.89 | 2.75 | 1.06 | 156.88 |
| OH | 25.23 | 2.80 | 1.05 | 142.08 |
| OK | 28.22 | 3.03 | 1.05 | 139.50 |
| OR | 35.08 | 2.69 | 1.06 | 121.18 |
| PA | 35.69 | 2.87 | 1.05 | 128.34 |
| RI | 56.23 | 2.86 | 1.05 | 110.35 |
| SC | 18.76 | 2.81 | 1.06 | 169.21 |
| SD | 32.01 | 3.12 | 1.04 | 130.75 |
| TN | 13.62 | 2.71 | 1.07 | 192.83 |
| TX | 23.85 | 2.92 | 1.05 | 160.44 |
| UT | 12.58 | 2.71 | 1.06 | 191.04 |
| VA | 25.18 | 2.63 | 1.08 | 145.65 |
| VT | 18.91 | 2.55 | 1.08 | 145.18 |
| WA | 28.71 | 2.75 | 1.06 | 134.02 |
| WI | 17.87 | 2.81 | 1.06 | 156.19 |
| WV | 28.45 | 2.67 | 1.08 | 136.57 |
| WY | 23.48 | 2.92 | 1.06 | 143.63 |

Table 7.4. Median values, aggregated by state plus DC, of selected measures of the municipal-scale street networks for every city and town in the U.S. For definitions and interpretation of these measures, see Table 5.1 in chapter 5 and section 4.4 in chapter 4.

# 7.6. Analysis of Neighborhood-Scale Street Networks

Thus far, we have examined every urban street network in the United States at the metropolitan and municipal scales. This analysis has focused on the complexity of the network in terms of density, resilience, and connectedness. While the metropolitan scale captures the emergent character of the wider region's complex system, and the municipal scale captures planning decisions made by a single city government, the neighborhood scale best represents the scale of individual urban design interventions into the urban form. Further, this scale more commonly reflects individual designs, eras, and paradigms in street network development than the "many hands, many eras" evolution of form at larger scales.

| | mean | $\sigma$ | min | median | max |
|---|---|---|---|---|---|
| Area (km$^2$) | 5.322 | 15.463 | 0.008 | 1.738 | 323.306 |
| Avg of the avg neighborhood degree | 2.598 | 0.436 | <0.001 | 2.670 | 3.632 |
| Avg of the avg weighted n'hood degree | 0.031 | 0.041 | <0.001 | 0.029 | 2.991 |
| Avg circuity | 1.080 | 0.411 | 1.000 | 1.044 | 24.29 |
| Avg clustering coefficient | 0.044 | 0.055 | <0.001 | 0.034 | 1.000 |
| Avg weighted clustering coefficient | 0.010 | 0.027 | <0.001 | 0.005 | 0.799 |
| Intersection count | 173 | 379 | 0 | 76 | 8371 |
| Avg degree centrality | 0.130 | 0.270 | 0.001 | 0.054 | 4.000 |
| Edge density (km/km$^2$) | 17.569 | 7.095 | 0.025 | 18.152 | 59.939 |
| Avg edge length (m) | 142.279 | 59.182 | 8.447 | 133.848 | 2231.331 |
| Total edge length (km) | 71.369 | 166.566 | 0.017 | 29.880 | 3563.409 |
| Proportion of dead-ends | 0.170 | 0.131 | <0.001 | 0.145 | 1.000 |
| Proportion of 3-way intersections | 0.559 | 0.146 | <0.001 | 0.574 | 1.000 |
| Proportion of 4-way intersections | 0.275 | 0.176 | <0.001 | 0.234 | 1.000 |
| Intersection density (per km$^2$) | 49.497 | 28.330 | <0.001 | 46.430 | 444.355 |
| Average node degree | 4.675 | 0.836 | 0.545 | 4.736 | 7.283 |
| $m$ | 5201 | 1185 | 1 | 217 | 27289 |
| $n$ | 208 | 459 | 2 | 90 | 9327 |
| Node density (per km$^2$) | 58.677 | 31.802 | 0.063 | 55.626 | 499.900 |
| Max PageRank value | 0.055 | 0.086 | <0.001 | 0.026 | 0.889 |
| Min PageRank value | 0.010 | 0.041 | <0.001 | 0.002 | 0.500 |
| Self-loop proportion | 0.007 | 0.034 | <0.001 | <0.001 | 1.000 |
| Street density (km/km$^2$) | 9.744 | 4.085 | 0.013 | 9.882 | 33.737 |
| Average street segment length (m) | 143.66 | 60.02 | 7.38 | 134.88 | 2231.33 |
| Total street length (km) | 40.049 | 93.987 | 0.009 | 16.248 | 1960.643 |
| Street segment count | 288 | 656 | 1 | 119 | 14754 |
| Average streets per node | 2.925 | 0.408 | 1.000 | 2.944 | 4.026 |

Table 7.5. Selected summary stats for all the neighborhood-scale street networks. For definitions and interpretation of these measures, see Table 5.1 in chapter 5 and section 4.4 in chapter 4.

Table 7.5 presents summary statistics for this data set. Compared to the summary statistics presented at the metropolitan scale (Table 7.1) and the municipal scale (Table 7.3), here we see much greater variance. This is expected, given the smaller network sizes at the neighborhood scale. A few neighborhoods have no intersections within their Zillow-defined boundaries, resulting in a minimum intersection density of 0 across the data set. Meanwhile, the small neighborhood of Cottages North in Davis, California has the highest intersection density in the country, 444/km², largely as an artifact of its small area as the denominator.

Nationwide, the typical neighborhood averages 2.9 streets per node, reflecting the prevalence of 3-way intersections in the U.S., discussed earlier. The median proportions of each node type are 14.5% for dead-ends, 57.4% for 3-way intersections, and 23.4% for 4-way intersections. The typical neighborhood averages 135-meter street segment lengths and 46.4 intersections per km². At the neighborhood scale (sample size of 6,857 in this analysis) we again find the same strong linear relationship between total street length and the number of nodes in a network ($r^2$=0.98), as seen in Figure 7.9, contradicting the small-sample findings of Strano et al. (2013).

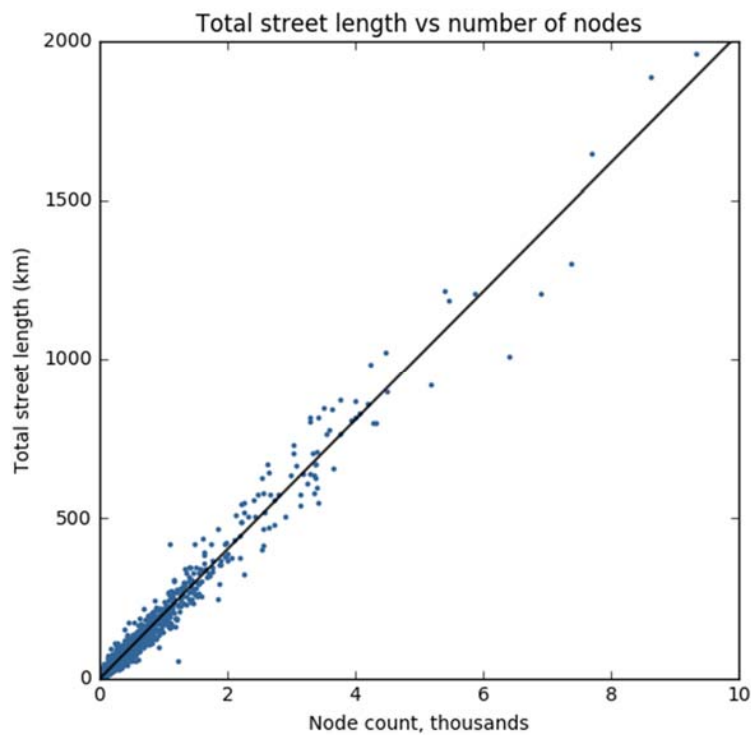

Figure 7.9. Relationship between total street length and number of nodes in neighborhood-scale street networks.

Due to the extreme values seen – resulting from the large variance in neighborhood size – we can filter the data set to examine only large neighborhoods (i.e., neighborhoods with area greater than the median value across the data set). In this filtered set, the five neighborhoods with the highest intersection density are all in central Philadelphia. Central neighborhoods are common at the top of this list, including Point Breeze, Philadelphia; Central Boston; Central City, New Orleans; Downtown Tampa; and Downtown Portland. The three neighborhoods with the lowest intersection density are on the outskirts of Anchorage, Alaska. In this filtered set, the neighborhoods with the greatest average number of streets per node tend to be older neighborhoods with orthogonal grids, such as Virginia Park, Tampa; Outer Sunset, San Francisco; and New Orleans' French Quarter. The neighborhoods with the *lowest* tend to be sprawling and often hilly suburbs far from the urban core, such as Scholl Canyon in Glendale, California or Sonoma Ranch in San Antonio, Texas.

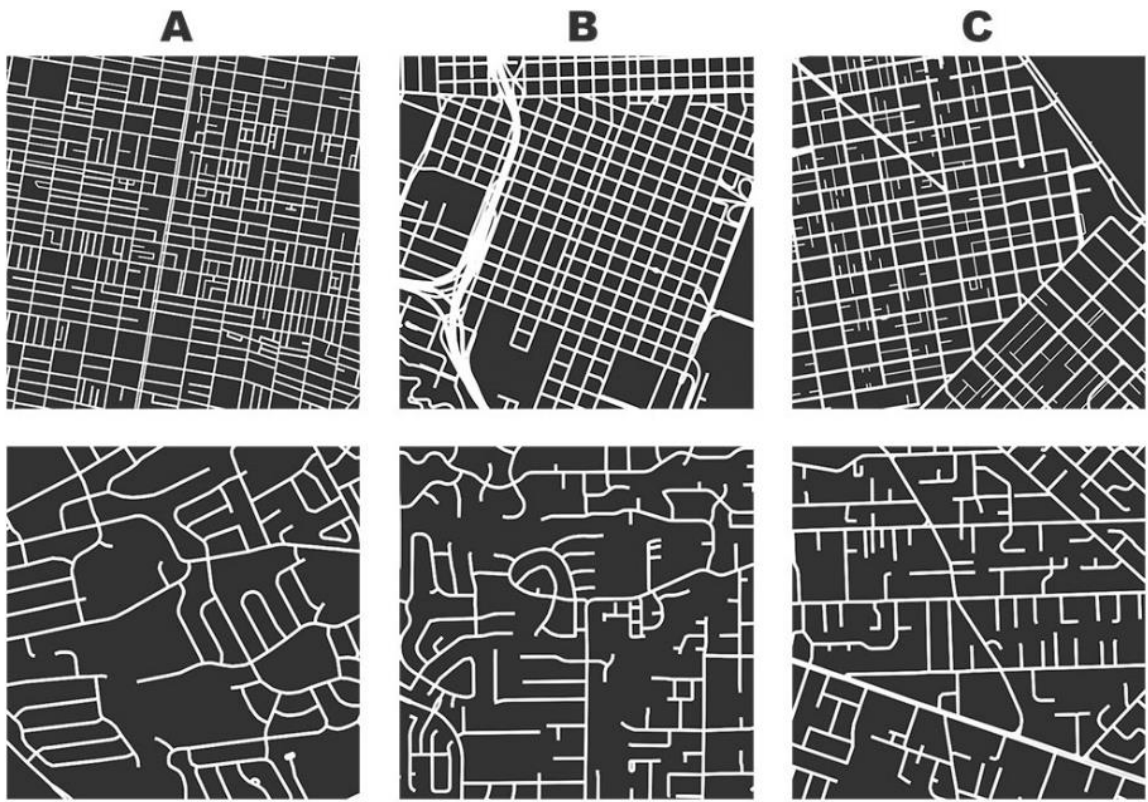

Figure 7.10. Square-mile comparisons of central cities and their suburbs. A: top, central Philadelphia; bottom, suburban King of Prussia. B: top, central Portland; bottom, suburban Beaverton. C: top, central San Francisco; bottom, suburban Concord.

For comparison, Figure 7.10 compares one square mile of the centers of Philadelphia, Portland, and San Francisco with one square mile of each of their suburbs. The connectedness and density of the central cities is clear, as is the disconnectivity of their suburbs. In fact, the suburbs have more in common with one another – despite being hundreds or thousands of miles apart – than they do with their central city neighbors, suggesting that land use and an era's prevailing design paradigm is paramount to geographical localism and regional context. The top row of Figure 7.10 represents an era of urban planning and development that preceded the automobile, while the bottom row reflects the exclusionary zoning and mid-late twentieth century era of automobility in residential suburb design – namely the "loops and lollipops" and the "lollipops on a stick" design patterns identified by Southworth and Ben-Joseph (1997).

## 7.7. Neighborhood-Scale Analysis of San Francisco

For clearer cross-sectional analysis, we single out the neighborhoods of San Francisco, California. This provides a more easily presented set of study sites as well as a more consistent geography to look across planning eras and design paradigms. Figure 7.11 shows these neighborhoods and underlying street networks. The Seacliff neighborhood, at the upper-left, features large parks, large lots, and large, expensive homes. It is one of San Francisco's master-planned residence parks developed in the aftermath of the 1906 earthquake. Inspired by the Garden Cities movement, Seacliff was designed to provide wealthy residents – through large lots, lush landscaping, and racially-restrictive covenants – a sense of suburban living while still being within the core city and in easy proximity to its downtown (Brandi 2014). Toward the upper-right, the central neighborhoods of San Francisco feature some of its oldest and densest development. Despite their hilly topography, planners draped an orthogonal grid street network across these neighborhoods, irrespective of terrain (Cole 2014).

In contrast, modern Diamond Heights was re-platted in the 1950s in harmony with its topography – under the auspices of the California Redevelopment Act – and has a street network that curves along its hillsides (Board of Supervisors 1955; see also Hu 2013; Siodla 2015). West of these hills, in the early twentieth century, urban planners opened the Twin Peaks Tunnel, the Sunset Tunnel, and the N-Judah line to expose large swaths of western San Francisco to development (Nolte 2009). Some of these new neighborhoods, such as the working-class Sunset District, were designed with a simple orthogonal grid and consistent, small lot sizes. Others, such as the wealthy St. Francis Wood (at the west end of West Twin Peaks) were designed as wealthy Garden Cities-inspired residence parks, in the mold of Seacliff. These histories, economic drivers, planning decisions, and design paradigms can still be seen inscribed in the urban form and its street network in Figure 7.11.

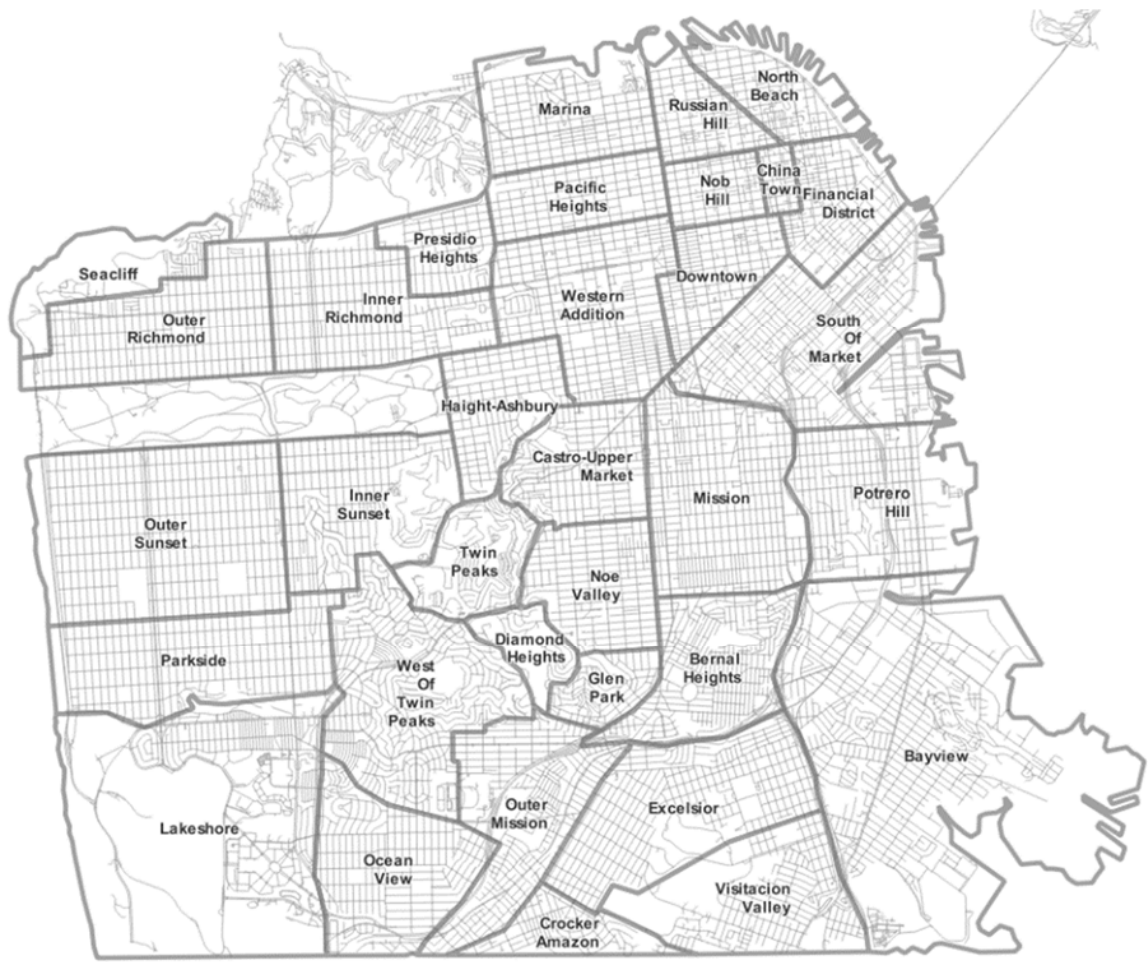

Figure 7.11. The neighborhoods and street network of San Francisco, California. Note that the Presidio and Golden Gate Park are excluded from any neighborhood's boundary, and thus from the quantitative analysis in this section.

| | Area (km²) | Avg circuity | Avg clustering coefficient | Intersect density (per km²) | Street density (km/km²) | Avg street length (m) | Streets per node |
|---|---|---|---|---|---|---|---|
| Bayview | 12.86 | 1.05 | 0.05 | 50.63 | 10.6 | 118.37 | 3.08 |
| Bernal Heights | 2.99 | 1.03 | 0.05 | 126.02 | 18.7 | 90.55 | 2.98 |
| Castro-Upper Market | 2.28 | 1.02 | 0.05 | 101.50 | 17.5 | 102.93 | 3.31 |
| Chinatown | 0.36 | 1.00 | 0.04 | 115.23 | 15.7 | 88.84 | 3.67 |
| Crocker Amazon | 1.20 | 1.02 | 0.05 | 90.27 | 13.4 | 97.93 | 3.05 |
| Diamond Heights | 0.89 | 1.19 | 0.16 | 48.11 | 8.5 | 130.97 | 3.08 |
| Downtown | 1.67 | 1.00 | 0.06 | 93.39 | 16.3 | 102.32 | 3.68 |
| Excelsior | 4.38 | 1.01 | 0.03 | 73.05 | 15.2 | 118.45 | 3.38 |
| Financial District | 1.81 | 1.01 | 0.07 | 66.80 | 11.9 | 104.73 | 3.64 |
| Glen Park | 0.96 | 1.03 | 0.01 | 109.85 | 16.7 | 102.27 | 3.15 |
| Haight-Ashbury | 2.00 | 1.02 | 0.04 | 75.95 | 14.3 | 116.50 | 3.54 |
| Inner Richmond | 3.57 | 1.01 | 0.02 | 72.76 | 14.7 | 114.51 | 3.54 |
| Inner Sunset | 3.49 | 1.07 | 0.06 | 62.16 | 13.9 | 133.73 | 3.33 |
| Lakeshore | 9.39 | 1.07 | 0.09 | 51.85 | 8.9 | 107.30 | 3.22 |
| Marina | 2.80 | 1.01 | 0.08 | 99.97 | 16.2 | 92.85 | 3.53 |
| Mission | 4.38 | 1.01 | 0.03 | 90.87 | 17.8 | 112.82 | 3.55 |
| Nob Hill | 0.95 | 1.00 | 0.01 | 81.71 | 17.2 | 119.93 | 3.73 |
| Noe Valley | 2.31 | 1.01 | 0.03 | 79.40 | 15.8 | 113.19 | 3.46 |
| North Beach | 1.76 | 1.04 | 0.05 | 61.28 | 12.1 | 112.24 | 3.21 |
| Ocean View | 3.42 | 1.03 | 0.04 | 76.65 | 15.6 | 121.03 | 3.22 |
| Outer Mission | 3.52 | 1.02 | 0.05 | 102.93 | 17.1 | 100.47 | 3.16 |
| Outer Richmond | 3.65 | 1.00 | 0.02 | 63.57 | 14.2 | 123.86 | 3.74 |
| Outer Sunset | 6.36 | 1.00 | 0.02 | 52.80 | 13.7 | 142.44 | 3.89 |
| Pacific Heights | 1.79 | 1.00 | 0.01 | 74.98 | 15.1 | 114.10 | 3.71 |
| Parkside | 4.03 | 1.00 | 0.02 | 56.86 | 13.3 | 132.24 | 3.66 |
| Potrero Hill | 3.73 | 1.02 | 0.03 | 64.17 | 14.5 | 128.20 | 3.25 |
| Presidio Heights | 1.26 | 1.02 | 0.07 | 78.64 | 14.6 | 114.17 | 3.38 |
| Russian Hill | 1.28 | 1.01 | 0.06 | 96.79 | 15.3 | 90.00 | 3.21 |
| Seacliff | 1.87 | 1.07 | 0.09 | 22.98 | 3.8 | 111.94 | 3.06 |
| South of Market | 5.68 | 1.03 | 0.05 | 67.42 | 14.9 | 133.00 | 3.20 |
| Twin Peaks | 1.76 | 1.17 | 0.10 | 61.52 | 13.3 | 143.77 | 2.80 |
| Visitacion Valley | 3.53 | 1.04 | 0.04 | 62.00 | 11.5 | 106.87 | 3.04 |
| West of Twin Peaks | 4.87 | 1.06 | 0.08 | 99.21 | 17.0 | 105.20 | 3.18 |
| Western Addition | 3.90 | 1.01 | 0.04 | 102.75 | 17.6 | 97.74 | 3.59 |

Table 7.6. Summary statistics for all San Francisco neighborhoods. For definitions and interpretation of these measures, see Table 5.1 in chapter 5 and section 4.4 in chapter 4.

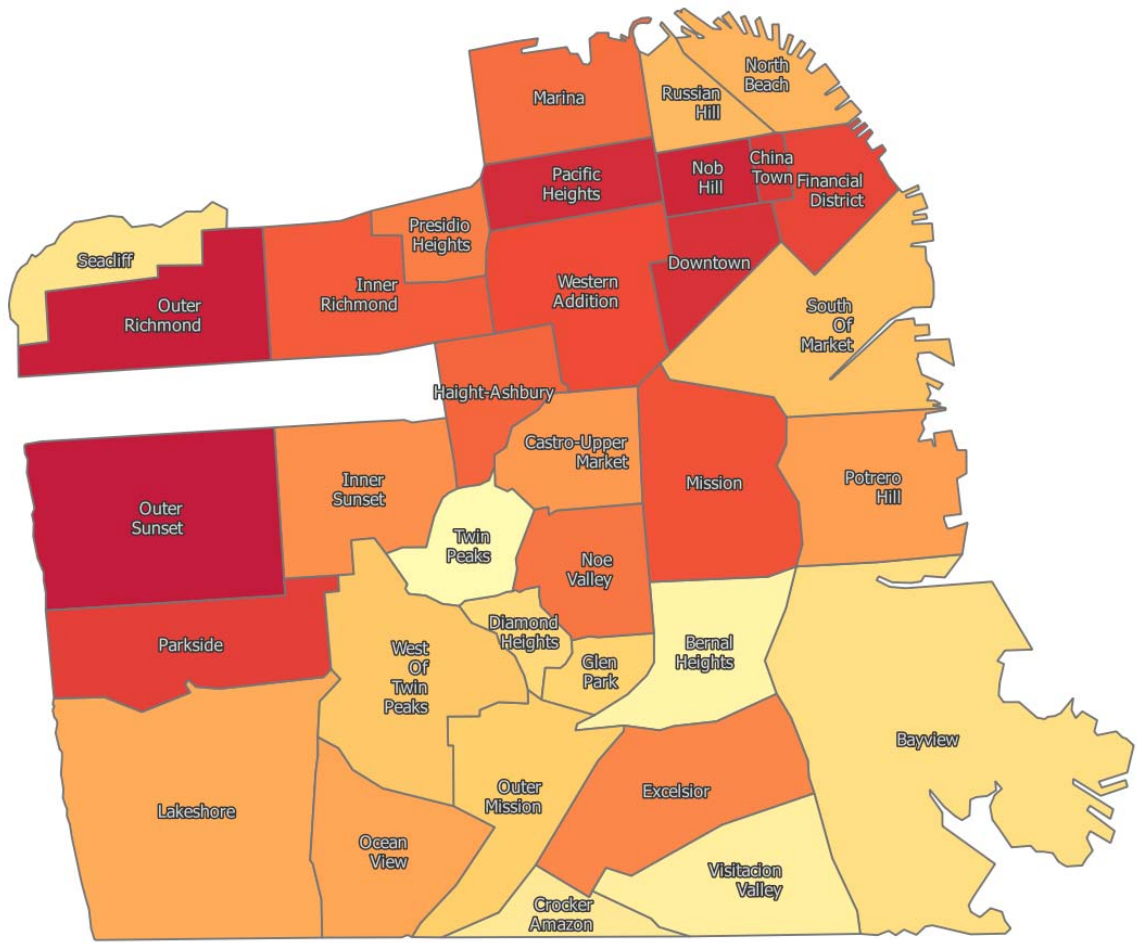

Figure 7.12. San Francisco neighborhoods colored by average number of streets per node (an indicator of connectedness), colored from lowest/least-connected (light yellow) to highest/most-connected (dark red). Compare with the street network detail in Figure 7.11.

We can use OSMnx to further examine the characteristics of these street networks. Looking at the summary stats for all San Francisco neighborhoods (Table 7.6), we see the orthogonal gridiron street networks of neighborhoods like Chinatown and Outer Sunset have average circuities of 1.0, indicating their street segments are no more circuitous than straight-line paths. Even the hilly street network of Nob Hill has an average circuity of 1.0, as its grid was laid out irrespective of the underlying terrain. However, the hilly neighborhoods of Twin Peaks and Diamond Heights have street networks that conform to the terrain, reflected in their circuity being 17% and 19%, respectively, greater than straight-line paths.

Diamond Heights and Seacliff have the lowest intersection densities (per km$^2$) and the lowest street densities (linear km per km$^2$), in part due to their circuitous, disconnected street networks, but also due to the fact that these small neighborhoods include large parks (Glen Canyon Park and Lands End, respectively) within their boundaries. Working-class Bernal Heights – with a street network designed by U.S. Army engineers in the 1860s (Mullins 2017) – has a dense mesh of streets and the highest intersection and street densities. Chinatown has the second highest intersection density and the Mission District has the second highest street density. This fine grain can be seen in the average street segment length: Chinatown, Bernal Heights, and Russian Hill have the shortest average street segment length, each approximately 90 meters. Twin Peaks has the longest (144 meters) due to its winding hilltop streets, followed by the Inner and Outer Sunset Districts due to their coarse-grain long blocks.

We can also examine these networks' complexity in terms of intersection types and connectedness. Figure 7.12 presents a map of San Francisco's neighborhoods by average number of streets per node (an indicator of connectedness), colored from lowest/least-connected to highest/most-connected (cf. Figure 7.11). Less connected neighborhoods such as Seacliff (3.1 streets per node, top left) and Twin Peaks (2.8 streets per node, center) immediately stand out. Similarly, the most connected neighborhoods by this measure, Outer Richmond (3.7 streets per node) and Outer Sunset (3.9 streets per node) lie directly south of Seacliff. These two neighborhoods are characterized by their typical 4-way intersections, as we saw in Figure 7.11.

A node's betweenness centrality represents the number of shortest paths in the network that pass through the node. The maximum betweenness centrality of any node in the network serves as a proxy for resilience: if a large number of shortest paths rely on a single node, the network is more prone to failure or inefficiency given a single point of failure. The Outer Sunset District has the lowest maximum betweenness centrality of any neighborhood – only 9.6% of all shortest paths pass through its most important node. By contrast, in Chinatown, 36% of shortest paths pass through its most important node, and in Twin Peaks it is 37%.

In Chinatown, this finding is the result of the small neighborhood comprising only a few streets *and* the fact that these streets are one-way, forcing paths through few routing options. In Twin Peaks, this is the result of the terrain and the disconnected network forcing paths through a small set of chokepoints that connect various sections of the network. In Ocean View, the neighborhood center has low betweenness centrality due to its disconnectedness: few shortest paths run through the center (Figure 7.13). However, in the Mission District, the neighborhood center has high betweenness centrality due to its orthogonal grid-like connectedness: many shortest paths run through the center. Note that as discussed in chapter 6, betweenness-centrality analyses for neighborhood-scale network subsets come with an important caveat: they only consider flows within the subset and thus demonstrate peripheral edge effects and artificially imposed centers (Gil 2016). Despite this caveat, these analyses do show the structure of the network *within* the encapsulated whole of a single neighborhood. Further, their limitations are ameliorated by the analyses presented earlier at broader scales, as well as by the average node connectivity analysis.

The average node connectivity is another proxy for network resilience. It represents the mean number of internally node-disjoint paths between each pair of nodes in the network. In other words, it indicates the expected number of nodes that must be removed to disconnect any randomly selected pair of non-adjacent nodes. Thus, a network with a higher average node connectivity is more resilient because more nodes must fail, on average, to disconnect a randomly selected pair of nodes. In San Francisco, the networks for Diamond Heights and Seacliff are not fully connected. Among neighborhoods with connected networks, Twin Peaks has the lowest average node connectivity: on average, only 1.05 nodes must be removed to disconnect a randomly selected pair of nodes (Table 7.7).

Conversely, Outer Richmond and Outer Sunset have the highest average node connectivity: on average, 2.8 and 3.2 nodes, respectively, must be removed to disconnect a randomly selected pair of nodes in these neighborhoods. This finding conforms to a qualitative assessment of the networks. As mentioned earlier, Twin Peaks has choke points – due to its terrain and disconnectivity – that separate sections of the network rely on to interface with each other. Outer Richmond and Outer Sunset, in contrast, are characterized by orthogonal grids with high connectedness and numerous fallback options should any single node fail.

Of note though, certain central orthogonal grid-like networks with many 4-way intersections such as Downtown, Chinatown, and the Financial District have surprisingly low average node connectivity: only 1.5, 1.3, and 1.6 nodes, respectively, must be removed to disconnect a randomly selected pair of nodes in these neighborhoods. This reflects the fact that these neighborhoods comprise primarily one-way streets. Although they have dense, highly connected networks, they are relatively un-resilient as they can be easily disconnected given that traffic cannot flow bi-directionally. These three neighborhoods also experience the greatest *increase* in average node connectivity if all the edges were converted to undirected: Chinatown's increases 87%, Downtown's increases 80%, and the Financial District's increases 75%. By contrast, the street network of Outer Sunset sees only a 6% increase due to it already comprising primarily bi-directional streets. This indicates that there are substantial possible resilience benefits in targeted conversion of one-way streets to bi-directional in Downtown, the Financial District, and Chinatown.

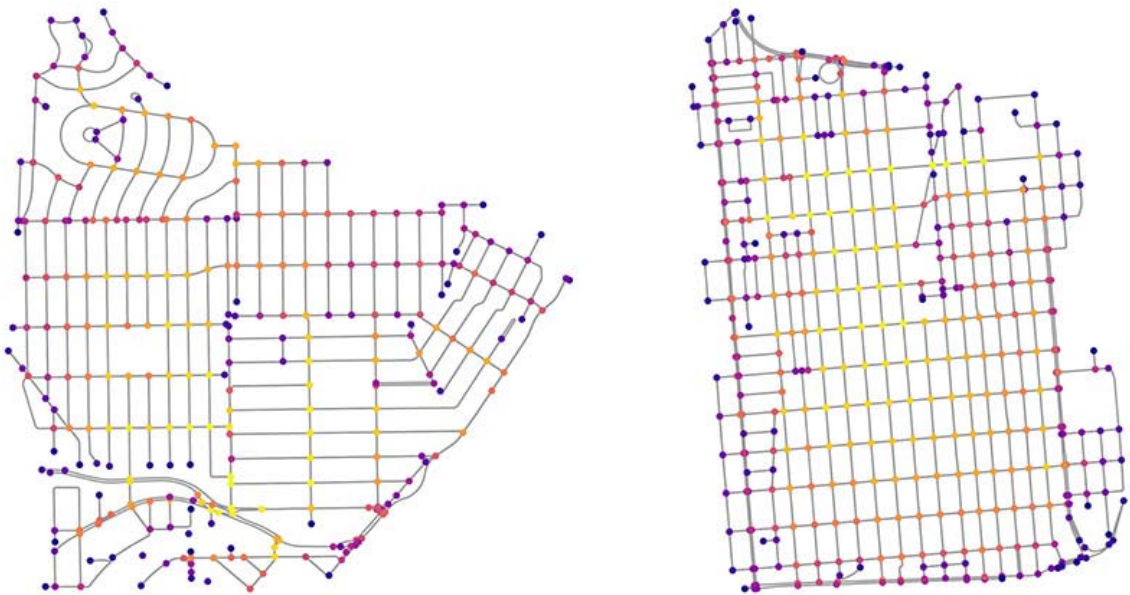

Figure 7.13. Relative node betweenness centralities for Ocean View (left) and the Mission District (right), colored from lowest (dark violet) to highest (light yellow) for flows originating from and traveling to nodes within the subset.

| Neighborhood | Max betweenness centrality | Avg node connectivity | Avg undirected node connectivity |
|---|---|---|---|
| Bayview | 0.161 | 1.740 | 2.076 |
| Bernal Heights | 0.199 | 1.609 | 1.942 |
| Castro-Upper Market | 0.241 | 1.978 | 2.447 |
| Chinatown | 0.362 | 1.349 | 2.526 |
| Crocker Amazon | 0.301 | 1.118 | 1.421 |
| Diamond Heights | 0.113 | 0.507 | 0.639 |
| Downtown | 0.207 | 1.533 | 2.755 |
| Excelsior | 0.206 | 2.315 | 2.482 |
| Financial District | 0.216 | 1.558 | 2.730 |
| Glen Park | 0.209 | 1.685 | 2.000 |
| Haight-Ashbury | 0.286 | 2.201 | 2.543 |
| Inner Richmond | 0.174 | 2.400 | 2.708 |
| Inner Sunset | 0.271 | 1.890 | 2.153 |
| Lakeshore | 0.194 | 1.461 | 2.329 |
| Marina | 0.186 | 2.231 | 2.719 |
| Mission | 0.146 | 2.158 | 2.775 |
| Nob Hill | 0.165 | 2.014 | 2.716 |
| Noe Valley | 0.218 | 2.187 | 2.392 |
| North Beach | 0.317 | 1.534 | 1.966 |
| Ocean View | 0.285 | 1.721 | 2.000 |
| Outer Mission | 0.279 | 1.625 | 1.941 |
| Outer Richmond | 0.204 | 2.771 | 3.073 |
| Outer Sunset | 0.096 | 3.167 | 3.360 |
| Pacific Heights | 0.150 | 2.502 | 2.895 |
| Parkside | 0.193 | 2.527 | 2.800 |
| Potrero Hill | 0.261 | 1.926 | 2.334 |
| Presidio Heights | 0.292 | 1.785 | 2.175 |
| Russian Hill | 0.254 | 1.838 | 1.982 |
| Seacliff | 0.150 | 0.949 | 0.987 |
| South of Market | 0.310 | 1.383 | 2.090 |
| Twin Peaks | 0.371 | 1.049 | 1.267 |
| Visitacion Valley | 0.262 | 1.498 | 1.830 |
| West of Twin Peaks | 0.232 | 1.915 | 2.311 |
| Western Addition | 0.213 | 1.996 | 2.848 |

Table 7.7. San Francisco neighborhoods, by indicators of resilience: maximum node betweenness centrality and average node connectivity (directed and undirected). For definitions and interpretation of these measures, see Table 5.1 in chapter 5 and section 4.4 in chapter 4.

## 7.8. Discussion

The pattern, texture, and grain of a street network is influenced by its development era, design paradigm, underlying topography, culture, and local economic conditions. The orthogonal grid is often traced back to Hippodamus of Miletus – whom Aristotle identified as the father of urban planning for his work in designing Piraeus, a port city outside of ancient Athens – but archaeologists have found its vestiges in earlier settlements around the world (Stanislawski 1946; Burns 1976; Cahill 2000; Paden 2001). In 1573, King Phillip II of Spain issued the Law of the Indies, systematizing how his colonists were to site new settlements and design street networks as rectilinear grids around a central plaza (Rodriguez 2005). In the United States, many east coast American cities planned their expansions around gridded street networks, including Philadelphia in 1682, Savannah in 1733, Washington in 1791, and New York in 1811 (Jackson 1985). During the Age of Enlightenment, Thomas Jefferson and associates drafted the Northwest Ordinances of 1784, 1785, and 1787 to divide the frontier into a regular grid of townships and parcels that guided urban development over the next 80 years (Jacobson 2002). The Jeffersonian ideal culminated in the Homestead Act of 1862, which divided the Midwest and Great Plains into square miles subdivided into 160-acre quarters for easy settlement, parceling, standardization, transportation, and a sense of urbane orderliness imposed on the frontier (Jackson 1985). The resulting street patterns guided by these doctrines, planning instruments, and land use policies can be examined through network analysis.

As defined in chapter 3, a complex spatial network is a network, embedded in space, that has a nontrivial topology. In other words, its structure and organization is neither fully regular nor fully random. Returning to the three categories of complexity discussed in chapter 4 (section 4.4.1), category I maximizes disorder and entropy, II maximizes a balance between structure and diversity, and III maximizes order. From this discussion, we developed a typology of measures for complexity in the context of urban design (section 4.5). Of the network measures, this empirical study has presented preliminary findings that particularly emphasize street network complexity in terms of density, resilience, and connectedness. Older, denser, and more self-organized networks, such as those at the heart of Boston or lower Manhattan are complex in terms of category I complexity. Sprawling, disconnected suburban neighborhoods rank low on all measures of complexity, with the exception that their high circuity can lend itself to disorder (i.e., category I). The orthogonal grid we see in the downtowns of Portland and San Francisco have high density (i.e., intersection and street densities), connectedness (i.e., average number of streets per node), and order (based on circuity and statistical dispersion of node types), but low resilience in the presence of one-way streets, measured by maximum betweenness centralities and average node connectivity increases when switching from one-way to bi-directional edges.

These latter neighborhoods are complex in terms of category II and category III complexity. Their connectedness, density, and orderliness balances at the midpoint embraced by category II. However, their extreme regularity of block sizes, streets per node, and orthogonality best represents category III. Recall from the discussion in chapter 3 (section 3.7) that a complex network is one with a topology that is neither fully regular nor fully random. Thus, the definition of a complex network stands in contradiction to the character of category III complexity, and raises an interesting question briefly introduced earlier: is maximum order not the antithesis of complexity? Are uniform, gridded street networks complex because of their topology, in spite of it, or not at all?

This is a broad question in the study of complexity, but one for which the urban planning literature may provide some insight. I propose that such orthogonal grid street networks *are* complex – not inherently because of their topology, but because of how that topology serves as a substrate that structures human dynamics. The orthogonal grid lends itself to platting and speculation (Hoyt 1933), navigation (Lynch 1960; Gell 1985; Sadalla and Montello 1989), the organization of symbolic, important, and memorable places (Lynch 1984; Kostof 1991; 1992), efficient human circulation (Institute of Transportation Engineers 2010), and resilience to decades of rapid technological change (Jackson 1985; Grant 2001). Taken in conjunction with density, streetscape, grain, and land use entropy, the *connectedness* of a grid supports route choice, convenience, and walkability – and in turn, the human dynamics of social mixing, activity, and encounter (Moudon and Untermann 1991; Jacobs 1995; Southworth and Ben-Joseph 1997; Guo 2009; Speck 2012; cf. Thisse 2014; Jabareen and Zilberman 2017). The complexity of the network, in all its various facets, influences and structures the *concordia discors* of complex human interactions and urban processes that run on it. This proposed link between structure and process forms a bridge between the discussion of dynamics in the first half of this dissertation and the analysis of form in the second, as contended in the discussion in chapter 3.

Another critical takeaway of this analysis is that *scale matters*. The median average circuity is lower across the neighborhoods data set than across the municipal set, which in turn is lower than across the urbanized areas set. Conversely, the median average number of streets per node is higher across the neighborhoods data set than across the municipal set, which in turn is higher than across the urbanized areas set. The median intersection density per km$^2$ is about 83% higher in the neighborhoods data set than in the municipal or urbanized areas sets. These findings make sense: the Zillow neighborhood boundaries focus on large, core cities with older and denser street networks. The municipal boundaries only include incorporated cities and towns – discarding small census-designated places and unincorporated communities. The urbanized area boundaries include far-flung sprawling suburbs.

The characteristics of an urban network for a city fundamentally depend on what *city* means: municipal boundaries, urbanized areas, or just certain neighborhoods? The first is a merely legal definition, but also captures the scope of city planning authority and decision-making for top-down interventions into a street network. The second captures a wider self-organized human system and its emergent built form, but tends to aggregate multiple non-homogeneous built forms together into a single unit of analysis. The third captures the nature of the local built form and lived experience, but at the expense of a broader view of the urban system and metropolitan-scale trip-taking. In short, multiple scales in concert provide planners and scholars a clearer view of the urban form and the topological and metric complexity of the street network than any single scale can.

We find a strong linear relationship, invariant across scales, between total street length, *L*, and the number of nodes, *n*, in a network that contradicts some previous findings in the literature that relied on small sample sizes and different geographic contexts. We also find that most networks empirically demonstrate a lognormal distribution of street segment lengths, contradicting some previous findings in the literature, as discussed earlier. However, we believe our empirical finding makes more sense theoretically *and* is supported by the large-sample data at multiple scales. An obvious exception to lognormal distribution lies in those networks that exhibit substantial uniformity across the entire network. At the neighborhood scale, examples include downtown neighborhoods with similarly consistent orthogonal grids, such as that of Portland, Oregon. At the municipal scale, examples include towns in the Great Plains that have orthogonal grids with consistent block sizes, platted at one time, and never subjected to sprawl.

These findings help to tell a story about the practice and history of planning. The spatial signatures of the Homestead Act, successive land use regulations, urban design paradigms, and planning instruments remain clearly visible today in these cities' urban forms and street networks due to path dependence. When comparing the median municipal street networks of each state, Nebraska has the lowest circuity, the highest average number of streets per node, the second shortest average street segment length, and the second highest intersection density for similar reasons. These preliminary findings point to how street networks across the Great Plains developed all at once but grew minimally afterwards – unlike, for instance, cities in California that were settled in a similar era but later subjected to sprawl.

This finding suggests future research could incorporate a temporal analysis that this present study does not do with its cross-sectional data. Returning to the typology of complexity measures in chapter 4 (section 4.5), this empirical analysis emphasized network *structure*. Expanding the study of complex urban form by examining the other types of complexity – and further linking structural complexity to the temporal complexity of dynamics and processes – lies ahead as critical future work.

In total, this chapter analyzed 497 urbanized areas' street networks, 19,655 cities' and towns' street networks, and 6,857 neighborhoods' street networks. These sample sizes were larger than those in any previous similar study. It looked at both metric and topological measures of the structure and complexity of these networks – particularly focusing on density, connectedness, and resilience. These preliminary empirical findings demonstrate the use of OSMnx as a new research tool. They suggest to urban planners new methods for acquiring and analyzing street network data, including new methods for evaluating network resilience and resilience gains with betweenness centralities and average node connectivities. Finally, this study has made all of these network datasets – for 497 urbanized areas, 19,655 cities and towns, and 6,857 neighborhoods – along with all of their attribute data and complexity measures available in an online public repository for other researchers to study and re-purpose (see Appendix).

# Chapter 8: Conclusion

# 8.1. Synopsis of the Dissertation

Cities are complex systems shaped through decentralized, bottom-up, self-organizing processes and top-down planning interventions. Humans both shape their environments (i.e., institutions, cultures, physical built form), and are in turn shaped by them. Cities – human ecosystems – comprise many interdependent and interacting components. Urban complexity is manifested through the self-organization of these components and the emergence of large-scale structure and characteristics. In particular, these components interact through networks – both virtual (Internet and telecommunication, flows of capital, etc.) and physical (street networks, rail networks, etc.).

This dissertation focused on street networks and their complexity, emphasizing density, resilience, and connectedness. These attributes influence the way an urban system's physical links can structure complex interactions, connections, and dynamics. This dissertation developed its theoretical framing over chapters 2 and 3. Chapter 2 introduced the background of the nonlinear paradigm by discussing systems, dynamics, self-similarity, and the nature of prediction in the presence of nonlinearity. These foundations set up the theoretical framework of complexity, cities, and the study of networks presented in chapter 3. Chapter 4 collated various measures of complexity from multiple research literatures into a typology of measures of the complexity of urban form, emphasizing the scale of urban design interventions. In particular, it presented several measures of network complexity and structure that were operationalized in chapters 5, 6, and 7.

Chapter 5 introduced OSMnx, a new tool to acquire, construct, correct, visualize, and analyze complex urban street networks. The current tool landscape does not offer a straightforward method of collecting street network data and conducting analysis consistently for any study site in the world. OSMnx fills this gap, allowing researchers to download and analyze street networks, building footprints, and elevation data. Chapter 6 applied OSMnx empirically in a small case study of street networks in Portland, Oregon to demonstrate the tool's usage. Chapter 7 then expanded the empirical application of OSMnx to a large study of 27,000 urban street networks at various scales across the United States. This analysis addressed current shortcomings in the research literature by using large sample sizes, clearly defined extents and topologies, and non-planar directed graphs. It presented wide-ranging empirical findings on U.S. urban form.

## 8.2. Summary of Key Contributions

### 8.2.1. Contribution to the Literature

Across its six substantive chapters, this dissertation makes various theoretical, methodological, and empirical contributions to the urban planning research literature including new software tools, typologies of measures, and empirical findings. This subsection addresses them in order.

Chapter 2 makes one methodological and one theoretical contribution to the literature. The former contribution is methodological as this chapter presents Pynamical, a new tool developed by this author to visualize and explore nonlinear dynamical systems' behavior. Comparable tools usually must be developed from scratch or rely on expensive commercial software such as MATLAB. Developing tools for exploring, understanding, and visualizing dynamical systems in Python makes them available to a much wider audience of systems analysts, researchers, and students. Pynamical provides a fast, simple, reusable, extensible, free, and open-source new means for exploring system behavior – particularly for the qualitative analysis of such systems in research and pedagogy. The latter contribution reviews the theory of nonlinearity and the qualitative analysis of nonlinear dynamical systems' behavior for an interdisciplinary body of urban scholars and planners. Most formal treatments of chaos and nonlinear dynamics in the scholarly literature are densely technical and geared toward an audience of mathematicians and physicists. Instead, chapter 2 offers a step-by-step introduction to dynamical systems, for a broad social science audience, to provide a strong and unambiguous footing for forays into complexity studies.

Similarly, chapter 3 offers a theoretical contribution to the planning literature by unpacking the key foundational concepts of complex systems and network science in a brief, straightforward manner targeted at planners. It argues that the interdisciplinary appeal of complexity in the social sciences has resulted in ambiguous terminology, internal inconsistencies, and overloaded concepts. This chapter provides explanatory examples of these ideas that are familiar to scholars and practitioners not already versed in the technical science of complexity. Complexity suggests how systems might self-organize structure, stability, and resilience. Through nonlinearity it problematizes certainty, prediction, and optimization. Finally, and most relevant to the present study, this chapter expounds the theory of networks and the methods of network analysis that form the foundation of the remaining chapters.

Chapter 4 unpacks the connections between the built form and the types and measures of its complexity. It primarily contributes a new typology of tools and metrics from different scientific disciplines to assess measures of complexity that apply to urban form and particularly to urban design's scale of intervention. In particular, the measures of network structure characterize the complexity of the circulation network in terms of density, resilience, and connectedness. Bridging between the earlier chapters and those that follow, these attributes influence the way an urban system's physical connections structure complex human interactions and dynamics. The analytical framework developed here is generalizable to empirical research of multiple neighborhood types and design standards.

Chapter 5 offers the primary methodological contribution of this dissertation – OSMnx, a new tool developed by this author to download, construct, correct, analyze, and visualize urban street networks using OpenStreetMap data (Boeing 2017h). Street network analysis in the urban planning literature suffers from challenges of usability, planarity, reproducibility, and sample sizes. To address these challenges, the primary methodological thrust of this study developed OSMnx to make the collection of data and creation and analysis of street networks simple, consistent, and automatable. OSMnx contributes five significant new capabilities for researchers: first, the automatic downloading of place boundaries and building footprints; second, the tailored and automated downloading and constructing of street networks from OpenStreetMap; third, the automatic correction and simplification of network topology; fourth, the ability to save street networks to disk as shapefiles, GraphML, or SVG files; and fifth, the ability to analyze street networks, calculate routes, visualize the networks, and calculate network metrics and statistics. These metrics and statistics include both those common in urban design and transportation studies, and metrics that measure the complexity of the network. Moreover, OSMnx allows street network data collection for anywhere in the world that OpenStreetMap has data. One can thus easily acquire street networks for places where such data might otherwise be inconsistent or prohibitively difficult to come by.

OSMnx enables researchers to download spatial data such as political boundaries, building footprints, elevation data, and complex street networks. It makes the acquisition, construction, and analysis of urban street networks easy, consistent, and reproducible for powerful and consistent research, transportation engineering, and urban planning and design. In turn, it allows researchers to ask new questions about network resilience, accessibility, connectedness and segregation, walkability, market responses to built form variables, and the performance of alternative street layouts. OSMnx is built on top of NetworkX, matplotlib, and geopandas for rich network analysis capabilities, easy and beautiful visualizations, and accelerated spatial queries with R-tree spatial indexing.

This chapter presents several street network and urban form figure-ground visualizations and discusses how they reveal various planning histories, processes, and instruments – including modernism's inversion of traditional urban spatial order. Finally, chapter 5 adapts measures from traditional network analysis to make them better-suited to accurately describing the physical form of street intersections and network connectivity. Namely, it adapts abstract nodes and directed edges into faithful representations of intersections, dead-ends, and physical streets. However, work remains to be done with accurately representing divided roads, as will be discussed in section 8.3.

In turn, chapter 6 presents a small case study to simply but plainly demonstrate the usage of OSMnx for research. It collects three small half-kilometer sections of the street network in different neighborhoods in Portland, Oregon to perform a cross-sectional analysis. This scale of analysis and sample size are small, but they provide straightforward examples to tie together the network concepts presented in chapter 3, the network measures presented in chapter 4, and the methodological tool presented in chapter 5. This chapter thus serves to knit these preceding threads together.

This chapter also presents empirical findings of these three street network sections in Portland, Oregon and uses these quantitative measures to compare and contrast these network sections. It first introduces these neighborhoods from a qualitative and historical perspective, then explores their comparative quantitative measures of network complexity and structure. During the high modernist era of the mid-twentieth century, Portland's planners converted over 40 miles of streets to one-way, including nearly the entirety of Downtown Portland. This chapter's findings identify significant chokepoints in the suburban network and demonstrates how there could be substantial gains in network resilience if one-way streets in the dense, orthogonal downtown were converted to two-way streets.

Following up from this small case study, chapter 7 presents the primary empirical contribution of this dissertation: a large multi-scale analysis of 27,000 street networks. The empirical literature on street network analysis is growing ever richer, but suffers from some limitations. First, sample sizes tend to be fairly small due to data availability, gathering, and processing constraints. Second, reproducibility is difficult when the dozens of decisions that go into analysis – such as spatial extents, topological simplification and correction, definitions of nodes and edges, etc. – are ad hoc or only partly reported. Third, and related to the first two, studies frequently oversimplify to planar or undirected primal graphs for tractability, or use dual graphs despite the loss of geographic and metric information. Fourth, the current landscape of tools and methods offers no ideal technique that balances usability, customizability, reproducibility, and scalability in acquiring, constructing, and analyzing network data.

This fourth limitation was addressed by introducing OSMnx and then demonstrating its use in a small case study of Portland, Oregon in chapters 5 and 6. Chapter 7 in turn addressed the first three limitations by conducting an analysis of street networks at multiple scales, with large sample sizes, with clearly defined network definitions and extents for reproducibility, and using non-planar, directed graphs. In particular, it examined urban street networks – represented as primal, non-planar, weighted multidigraphs with possible self-loops – through the framework of complexity developed in this dissertation, focusing on structure, density, connectedness, centrality, and resilience.

Most studies in the street network literature that conduct topological and/or metric analysis tend to have sample sizes ranging around 5 to 50 networks. This chapter instead conducted a large analysis of 27,000 urban street networks at multiple overlapping scales across the United States. Namely, it examined the street networks of every U.S. incorporated city and town, urbanized area, and Zillow-defined neighborhood. In total, the study presented in this chapter uses OSMnx to download, construct, and analyze 497 urbanized areas' street networks, 19,655 cities' and towns' street networks, and 6,857 neighborhoods' street networks. It uses these street networks to conduct four analyses: at the metropolitan scale, at the municipal scale, at the neighborhood scale, and a case study looking deeper at the neighborhood-scale street networks in the city of San Francisco.

Chapter 7 presented preliminary empirical findings that examine street network complexity through the lens of density, resilience, and connectedness. We find that the typical American urban area has approximately 26 intersections/km², 2.8 streets connected to the average node, 160m average street segment lengths, and a network that is 7.4% more circuitous than straight-line streets would be. The typical city has approximately 25 intersections/km², 2.9 streets connected to the average node, 145m average street segment lengths, and a network that is 5.5% more circuitous than straight-line streets would be. The typical Zillow neighborhood has approximately 46 intersections/km², 2.9 streets connected to the average node, 135m average street segment lengths, and a network that is 4.4% more circuitous than straight-line streets would be. At all three scales, 3-way intersections are by far the most prevalent intersection type across the U.S.

Downtown Portland's and San Francisco's orthogonal grids exhibit high intersection and street densities, high connectedness in terms of the average number of streets per node, and high order. However, they also exhibit low resilience (for traffic that must obey edge directedness) due to the significant presence of one-way streets. This resilience was characterized by the maximum betweenness centralities and the average node connectivity increases when switching from one-way to reciprocal edges in both directions. Returning once again to the categories of complexity discussed in chapter 4 (defined in the framework presented in section 4.4.1), these downtowns are complex in terms of category II and category III complexity.

However, older and more self-organized networks, such as those in lower Manhattan or central Boston, are more complex in terms of category I complexity – and possibly category II complexity in that their messiness is structured into a mesh of physical connections. However, with the exception of high circuity lending itself to category I disorder, the sprawling and disconnected suburban neighborhoods rank low on all categories of complexity. This discussion argued that street networks can be complex either inherently because of their form and topology, or indirectly through how that topology structures human dynamics. This bridge serves as a preliminary link between the theory of dynamics in the early chapters with the empirical analysis of form and structure in the latter chapters. However, some gaps remain for future research, as we will discuss shortly.

In chapter 7 we also find that *scale* is critically important in analyses of street networks. However, invariant to scale, we find a strong linear relationship between total street length and the number of nodes in a network. This provides new evidence that contradicts some previous findings in the literature that relied on purely theoretical models or small sample sizes. We also find that most networks empirically demonstrate a lognormal distribution of street segment lengths. An obvious exception to lognormal distribution lies in those networks that exhibit substantial uniformity across the entire network, such as the consistent orthogonal grid of downtown Portland, Oregon. At the municipal scale, towns in the Great Plains typically have orthogonal grids with consistent block sizes, platted at one time, and never subjected to sprawl. Comparing median street networks of each state, Nebraska has the lowest circuity, the highest average number of streets per node, the second shortest average street segment length, and the second highest intersection density, for similar reasons. Through path dependence – a hallmark of complex systems – the spatial signature of urban design paradigms and planning instruments remains etched into the urban form across the United States. Street networks and other structural and configurational aspects of the urban form possess the potential to help knit cities and people together – or segregate them into enclaves (cf. Holston 1989; Caldeira 1996a; 1996b; Chapple 2006).

In conclusion, an adaptation of chapter 2 is currently in press to be published as a journal article (Boeing 2016c). Chapters 5 and 6 have recently been conflated and submitted as a journal article, now under review (Boeing 2017c). Chapters 4 and 7 are each in various stages of the article manuscript preparation process (Boeing 2017a; 2017b). According to its repository statistics as of this writing, OSMnx has been downloaded over 10,000 times since its release (Continuum Analytics 2017). Finally, this study has made these network datasets (Boeing 2017f) and their attribute datasets (Boeing 2017g) available in a public online repository for other researchers to study and re-purpose (see Appendix).

### 8.2.2. Contribution to Planning Practice

This dissertation makes methodological and empirical contributions to urban planning practice. This subsection addresses them in order.

The first, and primary, contribution is OSMnx. This software can be easily used by planning practitioners and transportation engineers to collect and analyze urban street networks. The current tool landscape does not provide a simple, consistent, flexible, and scalable option for this type of work. Some existing tools, such as the Urban Network Analysis Toolkit discussed in chapter 5, require extremely expensive ArcGIS licenses and pre-existing local data sets. OSMnx is free, open-source, and can collect street network data from anywhere in the world via various flexible methods – particularly useful for planners working in the Global South. Furthermore, as OpenStreetMap data is publicly editable, local planners may add local data directly to OpenStreetMap, then use OSMnx to immediately construct the data into a graph-theoretic object for network analysis. This software makes it much easier to conduct common planning analyses such as intersection density, the spatial distribution of intersection types, average (linear) block length, connectedness, resilience, shortest-path routing, and accessibility. Street networks can be saved as ESRI shapefiles, GraphML files, or SVG files for urban design work with tools like Adobe Illustrator.

Additionally, the preliminary empirical findings in this dissertation suggest to planning practitioners several methods for making qualitative assessments of urban resilience more concrete from a quantitative perspective. In particular, we found that maximum betweenness centralities indicate brittle choke points in urban networks at various scales, particularly in the discussion of Portland in chapter 6 and the discussion of San Francisco in chapter 7. We also saw how the increase in average node connectivity when switching from a directed graph to an undirected graph representation of the street network can serve as an indicator of which areas would gain the greatest efficiency and resilience benefits from making one-way streets bi-directional.

Finally, this dissertation has drawn from the complexity sciences to embed the study of complex networks in the practice of urban planning and design. It has demonstrated various facets and measures of complexity relevant to the planning discipline. With the introduction and demonstration of OSMnx, it also has made these analytics readily available to planners without requiring technical and computational expertise. Instead, with just two or three lines of simple code, planners can download street networks anywhere in the world, analyze, and visualize them. For example, the figure-ground visualizations presented in chapter 5 help planners examine the physical outcomes of planning and informal urbanization. They also serve as a simple tool for communicating planned and emergent phenomena – such as density, connectedness, pattern, texture, scale, and grain – in a clear and immediate manner to laymen. Moreover, planners can use the complexity measures here as a rubric for more resilient and efficient circulation networks and better street investment decisions.

# 8.3. Future Research

### 8.3.1. Prospects and Challenges

The emerging methods of computational data science, visualization, network science, and "big data" analysis are drastically broadening the scope of urban design's traditional toolbox. Such methods may yield new insights and rigor in urban form/design research, but they may also promulgate the weaknesses of reductionism and scientism by ignoring the theory, complexity, and qualitative nuance of human experience crucial to urbanism. The tools we use shape the kinds of questions we can even ask about cities. A critical – but certainly not new – question remains in how such methods might stake out a nuanced place in research and practice. Over 60 years ago, Lévi-Strauss argued that "the confidence now shown by so many social scientists in mathematical models is due not so much to the results they themselves have secured by those methods as to the enormous assistance that mathematics has provided in other fields, and particularly physics" (1954, p. 583). Today, the dissemination of quantitative network science into the social sciences offers an exciting opportunity to study the dynamics and structure of cities and urban form. But paths forward must consider cities as uniquely *human* complex systems, inextricably bound up with politics, privilege, power relations, and planning decisions. Quantitative scholars and practitioners cannot assume a purely objective, dispassionate, technocratic stance without resorting to naïveté or disingenuity.

This dissertation developed measures and methods for analyzing the complexity of the urban form. Future work can use OSMnx to continue collecting and analyzing massive street network data sets, particularly to examine housing costs and accessibility, transportation and streetscape policy, one-way versus two-way conversions, public investments in bicycling infrastructure, and the connectivity and resilience of new neighborhoods developed according to the LEED-ND standards (U.S. Green Building

Council 2012; Boeing et al. 2014). Given the worldwide coverage of OpenStreetMap data and its use as a digital repository by NGOs and humanitarian organizations, OSMnx could be used to study cities in the Global South to examine urbanization, slums, and percolation of circulation networks into informal settlements. Ultimately, such studies could improve our understanding of the structure and character of cities through infrastructure – including roads, housing, and water – around the world to evaluate accessibility, connectivity, resilience, and spatial justice.

### 8.3.2. Future Methodological Research

Future methodological research includes developing a decision-tree algorithm to infer street width from various attributes of the network edge. Figure 5.9 utilized OpenStreetMap's roadway type attribute data, in conjunction with an ad hoc approximation of local street widths in each geography, to map from roadway type to an inferred width in meters to a width in pixels given the dimensions of the image raster. OpenStreetMap also has attributes for street width (though this is frequently null) and number of lanes (though lane widths vary from place to place). A better future algorithm might first check street width; if it is not present it would estimate the width from the number of lanes; if this too is not present, it would fall back on some default value for the roadway type.

OpenStreetMap represents divided roads as two side-by-side one-way edges (with different spatial locations and unique IDs) running in opposite directions – as does TIGER/Line and essentially every other set of GIS streets. This does, however, cause some inconsistencies with counts, measures, and visual representation. For example, the intersection of a divided road and another street essentially becomes two intersections – i.e., two separate one-way edges intersecting with the street. Collapsing divided roads into a single spatial entity is a nontrivial computational task. One might draw a new centerline in the middle of the two reciprocal edges, but this could 1) be difficult if the road curves and 2) disrupt the topology of its connections to other streets. It might be simpler to collapse divided roads into a single non-spatial entity. In other words, allow it to remain two separate spatial entities for visual representation, but de-duplicate counts for intersection densities, total street length, etc. by street *name*.

BOEING

This method, however, will run into difficulties if the same two roads intersect multiple times (e.g., due to curvature) or if multiple roads with the same name exist in a network. For instance, there could be multiple towns in an urban area's street network with a road called "Main Street." Further, data quality becomes paramount in such an approach. Any inconsistencies in naming, spelling, or abbreviation (e.g., "Avenue" versus "Ave" versus "Av") could cause mismatches that require natural language processing or machine learning for better matching. Thus, to guarantee the best results of any such collapsing of divided roads, the researcher may be best suited by performing the task manually, comparing each edge to satellite photography, and ground-truthing the results against street-level imagery such as Google StreetView or an in-person site visit. However, due to the enormous workload required for such a process, we believe OSMnx presents the best balance of accuracy, flexibility, universality, scalability, and reproducibility currently available to computational street network researchers, even given these caveats. Finally, future work can enable the downloading of additional geospatial objects other than streets and buildings – for instance, trees (to support streetscape studies) and water sources (to support informal settlement studies).

### 8.3.3. Future Empirical Research

Future empirical research includes comparing multiple network types across multiple scales. How do the measures of network complexity vary when examining driving networks versus walking networks versus bikeable networks at different spatial scales? This could also help clarify the nature of network resilience and connectedness for driving versus walking by further examining a place such as San Francisco, where connectivity was designed to favor pedestrians over cars. This study has provided some simple preliminary findings on identifying the benefits of one-way to two-way conversions by examining betweenness centralities and average node connectivities. Future research can expand this into a more thorough study of conversions to target specific areas for policy purposes.

Additionally, cluster analyses can be performed across network variables to develop a taxonomy of physical types along with city vintage and urban development, in terms of changes to population and spatial extents (see also Louf and Barthélemy 2014). Such clustering could explain the physical form resulting from certain histories, policies, and local conditions – furthering the findings visualized in Figure 7.10. Another approach might take advantage of classification algorithms from machine learning, such as random forests, support vector machines, naïve Bayes, $k$-nearest neighbors, and neural networks (Wu et al. 2008; Kelleher et al. 2015; Hastie et al. 2016). This approach would require manually labeling some subset of the data set to serve as training data, using it to fit the model, then making predictions with the remaining test data. Manually labeling observations in a large data set is a nontrivial task, but could leverage distributed microtask workforces, such as Amazon's Mechanical Turk. The same approach could be used to add additional qualitative nuance from Google StreetView imagery. Furthermore, OSMnx can easily output built form variables to use in hedonic studies of housing costs or analyses of travel behavior and VMT. It can quickly and automatically produce useful transportation-design variables such as intersection densities, street grades and elevations, node types, and (proxies of) block sizes, but also more advanced measures of centrality, clustering, resilience, and node importance.

For instance, future work can further probe the link between the network's structure and its dynamics by investigating human behavior in terms of measures of network complexity. OSMnx provides a rich basket of urban forms variables to model human dynamics. In general, the difficult linkages between dynamical complexity and structural complexity often beguile urban researchers and are ripe for exploration. The empirical analysis in chapter 7 emphasized network structure, but examining the other types of complexity discussed in chapter 4 – and further linking structural complexity of the urban form to the temporal complexity of human dynamics and processes – lies ahead as critical future work.

# Appendix: Software and Data

This appendix provides information on where to find this dissertation's software, data, and technical documentation. The software tools developed as part of this project are open-source and freely available online to download and install. Likewise, the various street network datasets and their morphological measures are available in a public online repository.

Chapter 2 introduced a new software tool, Pynamical, to explore nonlinear dynamical systems' behavior. Pynamical is a Python package for modeling, simulating, visualizing, and animating discrete nonlinear dynamical systems and chaos. It uses pandas, numpy, and numba for numerical simulation, and matplotlib for visualization and animation to explore system behavior. This study used version 0.1.1 to produce the models, simulations, and visualizations in chapter 2. Pynamical is free, open source software. This software and all the code used to develop these models and visualizations are available in a public repository on GitHub at https://github.com/gboeing/pynamical and its documentation is available online at https://pynamical.readthedocs.io/.

Chapter 5 introduced a new software tool, OSMnx, to download, analyze, and visualize street network and building footprint data from OpenStreetMap. OSMnx makes the collection of data and creation and analysis of street networks easy, consistent, scalable, and automatable for any study site in the world. It allows researchers and practitioners to download place boundaries and building footprints, download and construct street networks from OpenStreetMap, correct network topology, save street networks to disk in various file formats, and analyze and visualize street networks, including calculating routes and metric and topological measures. This study used version 0.1 to conduct its street network data collection and analyses. OSMnx is free, open source software. It is available in a public repository on GitHub at https://github.com/gboeing/osmnx and its documentation is available online at https://osmnx.readthedocs.io/.

The street network datasets compiled and analyzed in chapters 6 and 7, as well as the datasets of their various measures, have been made available in a public online repository for other researchers to study and re-purpose. This includes the street network shapefiles and GraphML files for every urbanized area, every city and town, and every Zillow neighborhood, as well as data tables collating the networks' various topological and metric measures that were calculated with OSMnx. They are all available from the Harvard Dataverse at: https://dataverse.harvard.edu/dataverse/osmnx-street-networks.